\documentclass[fleqn,usenatbib]{mnras}

\usepackage{newtxtext,newtxmath}
\usepackage[flushleft]{threeparttable}
\usepackage{booktabs}
\usepackage{adjustbox}
\usepackage{multirow}
\usepackage{xcolor}

\usepackage[T1]{fontenc}
\usepackage{longtable}
\usepackage{xr}

\DeclareRobustCommand{\VAN}[3]{#2}
\let\VANthebibliography\thebibliography
\def\thebibliography{\DeclareRobustCommand{\VAN}[3]{##3}\VANthebibliography}

\usepackage{graphicx}	
\usepackage{amsmath}	

\newcommand{\revised}[1]{{#1}}

\newcommand{\fr}[1]{{\color{black}{#1}}}
\newcommand{\lrb}[1]{\left(#1\right)}

\newcommand{\Msun}{\mathrm{M_{\sun}}}
\newcommand{\vpec}{v_\mathrm{pec}}
\newcommand{\vpecp}{v_\mathrm{pec, p}}
\newcommand{\masyr}{\mathrm{mas~yr^{-1}}}
\newcommand{\gaia}{{\it Gaia}}
\newcommand{\inte}{{\it INTEGRAL\,}}
\newcommand{\kpc}{\mathrm{kpc}}
\newcommand{\kms}{\mathrm{km~s^{-1}}}
\newcommand{\mcomp}{M_\mathrm{1}}
\newcommand{\mnoncomp}{M_\mathrm{2}}
\newcommand{\mtot}{M_\mathrm{tot}}

\newcommand{\porb}{P_\mathrm{orb}}
\newcommand{\porbi}{P_\mathrm{orb, SN}}
\newcommand{\porbp}{P_\text{orb, post-SN}}
\newcommand{\nodata}{$\cdots$}
\newcommand{\pkv}[1][]{v_\mathrm{pec #1}^{z=0}}

\newcommand{\fdiv}{f_\mathrm{diverge}}
\newcommand{\frej}{f_\mathrm{rej}}
\newcommand{\ferr}{f_\mathrm{err}}
\newcommand{\rhop}{\rho_\mathrm{p}}
\newcommand{\rhos}{\rho_\mathrm{s}}
\newcommand{\galacticheight}{|z|}
\newcommand{\pvalue}{$p$-value}
\newcommand{\dpost}{d_\mathrm{\varpi}}
\newcommand{\dlit}{d_\mathrm{lit}}
\newcommand{\ruwe}{{\tt ruwe}}
\newcommand{\loo}{w_\mathrm{loo}}

\newcommand{\newpsrmatch}{7}
\newcommand{\totnumwdlit}{125}
\newcommand{\finalsamplesize}{85}
\newcommand{\finalsamplesizewapprox}{89}
\newcommand{\numwithradioastrometry}{14}

\defcitealias{Liu06}{L06}
\defcitealias{Liu07}{L07}
\defcitealias{Kalogera96}{K96}
\defcitealias{Manchester05}{ATNF}
\defcitealias{Brandt95}{B95}

\title[Mass-dependent peculiar velocities]{Evidence for mass-dependent peculiar velocities in compact object binaries: Towards better constraints on natal kicks}

\author[Zhao et al.]{Yue Zhao$^{1}$\thanks{E-mail: zhao.yue@soton.ac.uk},
Poshak Gandhi$^{1}$,
Cordelia Dashwood Brown$^{1}$,
Christian Knigge$^{1}$,
Phil A. Charles$^{1}$, 
\newauthor
Thomas J. Maccarone${^2}$, 
and Pornisara Nuchvanichakul$^{1,3}$
\\
$^{1}$School of Physics \& Astronomy, University of Southampton, Highfield, Southampton SO17 1BJ, UK\\
$^{2}$Department of Physics \& Astronomy, Texas Tech University, Lubbock TX, 79410-1051, USA\\
$^{3}$Department of Physics and Materials Science, Faculty of Science, Chiang Mai University, Chiang Mai, Thailand 50200
}

\date{Revised 2023 June 27; in original form 2023 April 5}

\pubyear{2023}

\begin{document}
\label{firstpage}
\pagerange{\pageref{firstpage}--\pageref{lastpage}}
\maketitle

\begin{abstract}
We compile a catalogue of low-mass and high-mass X-ray binaries, some recently reported binaries that likely host a neutron star (NS) or a black hole (BH), and binary pulsars (a pulsar and a non-degenerated companion) that have measured systemic radial velocities ($\gamma$). Using \gaia\ and radio proper motions together with $\gamma$, we integrate their Galactic orbits and infer their post-supernova (post-SN) 3D peculiar velocities ($\pkv$ at Galactic plane crossing); these velocities bear imprint of natal kicks that compact objects received at birth.
With the sample totalling \finalsamplesize\ objects, we model the overall distribution of $\pkv$ and find a two-component Maxwellian distribution with a low- ($\sigma_v \approx 21\,\kms$) and a high-velocity ($\sigma_v \approx 107\,\kms$) component. A further comparison between distributions of binary subgroups suggests that binaries hosting high-mass donors/luminous companions mostly have $\pkv \lesssim 100\,\kms$, while binaries with low-mass companions exhibit a broader distribution that extends up to $\sim 400\,\kms$. We also find significant anti-correlations of $\pkv$ with binary total mass ($\mtot$) and orbital period ($\porb$), at over 99\% confidence. Specifically, our fit suggests $\pkv\propto \mtot^{-0.5}$ and $\pkv\propto \porb^{-0.2}$. Discussions are presented on possible interpretation of the correlations in the context of kinematics and possible biases. The sample should enable a range of follow-up studies on compact object binary kinematics and evolution.
\end{abstract}

\begin{keywords}
stars: neutron -- stars: black holes -- supernovae: general -- parallaxes -- proper motions
\end{keywords}



\section{Introduction}
The deaths of massive stars can deliver an extra acceleration to their remnant black holes \citep[BHs;][]{BrandtPodsSig1995, Jonker04} and neutron stars \citep[NSs;][]{Trimble1971, Lyne1982, Lyne94} at the instant of supernovae (SNe). This impulse, termed a natal kick or NK, connects SN mechanisms to the kinematics of compact objects, providing a way of constraining multiple aspects of SN models. For example, NKs can be a result of recoil due to baryonic ejecta at the instant of SNe \citep{Blaauw61, Brandt95, Nelemans99}, or, in later stages, due to anisotropicity in gravitational attraction of compact object by the asymmetric ejecta \citep{Janka13, Janka17} --- NKs could therefore be related to ejecta mass and/or morphology, and some observational support for this scenario exists \citep{Holland-Ashford2017}. The recoil momentum can also be contributed by asymmetric neutrino emission, typically in highly magnetised and hot nuclear matter of nascent NSs, where neutrino emissivity and opacity is dependent on its direction relative to local magnetic fields \citep{Chugai84, Dorofeev85, Arras99}. More fundamentally, causes of asymmetries in ejecta or neutrino emission are related to intrinsic hydrodynamical processes \citep[see][for a review]{Lai04}. 

Observations of NSs or BHs or binaries that host them have put some constraints on NKs. Pulsars have long been observed to have high space velocities (mean velocity of $\approx 300\,\kms$) relative to their presumed OB progenitors \revised{\citep{Gunn70, Lyne94, Hobbs05, Kapil23}}, suggestive of high NKs, although a relatively low-velocity (mean $\approx 120\,\kms$) population is also present \citep{Brisken2003, Verbunt17, Igoshev20}. While isolated NSs might be observed as radio pulsars, constraining the kinematics of isolated BHs can only be done in rare microlensing events \citep{Sahu22, Lam22, Andrews22} or via accretion from the interstellar medium \citep{Fujita1998, Maccarone2005}. X-ray binaries (XRBs) that host a NS or BH are advantageous targets in this regard, because the luminous companion can be used to constrain orbital parameters and track systemic motion; inference on NKs can therefore be made using observed peculiar motions of BH and NS XRBs in Galactic potential \revised{\citep[e.g.,][]{Mirabel01, Gandhi19, Atri19, Fortin22, ODoherty23}} and results of binary evolutionary models \revised{\citep[e.g.,][]{Fragos09, Kimball22}}.

In addition to the kinematic inferences of NKs, there have also been indirect constraints from other perspectives. The eccentric ($e\gtrsim 0.3$) orbits of Be X-ray binaries are indicative of large NKs \citep{Verbunt95, Brandt95, vandenHeuvel00}, but there might also be a separate population of long-orbit and low-eccentricity high-mass X-ray binaries (HMXBs) that argue against a large NK \citep{Pfahl02b, Podsiadlowski04, Knigge11}. The retention fraction of NSs in globular clusters is another indicator of NK magnitude: even a moderate number of NSs (in XRBs) is at odds with large NKs derived from pulsar observations as they are hard to be retained by typical escape velocities of globular clusters \citep[$\lesssim 50\,\kms$;][]{Pfahl02a, Smits2006,Ivanova08}.

In this work, we compile a more comprehensive sample of binaries that host a NS or BH (so they have experienced only one NK), calculate their peculiar motion in Galactic potential, and investigate possible dependencies of their observed peculiar velocities on binary total mass and other orbital parameters. The sample includes binary pulsars (a binary that hosts a pulsar and a non-degenerate companion or sometimes a white dwarf), low- and high-mass XRBs, and non-interacting binaries hosting NSs or BHs. In Sec \ref{sec:methods}, we describe sample compilation, inference of distances, and calculation of peculiar velocities; in Sec \ref{sec:results} we describe statistical methods used in our analyses and present the results; in Sec \ref{sec:discussion}, we discuss possible interpretation of and possible biases in the results; and in Sec \ref{sec:conclusion}, we summarise the results and draw conclusions.

\section{Methodology}
\label{sec:methods}
\subsection{The raw sample}
To construct our sample, we start by concatenating compilations of different types of BH or NS binaries. For XRBs, we use the BlackCAT catalogue\footnote{\url{https://www.astro.puc.cl/BlackCAT/index.php}} \citep[][and references therein]{Corral-Santana16} and the catalogues of low-mass and high-mass XRBs (LMXBs and HMXBs) compiled by \cite{Liu07} and \cite{Liu06}, respectively (\citetalias{Liu06} and \citetalias{Liu07}, hereafter), and a recent HMXB catalogue compiled by \citet{Fortin23}. Additionally, we include binaries that have been recently suggested to host BHs or NSs, including 2MASS J05215658+4359220 \citep[][J052156]{Thompson19}, AS 386 \citep{Khokhlov18}, HD 96670 \citep{Gomez21}, 2XMM J125556.57+565846.4 \citep[][J125556]{Mazeh22}, Gaia BH1 \citep{El-Badry23a}, Gaia BH2 \citep{Tanikawa22, El-Badry23b}, LAMOST J112306.9+400736 \citep[][J112306]{Yi22}, 2MASS J15274848+3536572 \citep[][J15274848]{Lin23}, 2MASS J06163552+2319094 \citep[][J06163552]{Yuan22}, and LAMOST J235456.76+335625.7 \citep[][J235456]{Zheng22}. Finally, we also include a sample of binary pulsars selected from the Australia Telescope National Facility (ATNF version 1.67) pulsar catalogue\footnote{\url{https://www.atnf.csiro.au/research/pulsar/psrcat/}} \citep{Manchester05}, using {\tt type(binary)} to select pulsars with one or more companions.

\subsection{Binaries excluded}
\fr{We exclude XRBs in globular clusters from our sample because their present-day velocities do not reflect the kinematics and SN kicks at birth. Most XRBs in globular clusters are thought to form via dynamical encounters between compact objects and normal stars \citep[e.g.,][]{Fabian75, Hills76, Camilo05}, after which the systems have lost track of the SN impact. Even if there are XRBs in globular clusters that formed through primordial binary evolution (similar to field XRBs), their systemic motion could have been greatly modified by encounters after formation. We also do not consider double pulsars as their progenitors might have experienced two supernovae, which makes them distinct from the single-degenerate binaries in our sample.} 


\subsection{Binary type names}
Binaries in our compilation are labelled based on their compact objects, NS or BH, and the mass of the non-degenerate companions, either low-mass (LM) or high-mass (HM) according to classification in the literature; for non-interacting binaries (NIs), we consider those with non-degenerate companion mass below $2\,\Msun$ as LM, while those above as HM. Concretely, there are BH or NS low-mass and high-mass XRBs (labelled: BH- or NS-LMXBs or HMXBs), low-mass or high-mass non-interacting (NI) binaries that host BH or NS (BH or NS-HMNI/LMNI). Pulsars are labelled separately from the XRBs by the mass of their non-degenerate companions (as LM or HM-PSR), although they are evolutionarily related to NS XRBs.

\subsection{Binary masses and orbital periods}
\fr{We also include available component masses and orbital periods from the literature. For binary component masses, we use the following notation: in any binary, $\mcomp$ is used to denote mass of the compact object (either a BH or a NS), while $\mnoncomp$ refers to the mass of the non-degenerate component: for XRBs it refers to the mass donor; and for NIs, this is the luminous component. $\mtot = \mcomp + \mnoncomp$ is the total mass of a binary. For binaries that host a NS, we assume a canonical NS mass ($\mcomp$) of $1.4\,\Msun$ when there are no constraints in the literature. We denote orbital period by $\porb$, as distinct from the value at time of SN ($\porbi$) used in the discussion sections (Sec \ref{sec:discussion}). A list of notations is presented in Appendix \ref{sec:notations}.}

\subsection{Identification with \gaia}
\label{sec:identification-with-gaia}
We cross-match literature positions of our sample with source positions in \gaia\ Data Release 3 \citep{GaiaCollaboration16, GaiaCollaboration22}. Since position uncertainties are not consistently estimated and some are unavailable, we use a relatively large search radius of 2\arcsec\ to find potential counterparts. Inevitably, some binaries in our list have multiple \gaia\ sources within the search radius; we therefore further compare their literature astrometric (available for some PSRs and XRBs) and photometric properties (of the optical counterparts) to those measured by \gaia, manually removing unlikely matches. 


The HMXBs are cross-checked with \cite{Fortin22}, who complemented \citetalias{Liu06} with the latest \inte catalogue \citep{Bird16}, compiling a list of $58$ unambiguous \gaia\ identifications with Galactic HMXBs. For LMXBs, \cite{Gandhi19} reported a list of $18$ matches dynamically confirmed BH systems with \gaia\ DR2, including at most $16$ BH-LMXBs; all their matches are confirmed by our search; a larger catalogue of \gaia\ (DR2) counterparts to LMXBs is summarised in \cite{Arnason21}, where a total of 32 LMXBs are identified. \cite{Jennings18} cross-matched the binary pulsars from the ATNF with \gaia\ DR2, giving a total $22$ binary pulsars that have confident \gaia\ counterparts; for this work, we exclude the extragalactic PSR J0045--7319 \citep{Storm04}, and the triple system PSR J0337+1715 \citep{Ransom14}. Finally, we also cross-check our compilation of NS-LMXBs and binary pulsars with \citet{ODoherty23}.

The identified Gaia {\tt source\_id} values are further cross-checked with sources in the \texttt{gaiadr3.nss\_two\_body\_orbit} table. This table contains sources whose astrometric solution is compatible with a two-body system, compared to the single-body solutions in the main table, some bright sources could have much more constrained parameters with a two-body model. Three of our sources are also in this table, including Gaia BH1, Gaia BH2, and J125556; however, only Gaia BH1 and BH2 have derived parallaxes and proper motions. We thus adopt two-body solution for these two sources, while using solutions in the main table for the other binaries. Parallaxes from the main \gaia\ table are corrected for zeropoint offset using the {\sc gaiadr3\_zeropoint} package\footnote{\url{https://gitlab.com/icc-ub/public/gaiadr3_zeropoint}} \citep{Lindegren21} except for Gaia BH1 and BH2 whose parallaxes are from non-single star solutions.

\subsubsection{New binary pulsar matches with \gaia\ DR3}
Our positional match confirms all confident counterparts in \cite{Jennings18} and finds $\newpsrmatch$ new matches; four of these matches have a literature systemic radial velocity so are included for our calculation and analyses (Sec \ref{sec:final_sample}). We therefore further investigate the four pulsars by comparing literature (from ATNF) and \gaia\ astrometry, and by searching for identification in literature. Here we summarise the confident matches. 

{\bf PSR J1306--4035} is a millisecond pulsar (MSP) that was discovered by the SUrvey for Pulsars and Extragalactic Radio Bursts \citep[SUPERB; ][]{Keane18}. \citet{Linares18a} discovered a variable optical source (SSS J130656.3--403522) within the radio error circle ($\approx 7\arcsec$); the source is modulated at the same period ($P\approx 1.1~\mathrm{day}$) as its X-ray flux, confirming its association with the pulsar. We found that the \gaia\ source 6140785016794586752 is within the $0.36\arcsec$ error circle (combining positional error in \cite{Linares18a} and in \gaia), and the \gaia\ G-band magnitude ($18.09$) is consistent with the V-band magnitudes reported in \citet[][17.4--18.4]{Linares18a}; we therefore consider this match genuine.

{\bf PSR J1622--0315} is a MSP discovered in Fermi unassociated sources \citep{Sanpa-Arsa16}. The orbital light curve is modulated at half of its orbital period ($\approx 0.16~\mathrm{days}$), varying between $R=19.1$ and $19.4$. There is only one \gaia\ source, 4358428942492430336, in our search radius. This source is $0.07\arcsec$ away from the nominal position and has a \gaia\  G-band magnitude of 19.21, consistent with the observed light curve in \citet{Sanpa-Arsa16}. This match is therefore confident. 

{\bf PSR J1653--0158} is a fast-spinning ($P_\mathrm{spin}=1.97~\mathrm{ms}$) and very compact ($P_\mathrm{orb}=75~\mathrm{min}$) binary pulsar that was recently discovered to be associated with the {\it Fermi Large Area Telescope} (LAT) source 4FGL J1653.6--0158 \citep{Nieder20}. The pulsar nature is confirmed by gamma-ray pulsations with Einstein@Home \citep{Allen13}. Consistency in astrometry with \gaia\ has been confirmed by \citep{Nieder20}. We therefore also include this pulsar in our calculation.

{\bf PSR J2039--5617} is another pulsar confirmed by detection of gamma-ray pulsations \citep{Clark21}, where a \gaia\ DR2 counterpart was confirmed. In our search, this counterpart is $\approx 8~\mathrm{mas}$ from the Gaia DR3 source 6469722508861870080, and has consistent and better constrained astrometry compared to DR2. This match is therefore also robust.


\subsection{Radio inteferometric and timing astrometry}
There are a few of XRBs and binary pulsars with literature astrometry measured by radio observations (interferometric or timing), which might be more constrained compared to \gaia. We adopt radio measurement in our analyses if \gaia\ astrometry is not available or if radio parallax or any of the proper motion component is more constrained than \gaia\ measurement.

Of the selected binaries, $\numwithradioastrometry$ have known radio parallaxes and/or proper motions, ten of which have more constrained radio astrometry. These binaries are summarised in Table \ref{tab:selected_sources_continued} and are indicated with superscripts that point to the corresponding references.

\subsection{The final sample}
\label{sec:final_sample}
The analysis in this work (Sec \ref{sec:KS_test}) is built on 3D physical velocities, which in turn requires input of proper motion, systemic radial velocity ($\gamma$), and distance (or parallax). Therefore, from the list of binaries that have \gaia\ counterparts, we are specifically interested in those with already established $\gamma$, as well as
\begin{itemize}
    \item well-constrained \gaia\ parallaxes ($\varpi$) such that $0<\sigma_\varpi/\varpi < 0.2$, or
    \item already constrained distances in literature.
\end{itemize}

Including the two BH-LMXBs that only have radio parallaxes and proper motions, i.e., GRS 1915+105 and MAXI J1836--194, our final sample contains a total of \finalsamplesizewapprox\ binaries. There are $4$ binaries that have known $\gamma$ values but has no well-constrained parallaxes or established distances, including 2S 0921--630, GX 1+4, BW Cir, and GX 339--4. BW Cir and GX 339--4 have lower limits on $d_\mathrm{lit}$ (Table \ref{tab:d_lit_table}, available online), while 2S 0921--630 and GX 1+4 only have approximate distances. The latter are also listed in Table \ref{tab:d_lit_table} but are not used in modelling the distance prior (Sec \ref{sec:d_post}). All of these \finalsamplesizewapprox\, sources are summarised in Table \ref{tab:selected_sources_continued}. 


\begin{table*}
    \centering
    \caption{Selected \gaia\ counterparts.}
    \begin{adjustbox}{max width=\textwidth}
    \begin{threeparttable}
        \begin{tabular}{llcccccc}
            \toprule
            Name & Class    & Gaia DR3 & $\alpha$ (ICRS) & $\delta$ (ICRS) & $\varpi$ & $\mu_\alpha\cos\delta$ & $\mu_\delta$ \\
                 &          &          & (h:m:s) & ($^{\circ}:^{\arcmin}:^{\arcsec}$) & (mas) & ($\masyr$) & ($\masyr$) \\
            \midrule
    	2S 0921--630               & NS-LMXB   & 5250451391698996992 & 09:22:34.669 & $-$63:17:41.288 & $ 0.10\pm 0.02$ & $-3.16\pm 0.03$ & $ 4.25\pm 0.03$ \\ 
		4U 1254--69                & NS-LMXB   & 5844938121765636864 & 12:57:37.167 & $-$69:17:19.018 & $ 0.18\pm 0.16$ & $-5.34\pm 0.19$ & $-1.32\pm 0.21$ \\ 
		Cen X-4                    & NS-LMXB   & 6205715168442046592 & 14:58:21.936 & $-$31:40:08.406 & $ 0.55\pm 0.13$ & $ 0.84\pm 0.15$ & $-55.68\pm 0.13$ \\ 
		Sco X-1                    & NS-LMXB   & 4328198145165324800 & 16:19:55.061 & $-$15:38:25.215 & $ 0.47\pm 0.02$ & $-7.19\pm 0.03$ & $-12.33\pm 0.02$ \\ 
		4U 1636--536               & NS-LMXB   & 5930753870442684544 & 16:40:55.587 & $-$53:45:05.081 & $ 0.29\pm 0.12$ & $-5.90\pm 0.16$ & $-8.33\pm 0.12$ \\ 
		Her X-1                    & NS-LMXB   & 1338822021487330304 & 16:57:49.809 & $+$35:20:32.361 & $ 0.14\pm 0.01$ & $-1.21\pm 0.01$ & $-7.86\pm 0.02$ \\ 
		MXB 1659--298              & NS-LMXB   & 6029391608332996224 & 17:02:06.534 & $-$29:56:44.286 & $ 0.22\pm 0.23$ & $-1.62\pm 0.52$ & $-1.51\pm 0.33$ \\ 
		4U 1700+24                 & NS-LMXB   & 4571810378118789760 & 17:06:34.503 & $+$23:58:18.550 & $ 1.87\pm 0.05$ & $-8.73\pm 0.04$ & $-5.57\pm 0.05$ \\ 
		GX 1+4                     & NS-LMXB   & 4110236324513030656 & 17:32:02.149 & $-$24:44:44.161 & $-0.02\pm 0.07$ & $-3.52\pm 0.08$ & $-1.99\pm 0.06$ \\ 
		4U 1735--444               & NS-LMXB   & 5955379701104735104 & 17:38:58.233 & $-$44:27:00.820 & $ 0.12\pm 0.10$ & $-3.47\pm 0.12$ & $-7.54\pm 0.07$ \\ 
		XTE J1814--338             & NS-LMXB   & 4039437121598306176 & 18:13:39.102 & $-$33:46:23.865 & $ 1.07\pm 0.48$ & $ 2.80\pm 0.47$ & $-3.99\pm 0.40$ \\ 
		2A 1822--371               & NS-LMXB   & 6728016172687965568 & 18:25:46.806 & $-$37:06:18.570 & $ 0.18\pm 0.03$ & $-9.15\pm 0.04$ & $-2.53\pm 0.03$ \\ 
		Ser X-1                    & NS-LMXB   & 4283919201304278912 & 18:39:57.544 & $+$05:02:09.433 & $-0.00\pm 0.27$ & $-1.14\pm 0.36$ & $-3.63\pm 0.27$ \\ 
		Aql X-1                    & NS-LMXB   & 4264296556603631872 & 19:11:16.055 & $+$00:35:05.786 & $ 0.21\pm 0.28$ & $-1.80\pm 0.58$ & $-5.13\pm 0.58$ \\ 
		Cyg X-2                    & NS-LMXB   & 1952859683185470208 & 21:44:41.152 & $+$38:19:17.061 & $ 0.10\pm 0.02$ & $-1.79\pm 0.02$ & $-0.32\pm 0.02$ \\ 
		IGR J00370+6122            & NS-HMXB   & 427234969757165952  & 00:37:09.632 & $+$61:21:36.479 & $ 0.29\pm 0.01$ & $-1.80\pm 0.01$ & $-0.53\pm 0.01$ \\ 
		2S 0114+650                & NS-HMXB   & 524924310153249920  & 01:18:02.694 & $+$65:17:29.841 & $ 0.22\pm 0.01$ & $-1.24\pm 0.01$ & $ 0.76\pm 0.01$ \\ 
		RX J0146.9+6121            & NS-HMXB   & 511220031584305536  & 01:47:00.210 & $+$61:21:23.662 & $ 0.37\pm 0.02$ & $-1.03\pm 0.02$ & $-0.08\pm 0.02$ \\ 
		LS I +61 303               & NS-HMXB   & 465645515129855872  & 02:40:31.664 & $+$61:13:45.590 & $ 0.40\pm 0.01$ & $-0.42\pm 0.01$ & $-0.26\pm 0.01$ \\ 
		X Persi                    & NS-HMXB   & 168450545792009600  & 03:55:23.076 & $+$31:02:45.010 & $ 1.67\pm 0.04$ & $-1.28\pm 0.05$ & $-1.87\pm 0.03$ \\ 
		XTE J0421+560              & NS-HMXB   & 276644757710014976  & 04:19:42.135 & $+$55:59:57.698 & $ 0.24\pm 0.01$ & $-0.47\pm 0.02$ & $-0.51\pm 0.01$ \\ 
		EXO 051910+3737.7          & NS-HMXB   & 184497471323752064  & 05:22:35.233 & $+$37:40:33.575 & $ 0.76\pm 0.03$ & $ 1.30\pm 0.04$ & $-4.00\pm 0.03$ \\ 
		1A 0535+262                & NS-HMXB   & 3441207615229815040 & 05:38:54.574 & $+$26:18:56.791 & $ 0.56\pm 0.02$ & $-0.59\pm 0.03$ & $-2.88\pm 0.02$ \\ 
		HD 259440                  & NS-HMXB   & 3131822364779745536 & 06:32:59.257 & $+$05:48:01.154 & $ 0.57\pm 0.02$ & $-0.03\pm 0.02$ & $-0.43\pm 0.02$ \\ 
		IGR J08408--4503           & NS-HMXB   & 5522306019626566528 & 08:40:47.780 & $-$45:03:30.138 & $ 0.45\pm 0.02$ & $-7.46\pm 0.02$ & $ 6.10\pm 0.02$ \\ 
		Vela X-1                   & NS-HMXB   & 5620657678322625920 & 09:02:06.854 & $-$40:33:16.751 & $ 0.51\pm 0.02$ & $-4.82\pm 0.01$ & $ 9.28\pm 0.02$ \\ 
		2FGL J1019.0--5856         & NS-HMXB   & 5255509901121774976 & 10:18:55.574 & $-$58:56:45.939 & $ 0.23\pm 0.01$ & $-6.45\pm 0.01$ & $ 2.26\pm 0.01$ \\ 
		Cen X-3                    & NS-HMXB   & 5337498593446516480 & 11:21:15.085 & $-$60:37:25.593 & $ 0.15\pm 0.01$ & $-3.12\pm 0.02$ & $ 2.33\pm 0.01$ \\ 
		1E 1145.1--6141            & NS-HMXB   & 5334851450481641088 & 11:47:28.546 & $-$61:57:13.389 & $ 0.12\pm 0.01$ & $-6.23\pm 0.01$ & $ 2.36\pm 0.01$ \\ 
		2S 1145--619               & NS-HMXB   & 5334823859608495104 & 11:48:00.007 & $-$62:12:24.877 & $ 0.49\pm 0.02$ & $-6.23\pm 0.02$ & $ 1.60\pm 0.02$ \\ 
		GX 301--2                  & NS-HMXB   & 6054569565614460800 & 12:26:37.548 & $-$62:46:13.294 & $ 0.28\pm 0.02$ & $-5.23\pm 0.02$ & $-2.07\pm 0.02$ \\ 
		1H 1253--761               & NS-HMXB   & 5837600152935767680 & 12:39:14.458 & $-$75:22:14.269 & $ 4.79\pm 0.03$ & $-27.34\pm 0.03$ & $-8.93\pm 0.04$ \\ 
		1H 1249--637               & NS-HMXB   & 6055103928246312960 & 12:42:50.236 & $-$63:03:31.112 & $ 2.30\pm 0.08$ & $-12.86\pm 0.07$ & $-3.68\pm 0.07$ \\ 
		1H 1255--567               & NS-HMXB   & 6060547335455660032 & 12:54:36.828 & $-$57:10:07.361 & $ 8.29\pm 0.12$ & $-28.39\pm 0.09$ & $-10.45\pm 0.11$ \\ 
		4U 1538--52                & NS-HMXB   & 5886085557746480000 & 15:42:23.352 & $-$52:23:09.643 & $ 0.18\pm 0.02$ & $-6.71\pm 0.01$ & $-4.11\pm 0.01$ \\ 
		4U 1700--37                & NS-HMXB   & 5976382915813535232 & 17:03:56.776 & $-$37:50:38.833 & $ 0.67\pm 0.03$ & $ 2.41\pm 0.03$ & $ 5.02\pm 0.02$ \\ 
		IGR J17544--2619           & NS-HMXB   & 4063908810076415872 & 17:54:25.272 & $-$26:19:52.588 & $ 0.42\pm 0.03$ & $-0.51\pm 0.03$ & $-0.67\pm 0.02$ \\ 
		LS 5039                    & NS-HMXB   & 4104196427943626624 & 18:26:15.064 & $-$14:50:54.378 & $ 0.53\pm 0.02$ & $ 7.43\pm 0.01$ & $-8.15\pm 0.01$ \\ 
		1H 2202+501                & NS-HMXB   & 1979911002134040960 & 22:01:38.206 & $+$50:10:04.629 & $ 0.90\pm 0.01$ & $ 2.36\pm 0.01$ & $-0.29\pm 0.01$ \\ 
		4U 2206+543                & NS-HMXB   & 2005653524280214400 & 22:07:56.229 & $+$54:31:06.356 & $ 0.32\pm 0.01$ & $-4.17\pm 0.02$ & $-3.32\pm 0.01$ \\ 
		A 0620--00                 & BH-LMXB   & 3118721026600835328 & 06:22:44.542 & $-$00:20:44.373 & $ 0.73\pm 0.12$ & $-0.44\pm 0.11$ & $-5.14\pm 0.10$ \\ 
		XTE J1118+480              & BH-LMXB   & 789430249033567744  & 11:18:10.764 & $+$48:02:12.208 & $ 0.18\pm 0.27$ & $-18.10\pm 0.16$ & $-6.69\pm 0.22$ \\ 
		GRS 1124--684              & BH-LMXB   & 5234956524083372544 & 11:26:26.588 & $-$68:40:32.900 & $ 0.19\pm 0.24$ & $-2.93\pm 0.24$ & $-1.39\pm 0.26$ \\ 
		MAXI J1305--704            & BH-LMXB   & 5843823766718594560 & 13:06:55.280 & $-$70:27:05.101 & $-0.50\pm 0.57$ & $-7.89\pm 0.62$ & $-0.16\pm 0.72$ \\ 
		BW Cir                     & BH-LMXB   & 5852296053620905472 & 13:58:09.710 & $-$64:44:05.292 & $ 1.28\pm 0.53$ & $-5.07\pm 0.63$ & $-2.11\pm 0.58$ \\ 
		4U 1543--475               & BH-LMXB   & 5986117923045619072 & 15:47:08.265 & $-$47:40:10.370 & $ 0.22\pm 0.06$ & $-7.54\pm 0.05$ & $-5.36\pm 0.05$ \\ 
		XTE J1550--564             & BH-LMXB   & 5884099221246440064 & 15:50:58.651 & $-$56:28:35.314 & $ 2.29\pm 2.58$ & $-6.49\pm 1.96$ & $-3.49\pm 1.74$ \\ 
		GRO 1655--40               & BH-LMXB   & 5969790961312131456 & 16:54:00.136 & $-$39:50:44.881 & $ 0.33\pm 0.05$ & $-4.33\pm 0.06$ & $-7.46\pm 0.05$ \\ 
		GX 339--4                  & BH-LMXB   & 5938658156448950528 & 17:02:49.381 & $-$48:47:23.166 & $ 0.21\pm 0.12$ & ${-3.95\pm 0.07}^a$ & ${-4.71\pm 0.06}^a$\\ 
		H 1705--250                & BH-LMXB   & 4112450294268643456 & 17:08:14.500 & $-$25:05:30.296 & $ 2.15\pm 1.67$ & $-6.97\pm 2.83$ & $-8.51\pm 1.38$ \\ 
		Swift J1753.5--0127        & BH-LMXB   & 4178766135477201408 & 17:53:28.291 & $-$01:27:06.313 & $ 0.17\pm 0.09$ & $ 0.99\pm 0.09$ & $-3.54\pm 0.09$ \\ 
		MAXI J1820+070             & BH-LMXB   & 4477902563164690816 & 18:20:21.939 & $+$07:11:07.184 & $ 0.40\pm 0.08$ & $-3.09\pm 0.09$ & $-6.29\pm 0.09$ \\ 
		MAXI J1836--194            & BH-LMXB   & \nodata             & 18:35:43.445$^b$ & $-$19:19:10.484$^b$ & \nodata & ${-2.30\pm 0.60}^c$ & ${-6.10\pm 1.00}^c$\\ 
		GRS 1915+105               & BH-LMXB   & \nodata             & 19:15:11.550$^d$ & $+$10:56:44.800$^d$ & \nodata & ${-2.86\pm 0.07}^e$ & ${-6.20\pm 0.09}^e$\\ 
		V404 Cyg                   & BH-LMXB   & 2056188624872569088 & 20:24:03.818 & $+$33:52:01.837 & ${ 0.42\pm 0.02}^f$& $-5.18\pm 0.08$ & $-7.78\pm 0.09$ \\ 
		HD 96670                   & BH-HMXB?  & 5337747731560719616 & 11:07:13.920 & $-$59:52:23.153 & $ 0.32\pm 0.03$ & $-6.75\pm 0.03$ & $ 1.69\pm 0.03$ \\ 
		V4641 Sgr                  & BH-HMXB   & 4053096388919082368 & 18:19:21.633 & $-$25:24:25.843 & $ 0.21\pm 0.03$ & $-0.78\pm 0.03$ & $ 0.43\pm 0.02$ \\ 
		SS 433                     & BH-HMXB?  & 4293406612283985024 & 19:11:49.562 & $+$04:58:57.751 & $ 0.13\pm 0.02$ & $-3.03\pm 0.02$ & $-4.78\pm 0.02$ \\ 
		Cyg X-1                    & BH-HMXB   & 2059383668236814720 & 19:58:21.671 & $+$35:12:05.684 & $ 0.47\pm 0.01$ & ${-3.80\pm 0.01}^g$ & ${-6.28\pm 0.02}^g$\\ 
		MWC 656                    & BH-HMXB?  & 1982359580155628160 & 22:42:57.298 & $+$44:43:18.209 & $ 0.51\pm 0.02$ & $-3.48\pm 0.02$ & $-3.16\pm 0.02$ \\ 
        \bottomrule
        \end{tabular}
    \end{threeparttable}
    \end{adjustbox}
    \label{tab:selected_sources_continued}
\end{table*}

\begin{table*}
    \centering
    \contcaption{Selected \gaia\ counterparts.}
    \begin{adjustbox}{max width=\textwidth}
    \begin{threeparttable}
    \begin{tabular}{llcccccc}
        \toprule
        Name & Class    & \gaia\ ID & $\alpha$ (ICRS) & $\delta$ (ICRS)            & $\varpi$ & $\mu_\alpha\cos\delta$ & $\mu_\delta$ \\
             &          &           &     (h:m:s)      & ($^{\circ}:^{\arcmin}:^{\arcsec}$) & (mas)    & ($\masyr$)             & ($\masyr$) \\
        \midrule
		PSR J0348+0432             & LM-PSR    & 3273288485744249344 & 03:48:43.640 & $+$04:32:11.458 & ${ 0.47\pm 0.47}^h$& ${ 4.04\pm 0.16}^h$ & ${ 3.50\pm 0.60}^h$\\ 
		PSR J1012+5307             & LM-PSR    & 851610861391010944  & 10:12:33.439 & $+$53:07:02.134 & $ 1.75\pm 0.29$ & ${ 2.64\pm 0.04}^i$ & ${-25.54\pm 0.04}^i$\\ 
		PSR J1023+0038             & LM-PSR    & 3831382647922429952 & 10:23:47.689 & $+$00:38:40.729 & ${ 0.73\pm 0.02}^j$& ${ 4.76\pm 0.03}^j$ & ${-17.34\pm 0.04}^j$\\ 
		PSR J1024--0719            & LM-PSR    & 3775277872387310208 & 10:24:38.661 & $-$07:19:19.819 & ${ 0.83\pm 0.13}^k$& ${-35.28\pm 0.01}^k$ & ${-48.23\pm 0.03}^k$\\ 
		PSR J1048+2339             & LM-PSR    & 3990037124929068032 & 10:48:43.416 & $+$23:39:53.392 & $ 0.50\pm 0.44$ & $-15.45\pm 0.35$ & $-11.62\pm 0.34$ \\ 
		XSS J12270--4859            & LM-PSR    & 6128369984328414336 & 12:27:58.717 & $-$48:53:42.708 & $ 0.49\pm 0.13$ & $-18.77\pm 0.11$ & $ 7.30\pm 0.09$ \\ 
		PSR B1259--63              & HM-PSR    & 5862299960127967488 & 13:02:47.637 & $-$63:50:08.632 & $ 0.46\pm 0.01$ & $-7.09\pm 0.01$ & $-0.34\pm 0.01$ \\ 
		PSR J1306--4035            & LM-PSR    & 6140785016794586752 & 13:06:56.272 & $-$40:35:23.399 & $ 0.34\pm 0.15$ & $-6.18\pm 0.14$ & $ 4.16\pm 0.11$ \\ 
		PSR J1311--3430            & LM-PSR    & 6179115508262195200 & 13:11:45.721 & $-$34:30:30.376 & $ 1.95\pm 0.97$ & $-6.13\pm 1.60$ & $-5.14\pm 0.68$ \\ 
		PSR J1417--4402            & LM-PSR    & 6096705840454620800 & 14:17:30.566 & $-$44:02:57.580 & $ 0.24\pm 0.05$ & $-4.76\pm 0.04$ & $-5.10\pm 0.05$ \\ 
		PSR J1431--4715            & LM-PSR    & 6098156298150016768 & 14:31:44.613 & $-$47:15:27.632 & $ 0.56\pm 0.13$ & $-11.82\pm 0.14$ & $-14.52\pm 0.15$ \\ 
		PSR J1622--0315            & LM-PSR    & 4358428942492430336 & 16:22:59.627 & $-$03:15:37.266 & $ 0.64\pm 0.30$ & $-13.18\pm 0.32$ & $ 2.30\pm 0.23$ \\ 
		PSR J1628--3205            & LM-PSR    & 6025344817107454464 & 16:28:07.002 & $-$32:05:49.021 & $ 0.69\pm 0.41$ & $-6.19\pm 0.47$ & $-21.43\pm 0.34$ \\ 
		PSR J1653--0158            & LM-PSR    & 4379227476242700928 & 16:53:38.053 & $-$01:58:36.895 & $ 1.79\pm 0.78$ & $-17.30\pm 0.95$ & $-3.24\pm 0.72$ \\ 
		PSR J1723--2837            & LM-PSR    & 4059795674516044800 & 17:23:23.179 & $-$28:37:57.595 & $ 1.11\pm 0.04$ & $-11.73\pm 0.04$ & $-24.05\pm 0.03$ \\ 
		PSR J1816+4510             & LM-PSR    & 2115337192179377792 & 18:16:35.934 & $+$45:10:33.847 & $ 0.22\pm 0.10$ & ${ 5.30\pm 0.80}^l$ & ${-3.00\pm 1.00}^l$\\ 
		PSR J2039--5617            & LM-PSR    & 6469722508861870080 & 20:39:34.968 & $-$56:17:09.276 & $ 0.51\pm 0.17$ & $ 3.86\pm 0.14$ & $-15.18\pm 0.12$ \\ 
		PSR J2129--0429            & LM-PSR    & 2672030065446134656 & 21:29:45.060 & $-$04:29:06.811 & $ 0.51\pm 0.07$ & $12.10\pm 0.07$ & $10.19\pm 0.06$ \\ 
		PSR J2215+5135             & LM-PSR    & 2001168543319218048 & 22:15:32.687 & $+$51:35:36.437 & $ 0.32\pm 0.23$ & $ 0.01\pm 0.24$ & $ 2.24\pm 0.24$ \\ 
		PSR J2339--0533            & LM-PSR    & 2440660623886405504 & 23:39:38.745 & $-$05:33:05.273 & $ 0.55\pm 0.18$ & $ 3.92\pm 0.20$ & $-10.28\pm 0.19$ \\ 
		J05215658                  & BH-HMNI   & 207628584632757632  & 05:21:56.591 & $+$43:59:21.899 & $ 0.41\pm 0.02$ & $-0.50\pm 0.02$ & $-3.46\pm 0.01$ \\ 
		J06163552                  & NS-LMNI   & 3425175331243738240 & 06:16:35.525 & $+$23:19:09.285 & $ 0.96\pm 0.02$ & $-1.43\pm 0.02$ & $-4.97\pm 0.01$ \\ 
		J112306.9                  & NS-LMNI   & 770431444010267392  & 11:23:06.909 & $+$40:07:36.366 & $ 3.19\pm 0.04$ & $-20.60\pm 0.03$ & $-24.28\pm 0.04$ \\ 
		J125556.57                 & NS-LMNI   & 1577114915964797184 & 12:55:56.553 & $+$56:58:46.498 & $ 1.73\pm 0.01$ & $-9.12\pm 0.01$ & $10.67\pm 0.01$ \\ 
		Gaia BH2                   & BH-LMNI   & 5870569352746779008 & 13:50:16.732 & $-$59:14:20.418 & $ 0.86\pm 0.02$ & $-10.48\pm 0.10$ & $-4.61\pm 0.06$ \\ 
		J15274848                  & NS-LMNI   & 1375051479376039040 & 15:27:48.437 & $+$35:36:57.376 & $ 8.48\pm 0.01$ & $-32.00\pm 0.01$ & $ 4.07\pm 0.01$ \\ 
		Gaia BH1                   & BH-LMNI   & 4373465352415301632 & 17:28:41.090 & $-$00:34:51.931 & $ 2.10\pm 0.02$ & $-7.70\pm 0.02$ & $-25.85\pm 0.03$ \\ 
		AS 386                     & BH-HMNI   & 2061686870220386816 & 20:10:54.197 & $+$38:18:09.284 & $ 0.17\pm 0.01$ & $-2.66\pm 0.01$ & $-4.88\pm 0.01$ \\ 
		J235456.76                 & NS-LMNI   & 2874966759081257728 & 23:54:56.765 & $+$33:56:25.697 & $ 7.86\pm 0.02$ & $23.21\pm 0.02$ & $-18.93\pm 0.01$ \\ 
        \bottomrule
    \end{tabular}
    \begin{tablenotes}
        \item $\varpi$ is the parallax after zeropoint correction except for Gaia BH1 and BH2.
        \item Classes: (NS)BH-(H)LMXB: (neutron star) black hole (high) low-mass X-ray binary; (NS)BH-(H)LMNI: non-interacting neutron star (black hole).
        \item binary with a (high) low-mass companions; (H)LM-PSR: binary pulsar with a (high) low-mass companion.
        \item References: $^a$: \cite{Atri19}, $^b$: \cite{Russell15}, $^c$: \cite{Russell14b}, $^d$: \cite{Dhawan00}, $^e$: \cite{Reid14b}, $^f$: \cite{Miller-Jones09}, $^g$: \cite{Miller-Jones21}, $^h$: \cite{Antoniadis13}, $^i$: \citetalias{Manchester05}, $^j$: \cite{Deller12}, $^k$: \cite{Reardon21}, $^l$: \cite{Stovall14}.
    \end{tablenotes}
    \end{threeparttable}
    \end{adjustbox}
\end{table*}


\subsection{The distances}
\label{sec:d_post}
We denote the distances estimated from trigonometric parallaxes by $\dpost$. For our estimation, we treat relatively nearby binaries ($\varpi > 2$) that have well-constrained ($\sigma_\varpi / \varpi < 0.2$) parallaxes separately from the other binaries. Since they have good-quality astrometric solution, we directly invert the parallax to get the distance, i.e., $\dpost = 1/\varpi$, and the uncertainty is simply propagated from the uncertainty in parallax, i.e., $\sigma_{\dpost} = \sigma_\varpi/\varpi^2$. For other binaries, we follow a Bayesian approach to infer distances from parallaxes. 

\subsubsection{Pulsars}
\cite{Lorimer06} derived a Galactic pulsar distribution in the vertical ($z$) and radial ($r$) directions (eq 10 and 11 in \citealt{Lorimer06}). Based on the distributions, \cite{Verbiest12} derived a distance prior for pulsars as:

\begin{equation}
    p(d|z, r) \propto d^2r^{1.9} \exp\left[-\frac{|z|}{E} - \frac{5(r - R_{\sun})}{R_{\sun}}\right].
    \label{eq:pulsar_prior}
\end{equation}
Here, $r$ is the Galactocentric distance projected in the Galactic plane, and $z$ is the vertical coordinate in cylindrical Galactocentric frame, which are both implicitly dependent on $d$. $E$ is the vertical scale height, for which
we choose $E=500~\mathrm{pc}$ as suggested by \cite{Verbiest12}.

We use a Gaussian likelihood \citep[see e.g.,][]{Bailer-Jones21} that centres on the inverse of the distance and spreads according to the uncertainty on parallax, i.e.,:

\begin{equation}
    p(\varpi | d, \sigma_\varpi) = \frac{1}{\sqrt{2\pi}\sigma_\varpi}\exp\left[-\frac{1}{2\sigma_\varpi^2}\left(\varpi - \frac{1}{d}\right)^2\right],
    \label{eq:gauss_likelihood}
\end{equation}
where the $\varpi$ and $\sigma_\varpi$ are parallax and the associated uncertainty, respectively. The posterior of distance is then given by Bayes' theorem:

\begin{equation}
    p(d | \varpi, \sigma_\varpi, z, r) \propto p(d|z, r) p(\varpi | d, \sigma_\varpi).
    \label{eq:d_posterior}
\end{equation}

\subsubsection{XRBs and others}
\label{sec:xrb_prior}
The remaining binaries are primarily XRBs and several NI binaries that likely host a BH or NS. To make a probabilistic estimate of their distances, we use the same likelihood as that for the pulsars (eq \ref{eq:gauss_likelihood}), and the exponential prior following \cite{Gandhi19}, which has the form of:

\begin{equation}
    p(d) = 
    \begin{cases}
         \dfrac{1}{2L^3}d^2\exp\left(-\dfrac{d}{L}\right) & r > 0 \\
         0 & r \leq 0
    \end{cases}.
    \label{eq:exponential_prior}
\end{equation}
This prior is characterised only by a scaling parameter $L$, which was found to be $2.17 \pm 0.12~\mathrm{kpc}$ by \cite{Gandhi19}. 

In this work, we use a more extended sample to model this prior. Since this prior is used for XRBs and non-interacting binaries with compact objects, we cross-check the list of sources in \citetalias{Liu06}, \citetalias{Liu07}, and BlackCAT for distances not inferred from parallaxes (denoted by $d_\mathrm{lit}$), including only values of $d_\mathrm{lit}$ that have reported uncertainties or lower limits. The final $d_\mathrm{lit}$ sample includes a total of $\totnumwdlit$ such binaries, which are summarised in Table \ref{tab:d_lit_table}. Most of the constrained values of $d_\mathrm{lit}$ have symmetric uncertainties, while for the asymmetric uncertainties, we use the central value ($(d_\mathrm{lit, low} + d_\mathrm{lit, up})/2$) for our modelling. For the constrained values, we draw 10,000 random distances for each binary in the sample as per a Gaussian distribution, with the standard deviation set to the maximum between upper and lower uncertainties; for the lower limits on $d_\mathrm{lit}$, we sample from a flat distribution between $d_\mathrm{lit, low}$ and $d_\mathrm{lit, low} + 5~\mathrm{kpc}$ --- this upper limit was used by \citet{Gandhi19} who also found that the result is not sensitive on choices of this limit. For each set of random distances, we compute the joint probability

\begin{equation}
    \mathcal{L}_n = \prod_i p(d_{i, n}|L),
    \label{joint_likelihood}
\end{equation}
where $p$ is given by eq (\ref{eq:exponential_prior}), $i$ indicates binaries in the sample, and $n$ indexes the random realisations. We search on a grid of $L$ for the value that maximises the joint probability --- considered a best-fitting scaling parameter. This procedure results in a distribution of $L$ with a median of $1.97^{+0.05}_{-0.05}~\mathrm{kpc}$ (errors correspond to 16th and 84th percentiles). 

With the constructed likelihood and priors, we make Bayesian distance inference for binaries with good astrometry ($\sigma_\varpi / \varpi<0.2$) and sample a total of $10^6$ random distances as per their posterior distributions using inverse transform sampling. In Table \ref{tab:vpec_and_distances}, we report medians and calculating a $68.27\%$ equal-tailed interval of the sampled distances for each binary. 

\subsection{The peculiar velocities}
\label{sec:vpec}
Peculiar velocity ($\vpec$) is the magnitude of 3-dimensional velocity relative to local Galactic rotation. For our calculation, we want to propagate uncertainties not only in source-specific parameters (parallax, proper motion, and radial velocity) but also in Galactic constants (Table \ref{tab:constants}) to the final $\vpec$; we therefore follow the formulation derived by \citet{Reid09}, assuming a Gaussian distribution for the source kinematic parameters, viz. proper motion and radial velocity, and for the Galactic constants. Distances are sampled from the posterior derived in Sec \ref{sec:d_post}; for binaries that have poorly constrained or invalid (zero or negative) \gaia\ parallaxes (Sec \ref{sec:final_sample}), we sample from a Gaussian distribution that centres at the distances inferred from dispersion measures from \citet{Yao17} with $20\%$ uncertainty as standard deviation for pulsars, and a Gaussian distribution that centres at the $d_\mathrm{lit}$ with symmetrised (averaged upper and lower error) uncertainty for the others (Table \ref{tab:d_lit_table}). In the last step of the calculation, the local Galactic rotation needs to be subtracted \citep{Reid09} from the Galactocentric 3D velocity; for this, we adjust the MWPotential2014 potential in the {\sc galpy} package\footnote{\url{http://github.com/jobovy/galpy}} \citep[version 1.8.3,][]{Bovy15} to adopt the Galactic constants used in this work (Table \ref{tab:constants}) and use it to find the local rotation velocity.

As mentioned in Sec \ref{sec:final_sample}, there are 4 binaries that have neither well-constrained \gaia\ parallaxes nor $d_\mathrm{lit}$, but two of them have lower limits on $d_\mathrm{lit}$ ($d_\mathrm{lit, low}$), i.e., BW Cir and GX 339--4; for these two binaries, we sample from the same distribution as that in modelling the exponential prior (Sec \ref{sec:d_post}), i.e., a uniform distribution between $d_\mathrm{lit, low}$ and $d_\mathrm{lit, low}+5~\kpc$, while for the other two, we sample from a Gaussian distribution that centres at the nominal $d_\mathrm{lit}$ with a $20\%$ uncertainty. $\vpec$ is calculated for these 4 binaries but should be treated with care. Indeed, physical velocities have a non-negligible dependence on distances; these 4 binaries are therefore not included in further statistical analyses.

With all the randomised parameters, we compute a sample of $10^6$ $\vpec$ values for each binary. Because the calculation uses present-day astrometric parameters, we refer to the calculated $\vpec$ as present-day peculiar velocity, denoted by $\vpecp$. We report median values and errors corresponding to $16$th and $84$th percentiles of $\vpecp$ in Table \ref{tab:vpec_and_distances}.

\subsection{Potential peculiar velocity at birth}
\label{sec:sec_pbpv}
The calculated $\vpecp$ could deviate from the $\vpec$ at birth depending on the binary's position and kinematics under the influence of Galactic potential. We therefore also calculate the potential birth peculiar velocities (denoted by $\pkv$) following the methods used by \cite{Atri19}, who referred to these as ``potential kick velocities''. In short, for each binary, we set up a total of 5,000 orbit instances by randomly drawing proper motions and $\gamma$ values from a Gaussian and distances from the posteriors calculated in Sec \ref{sec:d_post}. We then integrate these orbits of the binaries backward in time for $10~\mathrm{Gyr}$ using {\sc galpy} and record the $\vpec$ values every time the binary passes the Galactic plane (i.e., at $z=0$). The plane crossing epochs are found by linearly interpolating the $z(t)$ component of the orbit near the Galactic plane, which are then fed to the time-dependent $\vpec$ functions to obtain the $\pkv$ distributions. The resulting $\pkv$ are the potential $\vpec$ values at birth under the assumption that the binaries are born in the Galactic plane. Similar to $\vpecp$, the $\pkv$s and the errors corresponding to 16th and 84th percentiles are summarised in Table \ref{tab:vpec_and_distances}. 

Given the same uncertainties in all other parameters, $\pkv$ involves time as an extra variable that could complicate the final distributions over $\vpecp$. However, calculation of $\pkv$ also reduces possible effects of time by bringing all binaries to the same epoch (the birth time) of their evolution. Moreover, $\pkv$ relates to the epoch near the birth, while using $\vpecp$ as the proxy could underestimate the effect of binary evolution on the kinematics (e.g., mass loss due to stellar winds, unconservative mass transfer, etc.). 

\begin{figure}
    \centering
    \includegraphics[width=\columnwidth]{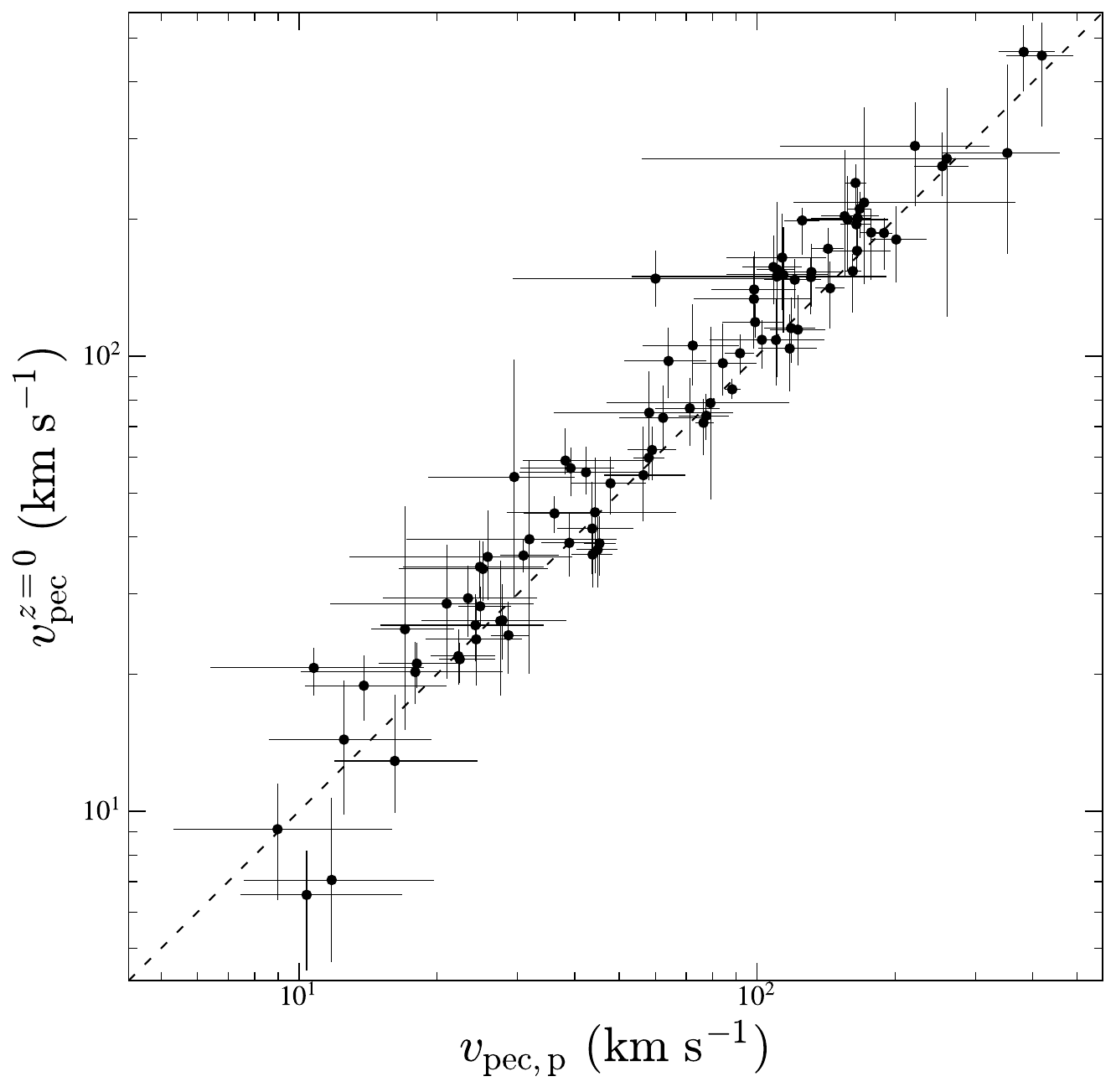}
    \caption{Median values of present-day peculiar velocity ($\vpecp$) vs. those of potential peculiar velocity at birth ($\pkv$) for binaries that have measured systemic radial velocity. The error bars correspond to 16th and 84th percentiles of the distributions. The dashed line indicates $\vpecp$=$\pkv$. Overall, $\pkv$ values agree with $\vpecp$ at $68\%$ level.}
    \label{fig:pbpv_vs_vpec}
\end{figure}

Our statistical analyses (Sec \ref{sec:pkv_distribution_analyses}, \ref{sec:KS_test}, \ref{sec:m_vs_v}, \ref{sec:porb_vs_pkv}) are performed on $\pkv$ and $\vpecp$, but in this paper, we only report results obtained with $\pkv$. Because $\vpecp$ are generally consistent with $\pkv$ at $68\%$ confidence level (Figure \ref{fig:pbpv_vs_vpec}), results using $\vpecp$ are in agreement with those using $\pkv$.

\begin{table*}
    \caption{Galactic and solar constants used in this work.}
    \centering
    \begin{tabular}{cccll}
    \toprule
    Name & Value & Unit & Ref & Description\\
    \midrule
    $E$                 & $0.5$           & $\kpc$ & [1] & Galactic scale height of pulsar distribution; used in eq (\ref{eq:pulsar_prior})\\
    $L$                 & $1.97\pm 0.05$  & $\kpc$ & [2] & Scaling parameter of the distance prior in eq (\ref{eq:exponential_prior})\\
    $R_{\sun}$          & $8.34 \pm 0.16$ & $\kpc$ & [3] & Galactocentric solar distance projected onto the plane \\
    $\Theta_{\sun}$     & $240 \pm 8$     & $\kms$ & [3] & Galactic rotation velocity at the location of the Sun \\
    $U_{\sun}$          & $10.7 \pm 1.8$  & $\kms$ & [3] & \\
    $V_{\sun}$          & $15.6 \pm 6.8$  & $\kms$ & [3] & Cartesian components of solar motion relative to the local standard of rest\\
    $W_{\sun}$          & $8.9 \pm 0.9$   & $\kms$ & [3] & \\
    \bottomrule
    \multicolumn{5}{l}{References: [1] \cite{Verbiest12}, [2] This work, [3] \cite{Reid14a}.}
    \end{tabular}
    \label{tab:constants}
\end{table*}

\begin{table*}
    \centering
    \caption{Binary properties: (1) Name. (2) Class. (3) Systemic radial velocity. (4) Distance (used in calculation). (5) Present-day peculiar velocity. (6) Potential peculiar velocity at birth. (7) Compact object mass. (8) Mass of non-degenerate companion. (9) Orbital period. }
    \begin{adjustbox}{width=0.98\textwidth}
    \begin{threeparttable}
        \begin{tabular}{llcllllll}
            \toprule
            Name & Class & $\gamma$ & $d$   & $\vpecp$ &  $\pkv$     & $\mcomp$ & $\mnoncomp$ & $\porb$ \\
                 &       & $(\kms)$ & (kpc) & $(\kms)$ & $(\kms)$ & $(\Msun)$  & $(\Msun)$ & $(\mathrm{d})$  \\
            (1)  & (2)   & (3)      & (4)   & (5)      & (6)  & (7) & (8) & (9) \\
            \midrule
				2S 0921$-$630        & NS-LMXB  & $  44.4\pm  2.4$ [1]   & $8.5^\ast$                  & $38.2^{+18.6}_{-7.4}$     & $58.9^{+10.4}_{-3.9}$     & $1.4^{+0.1}_{-0.1}$ [2]   & $0.3^{+0.0}_{-0.0}$ [2]      & $9.00$ [3]   \\
				4U 1254$-$69         & NS-LMXB  & $ 183.0\pm  3.0$ [4]   & ${13.0^{+3.0}_{-3.0}}^\dag$ & $155.3^{+29.0}_{-17.6}$   & $202.9^{+81.1}_{-52.9}$   & $1.5^{+0.3}_{-0.3}$ [4]   & $0.5^{+0.1}_{-0.1}$ [4]      & $0.16$ [5]   \\
				Cen X-4              & NS-LMXB  & $ 194.5\pm  0.2$ [6]   & ${1.4^{+0.3}_{-0.3}}^\dag$  & $418.7^{+72.1}_{-68.5}$   & $457.2^{+84.1}_{-137.8}$  & $1.9^{+0.4}_{-0.8}$ [6]   & $0.3^{+0.2}_{-0.1}$ [6]      & $0.63$ [6]   \\
				Sco X-1              & NS-LMXB  & $-113.8\pm  0.6$ [7]   & $2.2^{+0.1}_{-0.1}$         & $167.5^{+10.3}_{-9.8}$    & $210.6^{+19.0}_{-29.0}$   & $1.4^\ast$                & $0.4^\ast$ [8]               & $0.79$ [9]   \\
				4U 1636$-$536        & NS-LMXB  & $ -34.0\pm  5.0$ [10]  & ${6.0^{+0.5}_{-0.5}}^\dag$  & $164.7^{+12.6}_{-12.6}$   & $194.6^{+15.0}_{-25.7}$   & $1.4^\ast$                & $0.5^\ast$ [10]              & $0.16$ [10]  \\
				Her X-1              & NS-LMXB  & $ -65.0\pm  2.0$ [11]  & $7.1^{+0.8}_{-0.6}$         & $125.5^{+11.3}_{-10.9}$   & $198.5^{+13.2}_{-32.0}$   & $1.5^{+0.3}_{-0.3}$ [11]  & $2.3^{+0.3}_{-0.3}$ [11]     & $1.70$ [11]  \\
				MXB 1659$-$298       & NS-LMXB  & $ -49.0\pm 16.0$ [12]  & ${9.0^{+2.0}_{-2.0}}^\dag$  & $259.7^{+92.1}_{-203.6}$  & $271.4^{+116.9}_{-149.3}$ & $2.1^{+0.9}_{-0.9}$ [12]  & $0.6^{+0.2}_{-0.2}$ [12]     & $0.30$ [13]  \\
				4U 1700+24           & NS-LMXB  & $ -47.4\pm  0.1$ [14]  & $0.5^{+0.0}_{-0.0}$         & $39.2^{+9.5}_{-8.8}$      & $56.7^{+6.3}_{-7.5}$      & $1.4^\ast$                & $1.6^{+0.1}_{-0.2}$ [14]     & $4391.00$ [14]  \\
				GX 1+4               & NS-LMXB  & $-176.7\pm  0.2$ [15]  & $4.3^\ast$                  & $189.3^{+8.1}_{-8.3}$     & $186.3^{+15.2}_{-31.8}$   & $1.4^\ast$                & $<1.2$ [15]                  & $1160.80$ [15]  \\
				4U 1735$-$444        & NS-LMXB  & $-140.0\pm  3.0$ [10]  & ${8.5^{+1.3}_{-1.3}}^\dag$  & $60.0^{+62.1}_{-30.7}$    & $147.9^{+22.7}_{-19.7}$   & $1.4^\ast$                & $0.5^\ast$ [10]              & $0.19$ [10]  \\
				XTE J1814$-$338      & NS-LMXB  & $ -30.0\pm 20.0$ [16]  & ${8.0^{+1.6}_{-1.6}}^\dag$  & $171.3^{+194.8}_{-51.1}$  & $217.3^{+134.3}_{-73.2}$  & $2.0^{+0.7}_{-0.5}$ [16]  & $0.2^{+0.1}_{-0.1}$ [16]     & $0.18$ [16]  \\
				2A 1822$-$371        & NS-LMXB  & $ -44.0\pm  5.0$ [17]  & $5.9^{+1.3}_{-0.9}$         & $253.7^{+36.2}_{-33.2}$   & $261.3^{+48.3}_{-36.4}$   & $2.0^{+0.4}_{-0.4}$ [18]  & $0.5^{+0.1}_{-0.1}$ [18]     & $0.23$ [19]  \\
				Ser X-1              & NS-LMXB  & $  92.0\pm  4.0$ [4]   & ${7.7^{+0.9}_{-0.9}}^\dag$  & $110.0^{+30.2}_{-31.3}$   & $108.4^{+25.7}_{-22.3}$   & $1.4^\ast$                & $0.1^\ast$ [4]               & $0.09$ [4]   \\
				Aql X-1              & NS-LMXB  & $ 104.0\pm  3.0$ [20]  & ${5.2^{+0.8}_{-0.8}}^\dag$  & $42.3^{+13.2}_{-12.0}$    & $55.5^{+7.8}_{-5.9}$      & $1.4^\ast$                & $<0.8$ [20]                  & $0.79$ [20]  \\
				Cyg X-2              & NS-LMXB  & $-207.8\pm  0.3$ [21]  & $9.7^{+1.9}_{-1.4}$         & $164.0^{+9.0}_{-8.5}$     & $240.0^{+23.5}_{-48.1}$   & $1.7^{+0.2}_{-0.2}$ [21]  & $0.6^{+0.1}_{-0.1}$ [21]     & $9.84$ [21]  \\
				IGR J00370+6122      & NS-HMXB  & $ -77.3\pm  1.4$ [22]  & $3.4^{+0.1}_{-0.1}$         & $23.4^{+9.7}_{-8.1}$      & $29.4^{+5.3}_{-5.3}$      & $1.4^\ast$                & $10.0^{+5.0}_{-5.0}$ [22]    & $15.66$ [23]  \\
				2S 0114+650          & NS-HMXB  & $ -58.2\pm  0.5$ [24]  & $4.5^{+0.2}_{-0.2}$         & $22.4^{+4.4}_{-2.2}$      & $21.6^{+1.9}_{-2.4}$      & $1.4^\ast$                & $16.0^{+5.0}_{-5.0}$ [25]    & $11.59$ [26]  \\
				RX J0146.9+6121      & NS-HMXB  & $ -37.0\pm  4.3$ [27]  & $2.8^{+0.2}_{-0.2}$         & $11.8^{+7.9}_{-4.2}$      & $7.1^{+3.6}_{-2.4}$       & $1.4^\ast$                & $11.0^{+2.0}_{-2.0}$ [28]    & $330.00^\ast$ [27]  \\
				LS I +61 303         & NS-HMXB  & $ -41.4\pm  0.6$ [29]  & $2.5^{+0.1}_{-0.1}$         & $9.0^{+7.0}_{-3.6}$       & $9.1^{+2.3}_{-2.7}$       & $1.4^\ast$                & $12.5^{+2.5}_{-2.5}$ [30]    & $24.50$ [31]  \\
				X Persi              & NS-HMXB  & $   1.0\pm  0.9$ [32]  & $0.6^{+0.0}_{-0.0}$         & $17.9^{+9.8}_{-7.8}$      & $20.2^{+3.4}_{-3.0}$      & $1.4^\ast$                & $14.0^{+3.0}_{-3.0}$ [32]    & $250.30$ [33]  \\
				XTE J0421+560        & NS-HMXB  & $ -51.0\pm  2.0$ [34]  & $4.1^{+0.3}_{-0.2}$         & $18.1^{+4.6}_{-3.2}$      & $21.1^{+2.3}_{-2.5}$      & $1.4^\ast$                & \nodata                      & $19.41$ [34]  \\
				EXO 051910+3737.7    & NS-HMXB  & $ -20.5\pm  4.4$ [35]  & $1.3^{+0.1}_{-0.1}$         & $24.4^{+6.4}_{-5.5}$      & $23.8^{+5.4}_{-5.0}$      & $1.4^\ast$                & \nodata                      & \nodata \\
				1A 0535+262          & NS-HMXB  & $ -30.0\pm  4.0$ [36]  & $1.8^{+0.1}_{-0.1}$         & $44.9^{+4.6}_{-4.5}$      & $37.5^{+6.7}_{-6.5}$      & $1.6^{+0.6}_{-0.6}$ [36]  & $7.5^{+2.5}_{-2.5}$ [36]     & $111.00$ [37]  \\
				HD 259440            & NS-HMXB  & $  36.9\pm  0.8$ [38]  & $1.8^{+0.1}_{-0.1}$         & $10.4^{+6.4}_{-2.9}$      & $6.5^{+1.6}_{-2.1}$       & $1.4^\ast$                & $16.1^{+2.9}_{-2.9}$ [39]    & $308.00^\ast$ [38]  \\
				IGR J08408$-$4503    & NS-HMXB  & $  15.3\pm  0.5$ [40]  & $2.2^{+0.1}_{-0.1}$         & $38.9^{+6.8}_{-5.1}$      & $38.8^{+6.1}_{-6.1}$      & $1.4^\ast$                & $33.0^\ast$ [40]             & $9.54$ [40]  \\
				Vela X-1             & NS-HMXB  & $  -3.4\pm  1.0$ [41]  & $2.0^{+0.1}_{-0.1}$         & $59.0^{+7.7}_{-6.8}$      & $62.1^{+7.9}_{-8.8}$      & $2.1^{+0.2}_{-0.2}$ [42]  & $26.0^{+1.0}_{-1.0}$ [42]    & $8.96$ [41]  \\
				2FGL J1019.0$-$5856  & NS-HMXB  & $  30.4\pm  1.3$ [43]  & $4.3^{+0.2}_{-0.2}$         & $30.9^{+6.0}_{-3.4}$      & $36.4^{+3.2}_{-3.0}$      & $1.4^\ast$                & $23.2^\ast$ [44]             & $16.54$ [45]  \\
				Cen X-3              & NS-HMXB  & $  39.0\pm  3.0$ [46]  & $6.9^{+0.7}_{-0.6}$         & $102.4^{+4.2}_{-3.9}$     & $108.5^{+11.2}_{-14.6}$   & $1.6^{+0.1}_{-0.1}$ [42]  & $24.0^{+1.0}_{-1.0}$ [42]    & $2.09^\ast$ [47]  \\
				1E 1145.1$-$6141     & NS-HMXB  & $ -13.0\pm  3.0$ [48]  & $8.3^{+0.8}_{-0.6}$         & $56.5^{+13.2}_{-10.2}$    & $54.7^{+15.3}_{-11.3}$    & $1.7^{+0.3}_{-0.3}$ [48]  & $14.0^{+4.0}_{-4.0}$ [48]    & $14.37$ [49]  \\
				2S 1145$-$619        & NS-HMXB  & $ -17.0\pm  7.4$ [50]  & $2.1^{+0.1}_{-0.1}$         & $16.2^{+8.4}_{-4.3}$      & $12.9^{+5.1}_{-3.0}$      & $1.4^\ast$                & $13.0^{+2.0}_{-2.0}$ [51]    & $187.50^\ast$ [47]  \\
				GX 301$-$2           & NS-HMXB  & $   4.1\pm  2.4$ [52]  & $3.6^{+0.2}_{-0.2}$         & $58.0^{+4.7}_{-4.3}$      & $59.8^{+6.7}_{-6.5}$      & $1.9^{+0.6}_{-0.6}$ [52]  & $43.0^{+10.0}_{-10.0}$ [52]  & $41.50$ [52]  \\
				1H 1253$-$761        & NS-HMXB  & $ -20.0\pm  7.4$ [50]  & $0.2^{+0.0}_{-0.0}$         & $27.5^{+10.9}_{-9.0}$     & $26.2^{+9.3}_{-8.3}$      & $1.4^\ast$                & $7.5^\ast$ [53]              & \nodata \\
				1H 1249$-$637        & NS-HMXB  & $  22.0\pm  7.0$ [50]  & $0.4^{+0.0}_{-0.0}$         & $21.0^{+11.5}_{-9.3}$     & $28.6^{+9.9}_{-9.0}$      & $1.4^\ast$                & \nodata                      & \nodata \\
				1H 1255$-$567        & NS-HMXB  & $  13.0\pm  3.7$ [50]  & $0.1^{+0.0}_{-0.0}$         & $12.5^{+7.0}_{-3.9}$      & $14.4^{+5.0}_{-4.6}$      & $1.4^\ast$                & \nodata                      & \nodata \\
				4U 1538$-$52         & NS-HMXB  & $-148.1\pm  0.9$ [54]  & $5.7^{+0.5}_{-0.5}$         & $77.5^{+9.2}_{-10.0}$     & $73.9^{+8.6}_{-8.5}$      & $1.0^{+0.2}_{-0.2}$ [42]  & $16.0^{+2.0}_{-2.0}$ [42]    & $3.71$ [55]  \\
				4U 1700$-$37         & NS-HMXB  & $ -60.0\pm 12.0$ [56]  & $1.5^{+0.1}_{-0.1}$         & $71.3^{+11.7}_{-11.5}$    & $76.7^{+12.8}_{-13.4}$    & $2.0^{+0.2}_{-0.2}$ [42]  & $46.0^{+5.0}_{-5.0}$ [42]    & $3.41$ [57]  \\
				IGR J17544$-$2619    & NS-HMXB  & $ -46.8\pm  4.0$ [58]  & $2.4^{+0.2}_{-0.1}$         & $43.7^{+4.6}_{-4.4}$      & $36.6^{+6.0}_{-5.6}$      & $1.4^\ast$                & $23.0^{+2.0}_{-2.0}$ [59]    & $12.17$ [58]  \\
				LS 5039              & NS-HMXB  & $  17.2\pm  0.7$ [30]  & $1.9^{+0.1}_{-0.1}$         & $88.2^{+3.7}_{-3.2}$      & $84.5^{+4.7}_{-4.1}$      & $3.7^{+1.3}_{-1.0}$ [30]  & $22.9^{+3.4}_{-2.9}$ [30]    & $3.91$ [30]  \\
				1H 2202+501          & NS-HMXB  & $ -16.8\pm  2.5$ [60]  & $1.1^{+0.0}_{-0.0}$         & $28.7^{+3.2}_{-2.5}$      & $24.3^{+4.5}_{-4.3}$      & $1.4^\ast$                & \nodata                      & \nodata \\
				4U 2206+543          & NS-HMXB  & $ -61.5\pm  1.6$ [61]  & $3.1^{+0.1}_{-0.1}$         & $27.8^{+6.6}_{-4.1}$      & $26.3^{+5.3}_{-4.7}$      & $1.4^\ast$                & $23.5^{+14.5}_{-8.0}$ [61]   & $9.55$ [61]  \\
				A 0620$-$00          & BH-LMXB  & $   8.5\pm  1.8$ [62]  & $1.5^{+0.3}_{-0.2}$         & $43.7^{+9.9}_{-7.1}$      & $41.8^{+11.2}_{-8.6}$     & $6.6^{+0.3}_{-0.3}$ [63]  & $0.5^{+0.1}_{-0.1}$ [63]     & $0.32$ [64]  \\
				XTE J1118+480        & BH-LMXB  & $   2.7\pm  1.1$ [65]  & ${1.7^{+0.1}_{-0.1}}^\dag$  & $142.9^{+11.4}_{-11.2}$   & $172.0^{+18.9}_{-28.4}$   & $7.5^{+0.7}_{-0.6}$ [66]  & $0.2^{+0.1}_{-0.1}$ [66]     & $0.17$ [64]  \\
				GRS 1124$-$684       & BH-LMXB  & $  14.2\pm  6.3$ [67]  & ${5.0^{+0.7}_{-0.7}}^\dag$  & $118.6^{+15.5}_{-15.2}$   & $115.1^{+19.3}_{-18.5}$   & $11.0^{+2.1}_{-1.4}$ [68] & $0.9^{+0.2}_{-0.1}$ [68]     & $0.43$ [69]  \\
				MAXI J1305$-$704     & BH-LMXB  & $  -9.0\pm  5.0$ [70]  & ${7.7^{+1.6}_{-1.6}}^\dag$  & $58.1^{+30.4}_{-22.1}$    & $75.1^{+17.4}_{-10.0}$    & $8.9^{+1.6}_{-1.0}$ [70]  & $0.4^{+0.2}_{-0.2}$ [70]     & $0.40$ [70]  \\
				BW Cir               & BH-LMXB  & $ 102.0\pm  4.0$ [71]  & $\geq25.0$                  & $352.0^{+106.9}_{-98.8}$  & $279.3^{+156.9}_{-111.3}$ & $>7.6$ [72]               & $>0.9$ [72]                  & $2.54$ [72]  \\
				4U 1543$-$475        & BH-LMXB  & $ -87.0\pm  3.0$ [73]  & ${7.5^{+0.5}_{-0.5}}^\dag$  & $99.0^{+14.5}_{-15.1}$    & $118.7^{+11.1}_{-9.1}$    & $9.4^{+1.0}_{-1.0}$ [74]  & $2.6^{+0.6}_{-0.5}$ [74]     & $1.12$ [74]  \\
				XTE J1550$-$564      & BH-LMXB  & $ -68.0\pm 19.0$ [75]  & ${4.5^{+0.5}_{-0.5}}^\dag$  & $79.1^{+38.2}_{-32.3}$    & $78.8^{+37.1}_{-30.4}$    & $11.7^{+3.9}_{-3.9}$ [76] & $0.3^{+0.1}_{-0.1}$ [76]     & $1.54$ [76]  \\
				GRO 1655$-$40        & BH-LMXB  & $-167.1\pm  0.6$ [77]  & $3.2^{+0.6}_{-0.4}$         & $161.8^{+7.4}_{-5.7}$     & $153.6^{+13.9}_{-28.9}$   & $6.0^{+0.4}_{-0.4}$ [78]  & $2.5^{+0.3}_{-0.3}$ [78]     & $2.62$ [79]  \\
				GX 339$-$4           & BH-LMXB  & $  26.0\pm  2.0$ [80]  & $\geq5.0$                   & $165.8^{+10.7}_{-34.7}$   & $201.0^{+33.0}_{-37.2}$   & $5.9^{+3.6}_{-3.6}$ [80]  & $1.1^{+1.7}_{-0.8}$ [80]     & $1.76$ [80]  \\
				H 1705$-$250         & BH-LMXB  & $ -54.0\pm 15.0$ [81]  & ${8.6^{+2.1}_{-2.1}}^\dag$  & $221.2^{+100.8}_{-109.0}$ & $289.4^{+71.3}_{-75.9}$   & $6.4^{+1.5}_{-1.5}$ [81]  & $0.3^{+0.1}_{-0.1}$ [81]     & $0.52$ [82]  \\
				Swift J1753.5$-$0127 & BH-LMXB  & $   6.0\pm  6.0$ [83]  & ${6.0^{+2.0}_{-2.0}}^\dag$  & $110.4^{+81.4}_{-57.1}$   & $149.5^{+68.2}_{-59.6}$   & $>7.4$ [84]               & $0.5^\ast$ [84]              & $0.14$ [85]  \\
				MAXI J1820+070       & BH-LMXB  & $ -21.6\pm  2.3$ [86]  & $2.8^{+0.8}_{-0.5}$         & $72.4^{+18.7}_{-16.2}$    & $105.4^{+24.6}_{-19.4}$   & $8.4^{+0.8}_{-0.7}$ [87]  & $0.6^{+0.1}_{-0.1}$ [87]     & $0.69$ [86]  \\
				MAXI J1836$-$194     & BH-LMXB? & $  61.0\pm 15.0$ [88]  & $7.0^{+3.0}_{-3.0}$         & $98.4^{+32.2}_{-25.7}$    & $133.4^{+32.1}_{-29.8}$   & \nodata                   & \nodata                      & \nodata \\
				GRS 1915+105         & BH-LMXB  & $  12.3\pm  1.0$ [89]  & $8.6^{+2.0}_{-2.0}$         & $31.8^{+17.5}_{-14.6}$    & $39.6^{+19.6}_{-19.5}$    & $12.4^{+2.0}_{-1.8}$ [89] & $0.5^{+0.4}_{-0.3}$ [89]     & $33.83$ [90]  \\
				V404 Cyg             & BH-LMXB  & $  -0.4\pm  2.2$ [91]  & $2.4^{+0.1}_{-0.1}$         & $45.3^{+3.9}_{-3.5}$      & $38.7^{+5.7}_{-5.9}$      & $9.0^{+0.2}_{-0.6}$ [92]  & $0.6^{+0.1}_{-0.1}$ [92]     & $6.47$ [93]  \\
				HD 96670             & BH-HMXB? & $ -27.5\pm  0.0$ [94]  & $3.1^{+0.3}_{-0.3}$         & $24.3^{+10.0}_{-9.2}$     & $25.6^{+4.4}_{-4.3}$      & $6.2^{+0.9}_{-0.7}$ [94]  & $22.7^{+5.2}_{-3.6}$ [94]    & $5.28$ [94]  \\
				V4641 Sgr            & BH-HMXB  & $ 107.4\pm  2.9$ [76]  & $4.8^{+0.7}_{-0.5}$         & $91.8^{+6.6}_{-6.7}$      & $101.4^{+10.1}_{-10.3}$   & $6.4^{+0.6}_{-0.6}$ [95]  & $2.9^{+0.4}_{-0.4}$ [95]     & $2.82$ [96]  \\
                \bottomrule
        \end{tabular}
        \label{tab:vpec_and_distances}
    \end{threeparttable}
    \end{adjustbox}
\end{table*}

\begin{table*}
    \centering
    \contcaption{}
    \begin{adjustbox}{width=0.98\textwidth}
    \begin{threeparttable}
        \begin{tabular}{llcllllll}
            \toprule
            Name & Class & $\gamma$ & $d$   & $\vpecp$ &  $\pkv$     & $\mcomp$ & $\mnoncomp$ & $\porb$ \\
                 &       & $(\kms)$ & (kpc) & $(\kms)$ & $(\kms)$ & $(\Msun)$  & $(\Msun)$ & $(\mathrm{d})$  \\
            \midrule
				SS 433               & BH-HMXB? & $  69.0\pm  4.7$ [97]  & $7.4^{+1.4}_{-1.1}$         & $44.3^{+22.2}_{-15.9}$    & $45.3^{+14.4}_{-12.0}$    & $4.3^{+0.4}_{-0.4}$ [97]  & $11.3^{+0.6}_{-0.6}$ [97]    & $13.08^\ast$ [98]  \\
				Cyg X-1              & BH-HMXB  & $  -5.1\pm  0.5$ [99]  & $2.1^{+0.1}_{-0.1}$         & $22.3^{+4.6}_{-2.9}$      & $21.9^{+3.2}_{-3.0}$      & $21.2^{+2.2}_{-2.2}$ [100] & $40.6^{+7.7}_{-7.1}$ [100]   & $5.60$ [101] \\
				MWC 656              & BH-HMXB? & $  -2.8\pm  9.4$ [102] & $2.0^{+0.1}_{-0.1}$         & $25.8^{+13.7}_{-12.9}$    & $36.2^{+9.7}_{-7.0}$      & $4.1^{+1.4}_{-1.4}$ [102] & $7.8^{+2.0}_{-2.0}$ [103]    & $60.37$ [103] \\
				PSR J0348+0432       & LM-PSR   & $  -1.0\pm 20.0$ [104] & ${2.1^{+0.4}_{-0.4}}^\dag$  & $62.3^{+13.4}_{-12.3}$    & $73.1^{+13.0}_{-11.4}$    & $2.0^{+0.0}_{-0.0}$ [104] & $0.2^{+0.0}_{-0.0}$ [104]    & $0.10$ [104] \\
				PSR J1012+5307       & LM-PSR   & $  44.0\pm  8.0$ [105] & $0.6^{+0.1}_{-0.1}$         & $84.2^{+15.7}_{-11.9}$    & $96.4^{+21.5}_{-14.7}$    & $1.6^{+0.2}_{-0.2}$ [105] & $0.2^{+0.0}_{-0.0}$ [105]    & $0.60$ [106] \\
				PSR J1023+0038       & LM-PSR   & $   1.0\pm  3.0$ [107] & $1.4^{+0.0}_{-0.0}$         & $131.1^{+6.7}_{-6.3}$     & $149.0^{+15.1}_{-25.3}$   & $1.6^{+0.2}_{-0.2}$ [108] & $0.2^{+0.0}_{-0.0}$ [108]    & $0.20$ [109] \\
				PSR J1024$-$0719     & LM-PSR   & $ 185.0\pm  4.0$ [110] & $1.2^{+0.2}_{-0.2}$         & $382.2^{+64.7}_{-45.8}$   & $465.7^{+68.7}_{-84.7}$   & $1.4^\ast$                & $0.4^\ast$ [111]             & \nodata \\
				PSR J1048+2339       & LM-PSR   & $ -24.0\pm  8.0$ [108] & ${2.0^{+0.4}_{-0.4}}^\dag$  & $157.7^{+35.8}_{-35.4}$   & $199.4^{+49.7}_{-46.0}$   & $>2.0$ [108]              & $>0.4$ [108]                 & $0.25$ [112] \\
				XSS J12270$-$4859    & LM-PSR   & $  67.0\pm  2.0$ [113] & ${1.6^{+0.3}_{-0.3}}^\dag$  & $131.4^{+21.1}_{-20.8}$   & $153.0^{+23.4}_{-24.5}$   & $1.4^\ast$                & $>0.3$ [108]                 & $0.29$ [113] \\
				PSR B1259$-$63       & HM-PSR   & $   0.0\pm  0.0$ [114] & $2.2^{+0.1}_{-0.1}$         & $24.8^{+9.4}_{-8.0}$      & $34.4^{+4.9}_{-5.0}$      & $1.4^\ast$                & $23.0^\ast$ [114]            & $1236.72$ [114] \\
				PSR J1306$-$4035     & LM-PSR   & $  32.0\pm  1.8$ [115] & ${4.7^{+0.9}_{-0.9}}^\dag$  & $120.9^{+17.1}_{-17.3}$   & $147.1^{+16.0}_{-19.6}$   & $1.8^{+0.1}_{-0.0}$ [115] & $0.5^{+0.0}_{-0.0}$ [115]    & $1.10$ [116] \\
				PSR J1311$-$3430     & LM-PSR   & $  62.5\pm  4.5$ [117] & ${2.4^{+0.5}_{-0.5}}^\dag$  & $108.6^{+16.6}_{-15.7}$   & $156.8^{+27.4}_{-27.0}$   & $2.1^{+0.1}_{-0.1}$ [117] & $0.0^{+0.0}_{-0.0}$ [117]    & $0.07$ [118] \\
				PSR J1417$-$4402     & LM-PSR   & $ -15.3\pm  0.9$ [43]  & $4.5^{+1.1}_{-0.8}$         & $98.5^{+22.8}_{-19.0}$    & $140.0^{+29.4}_{-23.6}$   & $2.0^{+0.1}_{-0.1}$ [43]  & $0.3^{+0.0}_{-0.0}$ [43]     & $5.37$ [43]  \\
				PSR J1431$-$4715     & LM-PSR   & $ -91.0\pm  2.0$ [108] & ${1.6^{+0.3}_{-0.3}}^\dag$  & $122.8^{+18.4}_{-16.0}$   & $114.3^{+22.1}_{-19.0}$   & $1.4^\ast$                & $>0.1$ [119]                 & $0.45$ [119] \\
				PSR J1622$-$0315     & LM-PSR   & $-135.0\pm  6.0$ [108] & ${1.1^{+0.2}_{-0.2}}^\dag$  & $144.4^{+11.1}_{-10.3}$   & $140.9^{+20.1}_{-26.1}$   & $>1.4$ [108]              & $>0.1$ [108]                 & $0.16$ [120] \\
				PSR J1628$-$3205     & LM-PSR   & $  -4.0\pm  7.0$ [108] & ${1.2^{+0.2}_{-0.2}}^\dag$  & $114.1^{+28.4}_{-28.3}$   & $150.8^{+40.9}_{-38.3}$   & $1.4^\ast$                & $>0.2$ [121]                 & $0.21$ [108] \\
				PSR J1653$-$0158     & LM-PSR   & $-174.6\pm  5.1$ [122] & ${0.8^{+0.2}_{-0.2}}^\dag$  & $177.5^{+10.3}_{-9.4}$    & $186.8^{+22.8}_{-40.0}$   & $2.2^{+0.2}_{-0.1}$ [123] & $0.0^{+0.0}_{-0.0}$ [123]    & $0.05$ [123] \\
				PSR J1723$-$2837     & LM-PSR   & $  33.0\pm  2.0$ [108] & $0.9^{+0.0}_{-0.0}$         & $110.4^{+10.8}_{-10.6}$   & $155.0^{+13.8}_{-33.0}$   & $1.2^{+0.3}_{-0.2}$ [124] & $0.4^{+0.1}_{-0.1}$ [108]    & $0.62$ [125] \\
				PSR J1816+4510       & LM-PSR   & $ -99.0\pm  8.0$ [126] & ${4.4^{+0.9}_{-0.9}}^\dag$  & $165.3^{+30.6}_{-26.3}$   & $170.3^{+24.6}_{-19.1}$   & $>1.8$ [126]              & $>0.2$ [126]                 & $0.36$ [127] \\
				PSR J2039$-$5617     & LM-PSR   & $   6.0\pm  3.0$ [108] & ${1.8^{+0.4}_{-0.4}}^\dag$  & $113.5^{+27.8}_{-27.8}$   & $164.5^{+40.5}_{-38.3}$   & $1.3^{+0.3}_{-0.2}$ [128] & $0.2^{+0.0}_{-0.0}$ [128]    & $0.23$ [128] \\
				PSR J2129$-$0429     & LM-PSR   & $ -64.0\pm  2.0$ [129] & $2.1^{+0.4}_{-0.3}$         & $201.3^{+33.3}_{-24.2}$   & $180.3^{+33.0}_{-35.5}$   & $1.7^{+0.2}_{-0.2}$ [129] & $0.4^{+0.0}_{-0.0}$ [129]    & $0.64$ [129] \\
				PSR J2215+5135       & LM-PSR   & $  49.0\pm  8.0$ [130] & ${2.8^{+0.6}_{-0.6}}^\dag$  & $117.8^{+17.4}_{-17.4}$   & $103.8^{+19.7}_{-20.2}$   & $2.3^{+0.2}_{-0.1}$ [108] & $0.3^{+0.0}_{-0.0}$ [130]    & $0.17$ [108] \\
				PSR J2339$-$0533     & LM-PSR   & $ -49.0\pm  8.0$ [131] & ${1.1^{+0.2}_{-0.2}}^\dag$  & $64.1^{+13.5}_{-12.9}$    & $97.5^{+18.1}_{-16.7}$    & $1.6^{+0.3}_{-0.2}$ [132] & $0.3^{+0.1}_{-0.1}$ [108]    & $0.19$ [132] \\
				J05215658            & BH-HMNI  & $  -0.4\pm  0.1$ [133] & $2.5^{+0.1}_{-0.1}$         & $24.8^{+4.2}_{-2.6}$      & $28.2^{+2.9}_{-2.8}$      & $3.3^{+1.1}_{-0.4}$ [133] & $3.2^{+0.4}_{-0.5}$ [133]    & $83.20$ [133] \\
				J06163552            & NS-LMNI  & $  29.2\pm  0.7$ [134] & $1.0^{+0.0}_{-0.0}$         & $17.0^{+4.8}_{-2.6}$      & $25.1^{+21.7}_{-10.0}$    & $1.3^{+0.1}_{-0.1}$ [134] & $1.7^{+0.1}_{-0.1}$ [134]    & $0.87$ [134] \\
				J112306.9            & NS-LMNI  & $  -8.0\pm  2.0$ [135] & $0.3^{+0.0}_{-0.0}$         & $29.5^{+10.5}_{-10.3}$    & $54.2^{+44.0}_{-24.7}$    & $1.2^{+0.0}_{-0.0}$ [135] & $0.6^{+0.0}_{-0.0}$ [135]    & $0.27$ [135] \\
				J125556.57           & NS-LMNI  & $  -3.4\pm  0.3$ [136] & $0.5^{+0.0}_{-0.0}$         & $36.1^{+7.1}_{-5.2}$      & $45.1^{+4.0}_{-4.2}$      & $1.4^{+0.7}_{-0.3}$ [136] & $1.2^{+0.1}_{-0.1}$ [136]    & $2.76$ [136] \\
				Gaia BH2             & BH-LMNI  & $  -4.2\pm  0.2$ [137] & $1.2^{+0.0}_{-0.0}$         & $25.2^{+9.7}_{-8.7}$      & $34.1^{+5.0}_{-5.1}$      & $8.9^{+0.3}_{-0.3}$ [137] & $1.1^{+0.2}_{-0.2}$ [137]    & $1276.70$ [137] \\
				J15274848            & NS-LMNI  & $ -26.1\pm  0.7$ [138] & $0.1^{+0.0}_{-0.0}$         & $13.8^{+7.1}_{-3.5}$      & $18.9^{+3.1}_{-3.0}$      & $1.0^{+0.1}_{-0.1}$ [138] & $0.6^{+0.1}_{-0.1}$ [138]    & $0.26$ [139] \\
				Gaia BH1             & BH-LMNI  & $  45.5\pm  0.8$ [140] & $0.5^{+0.0}_{-0.0}$         & $76.4^{+4.0}_{-3.1}$      & $71.3^{+9.2}_{-10.8}$     & $9.8^{+0.2}_{-0.2}$ [140] & $0.9^{+0.1}_{-0.1}$ [140]    & $185.59$ [140] \\
				AS 386               & BH-HMNI  & $ -31.8\pm  2.6$ [141] & $5.9^{+0.5}_{-0.4}$         & $10.8^{+8.0}_{-4.4}$      & $20.6^{+2.2}_{-2.7}$      & $>7.0$ [141]              & $7.0^{+1.0}_{-1.0}$ [141]    & $131.27$ [141] \\
				J235456.76           & NS-LMNI  & $  41.0\pm  2.4$ [142] & $0.1^{+0.0}_{-0.0}$         & $47.9^{+9.4}_{-8.7}$      & $52.5^{+7.6}_{-7.7}$      & $>1.3$ [142]              & $0.7^{+0.1}_{-0.1}$ [142]    & $0.48$ [142] \\
            \bottomrule
        \end{tabular}
        \begin{tablenotes}
            \item References: [1] \cite{Jonker05}, [2] \cite{Steeghs07}, [3] \cite{Ashcraft12}, [4] \cite{Cornelisse13}, [5] \cite{DiazTrigo09}, [6] \cite{Shahbaz14}, [7] \cite{Steeghs02}, [8] \cite{Cherepashchuk21}, [9] \cite{LaSala85}, [10] \cite{Casares06}, [11] \cite{Reynolds97}, [12] \cite{Ponti18}, [13] \cite{Iaria18}, [14] \cite{Hinkle19}, [15] \cite{Hinkle06}, [16] \cite{Wang17}, [17] \cite{Casares03}, [18] \cite{Munoz-Darias05}, [19] \cite{Jonker01}, [20] \cite{MataSanchez17}, [21] \cite{Premachandra16}, [22] \cite{Gonzalez-Galan14}, [23] \cite{denHartog04}, [24] \cite{Grundstrom07a}, [25] \cite{Reig96}, [26] \cite{Crampton85}, [27] \cite{Sarty09}, [28] \cite{Reig97}, [29] \cite{Aragona09}, [30] \cite{Casares05}, [31] \cite{Gregory02}, [32] \cite{Grundstrom07b}, [33] \cite{Delgado-Marti01}, [34] \cite{Barsukova05}, [35] \cite{Gontcharov06}, [36] \cite{Hutchings84}, [37] \cite{Priedhorsky83a}, [38] \cite{Moritani18}, [39] \cite{Aragona10}, [40] \cite{Gamen15}, [41] \cite{Stickland97}, [42] \cite{Falanga15}, [43] \cite{Strader15}, [44] \cite{Waisberg15}, [45] \cite{An15}, [46] \cite{vanderMeer07}, [47] \cite{Nagase89}, [48] \cite{Hutchings87}, [49] \cite{Ray02}, [50] \cite{Kharchenko07}, [51] \cite{Stevens97}, [52] \cite{Kaper06}, [53] \cite{Waters89}, [54] \cite{Abubekerov04}, [55] \cite{Davison77}, [56] \cite{Gies86}, [57] \cite{Islam16}, [58] \cite{Nikolaeva13}, [59] \cite{Bikmaev17}, [60] \cite{Chojnowski17}, [61] \cite{Hambaryan22}, [62] \cite{GonzalezHernandez10}, [63] \cite{Cantrell10}, [64] \cite{GonzalezHernandez14}, [65] \cite{GonzalezHernandez08b}, [66] \cite{Khargharia13}, [67] \cite{Wu15}, [68] \cite{Wu16}, [69] \cite{GonzalezHernandez17}, [70] \cite{MataSanchez21}, [71] \cite{Casares04}, [72] \cite{Casares09}, [73] \cite{Orosz98}, [74] \cite{Orosz03}, [75] \cite{Orosz02}, [76] \cite{Orosz11}, [77] \cite{GonzalezHernandez08a}, [78] \cite{Shahbaz03}, [79] \cite{vanderHooft98}, [80] \cite{Heida17}, [81] \cite{Harlaftis97}, [82] \cite{Remillard96}, [83] \cite{Neustroev14}, [84] \cite{Shaw16}, [85] \cite{Zurita08}, [86] \cite{Torres19}, [87] \cite{Torres20}, [88] \cite{Russell14b}, [89] \cite{Reid14b}, [90] \cite{Steeghs13}, [91] \cite{Casares94}, [92] \cite{Khargharia10}, [93] \cite{Casares19}, [94] \cite{Gomez21}, [95] \cite{MacDonald14}, [96] \cite{Orosz01}, [97] \cite{Picchi20}, [98] \cite{Crampton81}, [99] \cite{Gies08}, [100] \cite{Miller-Jones21}, [101] \cite{LaSala98}, [102] \cite{Casares12}, [103] \cite{Williams10}, [104] \cite{Antoniadis13}, [105] \cite{Callanan98}, [106] \cite{Nicastro95}, [107] \cite{Thorstensen05}, [108] \cite{Strader19}, [109] \cite{Archibald13}, [110] \cite{Bassa16}, [111] \cite{Kaplan16}, [112] \cite{Deneva16}, [113] \cite{deMartino14}, [114] \cite{Miller-Jones18}, [115] \cite{Swihart19}, [116] \cite{Linares18a}, [117] \cite{Romani12b}, [118] \cite{Romani12a}, [119] \cite{Bates15}, [120] \cite{Sanpa-Arsa16}, [121] \cite{Ray12}, [122] \cite{Romani14}, [123] \cite{Nieder20}, [124] \cite{vanStaden16}, [125] \cite{Crawford13}, [126] \cite{Kaplan13}, [127] \cite{Stovall14}, [128] \cite{Clark21}, [129] \cite{Bellm16}, [130] \cite{Linares18b}, [131] \cite{Romani11}, [132] \cite{Pletsch15}, [133] \cite{Thompson19}, [134] \cite{Yuan22}, [135] \cite{Yi22}, [136] \cite{Mazeh22}, [137] \cite{El-Badry23b}, [138] \cite{Lin23}, [139] \cite{Jayasinghe19}, [140] \cite{El-Badry23a}, [141] \cite{Khokhlov18}, [142] \cite{Zheng22} 
            \item $^\dag$: literature distance ($d_\mathrm{lit}$; Table \ref{tab:d_lit_table}) used because the 
            parallax is not well-constrained or not available (Sec \ref{sec:final_sample}); for PSRs, these are distances inferred from the \citet{Yao17} dispersion measures.
            \item $^\ast$: values that have no literature constraints.
        \end{tablenotes}
    \end{threeparttable}
    \end{adjustbox}
\end{table*}

\section{Analysis and results}
\label{sec:results}

\subsection{Comparing with literature distances}
\begin{figure}
    \centering
    \includegraphics[width=\columnwidth]{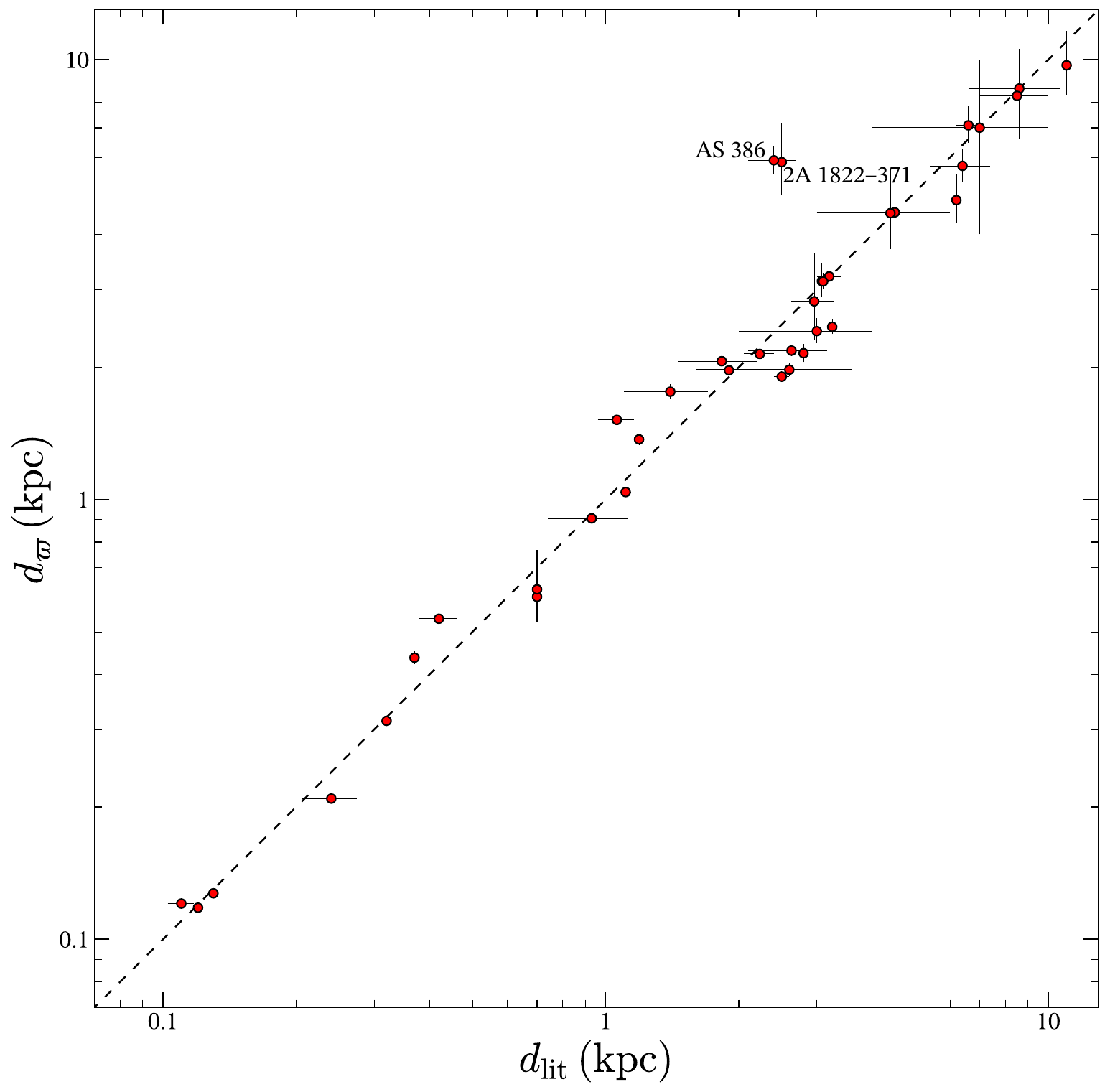}
    \caption{A comparison between literature distances ($\dlit$) and parallax-inferred distances ($\dpost$), showing that they agree for most binaries. The error bars on $\dpost$ correspond to $68\%$ confidence intervals. The two obvious outliers are labelled with their names.}
    \label{fig:d_lit_vs_d_post}
\end{figure}

\fr{Distances are clearly crucial in interpreting our peculiar velocities, and so we compare the well-constrained parallax-inferred distances ($\dpost$; Sec \ref{sec:final_sample}), with values in the literature ($\dlit$; Table \ref{tab:d_lit_table}) if available. In Figure \ref{fig:d_lit_vs_d_post}, we can see that $\dpost$ agrees with $\dlit$ for most binaries. However, there are two clear outliers,  2A 1822$-$371 and AS 386, that have somewhat smaller $\dlit$, despite both sources having an acceptable \gaia\ Renormalised Unit Weight Error (\ruwe) value of 0.9 and 1.2, respectively (for a satisfactory fit to a single-star model, \ruwe\ is expected to be close to $1$).\footnote{\gaia\ pipeline flags including \ruwe\ and {\tt astrometric excess noise} provide an estimate of the quality of the astrometric fit, but these are degenerate with non-single star solutions \citep[cf., ][]{Belokurov20, Gandhi22}, so we purposely do not preselect on these flags.}

The distance to 2A 1822$-$371 was estimated to be between $2$ and $3\,\kpc$\ based on the emitting region area at peak brightness of the light curve, while assuming a black-body for the inner emitting region gives an upper limit of $\approx 5\,\kpc$\ \citep{Mason82}, which is marginally consistent with our estimate at the $68\%$ level.

\citet[][K18]{Khokhlov18} estimated a distance of $2.4\pm 0.3\,\kpc$\ for AS 386, which is less than half of its $\dpost$ ($5.9^{+0.5}_{-0.4}\,\kpc$). The K18 estimate is based on a fitted extinction law and an estimated reddening of $\approx 0.9$. However, with the updated 3D ``Bayestar19" dust map \citep{Green19}, we get a similar reddening at a larger distance of at least $4.7\,\kpc$\ (calculated using the {\sc dustmap} package; see \citealt{Green18}). We thus adopt our estimate of distance even though the astrometric solution is mildly argued against by its \ruwe\ value.}

\subsection{Distribution of peculiar velocities}
\label{sec:pkv_distribution_analyses}
With the calculated $\pkv$ samples, the first analysis we can do is to get an overview of their distribution among different subclasses. We therefore divide our compilation into four subgroups, including (1) binaries with BHs, (2) binaries with NSs, (3) binaries with LM companions, and (4) binaries with HM companions. With the calculated $\pkv$ distributions of individual binaries, we model the overall $\pkv$ distribution for these subgroups following a Bayesian approach. Given a binary class, we first construct 1,000 bootstrapped data sets for each subgroup by randomly sampling (with replacement) from the individual $\pkv$ distributions. The Bayesian model is constructed as follows: for the likelihood, we use a Maxwellian distribution:

\begin{equation}
    f(v | \sigma_v) = \sqrt{\frac{2}{\pi}}\frac{v^2}{\sigma_v^3}\exp\lrb{-\frac{v^2}{2\sigma_v^2}},
    \label{eq:single-maxwellian}
\end{equation}
where $\sigma_v$ is the scale parameter, or a two-component Maxwellian distribution:
\begin{equation}
    f(v | \sigma_{v, 1}, \sigma_{v, 2}, w) = w f(v | \sigma_{v, 1}) + (1 - w) f(v | \sigma_{v, 2}),
    \label{eq:two-maxwellian}
\end{equation}
where $w$ is a weighting factor that determines contribution of the single Maxwellians that are characterised by $\sigma_{v, 1}$ and $\sigma_{v, 2}$, respectively. We use uninformative (flat) priors for all parameters: for a two-component Maxwellian likelihood, a flat prior between $0$ and $100\,\kms$ ($U(0, 100)$) is used for low-velocity component, while $U(0, 500)$ is used to represent the high-velocity component; for the single-component Maxwellian, we choose $U(0, 500)$ for the only scale parameter $\sigma_v$. 

The established posterior given the bootstrapped data set is then sampled with a Markov Chain Monte Carlo (MCMC) algorithm using the {\sc pymc} (version 5.0.2) package \citep{Salvatier16} which applies the No-U-Turn Sampler \citep[NUTS;][]{Hoffman11}. We set up four chains and run each chain for 3,000 iterations including 1,000 for tuning; convergence is checked using the Gelman-Rubin diagonistic \citep[$\hat R$;][converged chains should have $\hat R \approx 1$]{Gelman92}. We check $\hat{R}$ for each bootstrapped data set, and consider simulations that have relative deviation of $\hat R$ from unity greater than 0.01 (i.e., $|\hat R - 1| / \hat R \geq 0.01$) as diverging chains.

\begin{figure}
    \centering
    \includegraphics[width=\columnwidth]{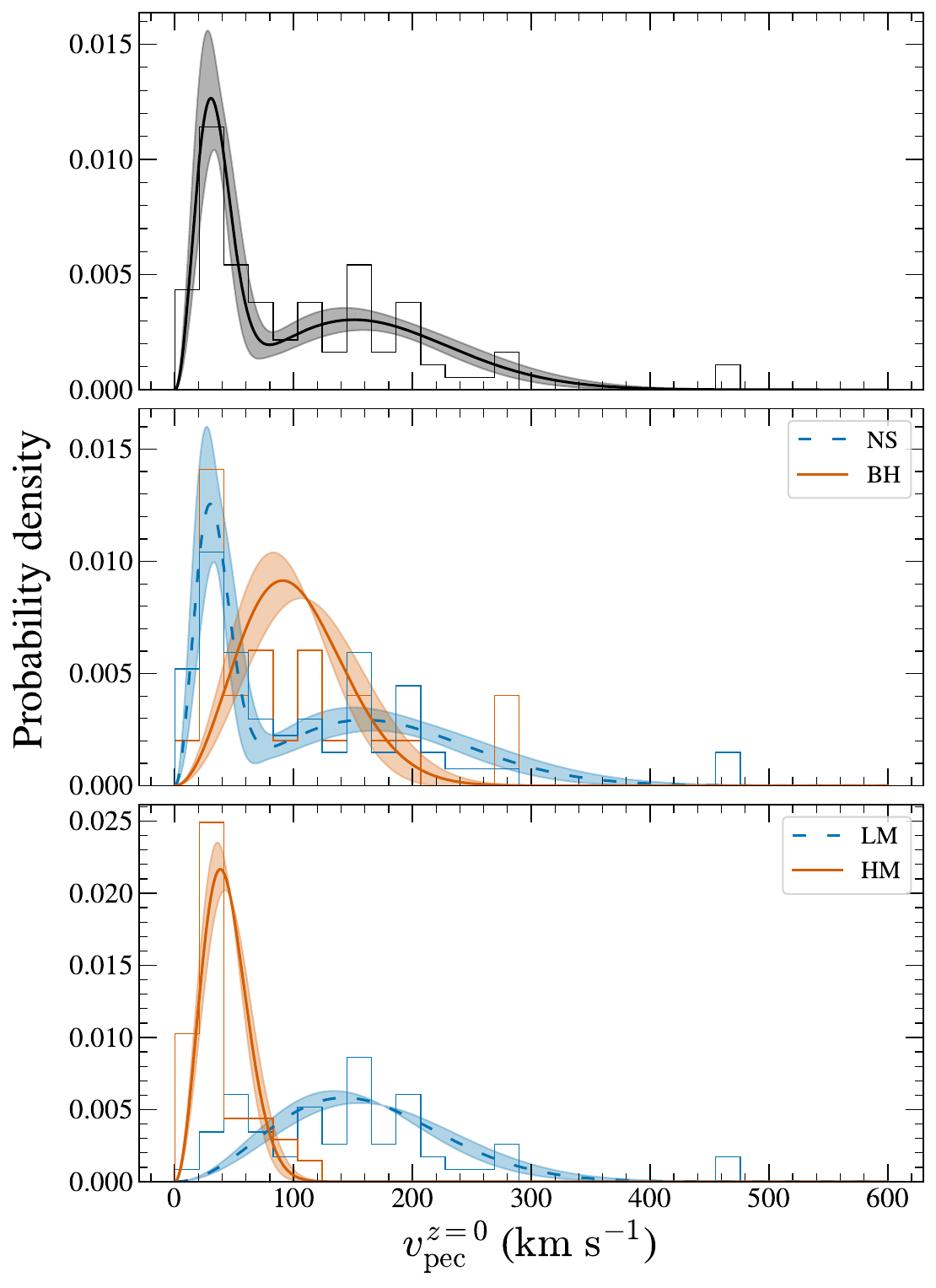}
    \caption{Distribution of $\pkv$ values of all binaries in Table \ref{tab:vpec_and_distances} (top), binaries that host a BH or NS (middle), and binaries that host a HM or LM non-degenerate companion (bottom). The solid lines in each panel represent the single or two-component Maxwellian model characterised by the median parameters, while the shaded band represents the $68\%$ credible region propagated from the parameter posteriors.}
    \label{fig:maxwellian_distribution}
\end{figure}

\revised{For model comparison, we use the {\sc arviz} package \citep{Arviz19} to compute the leave-one-out cross-validation \citep[LOO;][]{Vehtari15} for the bootstrapped data sets, which assigns a weight value ($\loo$) to each set of our fits (single vs. two-component Maxwellian) to represent the probability of each model being true. As a complementary diagnostic, we also compute the fraction of diverging chains for each parameter, which is denoted as $\fdiv$. Chains not converging could be a result of insufficient iterations, or simply because the assumed likelihood is not appropriate for the underlying population. In this regard, $\fdiv$ can be used as a diagnostic for the suitability of the specified model. Table \ref{tab:bayesian_fit_overall_distributions} summarises median and credible intervals of the posterior for the hyper-parameters, and in Figure \ref{fig:maxwellian_distribution}, we show the distribution of median $\pkv$ overplotted with the models.}

Our results show that a two-component Maxwellian distribution gives a sensible fit to all binaries as a whole, with $\hat{R}$ consistent within $10^{-4}$ of 1 for all simulations (i.e., $\fdiv=0$). \revised{All bootstrapped data sets strongly favour a two-component model ($\loo=1$) over a single Maxwellian ($\loo=0$). The posterior gives a low- and a high-$\pkv$ component that are characterised by $\sigma_{v, 1}=21\pm 3\,\kms$ and $\sigma_{v, 2} = 107^{+10}_{-9}\,\kms$, respectively (uncertainties correspond to $68\%$ credible level, and the same for other quoted parameters in this section). This two-component Maxwellian also applies to the NS subgroup, with a low- and high-velocity component at $\sigma_{v, 1} = 21^{+4}_{-3}\,\kms$ and $\sigma_{v, 2} = 113^{+12}_{-11}\,\kms$, and \revised{$\loo$ also favours two-components ($\loo=1$) over a single Maxwellian ($\loo=0$). The two-component model also gives a higher $\loo$ weight when applied to the BH subgroup}; however, it leads to divergence in over $44\%$ of the simulations; we thus conclude that the existence of a second component in BH subgroup is insignificant given the current sample.} 

\revised{Applying a two-component Maxwellian model to fit the LM and HM subgroups also leads to non-negligible divergence in over $26\%$ and $34\%$ of the simulations, respectively; whereas, a single Maxwellian fits the data without divergence, giving very distinct $\sigma_v$ at $101^{+8}_{-7}\,\kms$ and $27\pm 2\,\kms$ for LM and HM binaries, respectively. By visual inspection, the HM subgroup seems to be restricted below $100\,\kms$, while the LM group shows a broader distribution up to a velocity of $\approx 400\,\kms$ (Figure \ref{fig:maxwellian_distribution}). Despite the divergence fractions, the LOO diagnostic still favours a two-component ($\loo = 1$) over a single Maxwellian ($\loo \approx 10^{-15}$) for both LM and HM subgroups; this could due to biases in $\loo$ estimate for smaller sample sizes \citep{Vehtari15}, so more robust conclusion can be made with more future identification of systems in each class.} 

\begin{table*}
    \centering
    \caption{Parameters of Maxwellian distribution from MCMC sampling.}
    \begin{tabular}{llcccc}
    \toprule
    Class & Model & $\sigma_{v, 1}$ & $\sigma_{v, 2}$ & $w$ & $\fdiv^a$ \\
          &       & ($\kms$)   & ($\kms$)   &     & (\%)    \\
    \midrule
    \multirow{2}{*}{BH}  & Single Maxwellian & $64.2^{+8.5}_{-7.5}$   & \nodata & \nodata & $(0.0)$ \\ 
                         & Two Maxwellians   & $26.1^{+28.0}_{-7.0}$  & $87.4^{+29.1}_{-25.6}$ & $0.5^{+0.2}_{-0.2}$ & $(46, 45, 36)$ \\
    \\
    \multirow{2}{*}{NS}  & Single Maxwellian & $85.5^{+6.5}_{-6.1}$   & \nodata & \nodata & $(0)$ \\
                         & Two Maxwellians   & $20.9^{+3.8}_{-3.3}$   & $112.7^{+12.5}_{-10.9}$ & $0.4^{+0.1}_{-0.1}$ & $(0, 0, 0)$ \\
    \\
    \multirow{2}{*}{LM}  & Single Maxwellian & $100.8^{+7.9}_{-7.3}$  & \nodata & \nodata & $(0)$ \\
                         & Two Maxwellians   & $73.5^{+9.5}_{-24.7}$  & $177.3^{+80.1}_{-57.5}$ & $0.8^{+0.1}_{-0.4}$ & $(30, 29, 28)$ \\
    \\
    \multirow{2}{*}{HM}  & Single Maxwellian & $27.1^{+2.3}_{-2.1}$   & \nodata & \nodata & $(0)$ \\
                         & Two Maxwellians   & $15.9^{+6.4}_{-2.4}$   & $40.7^{+9.8}_{-12.0}$ & $0.6^{+0.1}_{-0.2}$ & $(49, 49, 47)$ \\
    \\
    \multirow{2}{*}{All} & Single Maxwellian & $80.5^{+5.3}_{-5.0}$   & \nodata & \nodata & $(0)$ \\
                         & Two Maxwellian    & $21.3^{+3.3}_{-2.8}$   & $106.7^{+10.4}_{-9.2}$ & $0.4^{+0.1}_{-0.1}$ & $(0, 0, 0)$ \\
    \bottomrule
    \multicolumn{6}{l}{$^a$ $\fdiv$ values in parentheses correspond to the parameters in column 3--5.} \\
    \end{tabular}
    \label{tab:bayesian_fit_overall_distributions}
\end{table*}

\subsection{Comparing distributions of peculiar velocities}
\label{sec:KS_test}
There is some difference between the LM and HM subgroups and possibly also between the BH and NS subgroups, as suggested by our MCMC fits (Sec \ref{sec:pkv_distribution_analyses}). These differences could relate to the possibility that different NK mechanisms apply to binaries with different compact objects; more importantly, it has been long suggested that observed recoil velocity is dependent on masses (component mass or total mass) following conservation of NK momentum, which might also lead to distinct distributions.

To investigate possible indication, we start by testing potential differences in distributions of $\pkv$ between subgroups. Since $\pkv$ has a rather strong dependence on distance ($\pkv \propto d$), we use the \finalsamplesize\, binaries that have relatively well-defined distances (either from literature or inferred from \gaia\ parallaxes). For each binary, we randomly draw (with replacement) 10,000 velocities from the $\pkv$ sample of Sec \ref{sec:vpec} and perform two-tailed Kolmogorov-Smirnov (K-S) tests comparing BH vs. NS and LM vs. HM binaries at a pre-defined significance level ($\alpha$) of $0.01$, using the {\tt ks\_2sample} function in the {\sc scipy} package\footnote{\url{https://scipy.org/}} \citep{Virtanen20}. 

\begin{figure*}
    \centering
    \includegraphics[width=0.8\textwidth]{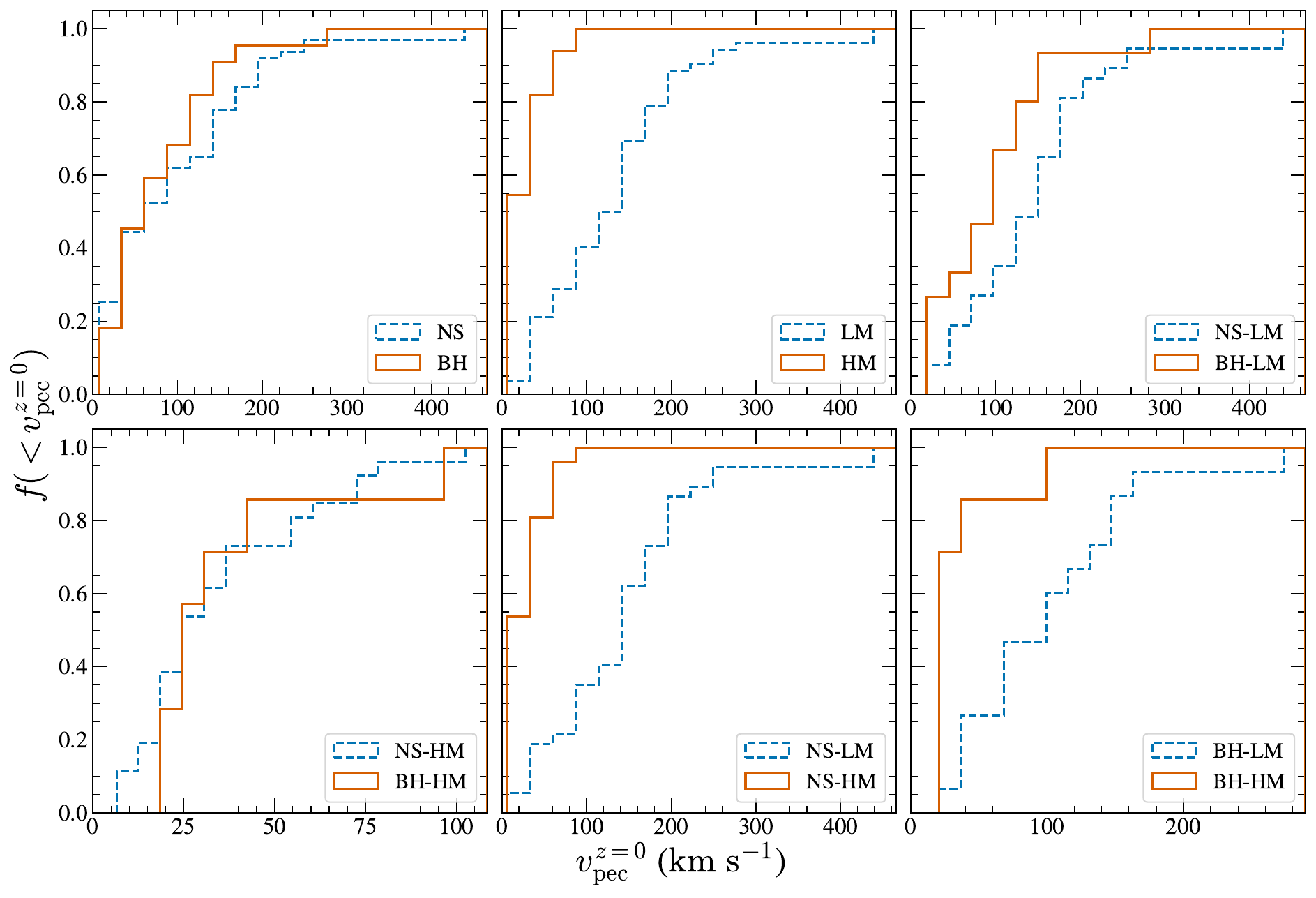}
    \caption{Cumulative distributions of median $\pkv$ for different binary subgroups.}
    \label{fig:ecdf_pbpv}
\end{figure*}

As a measure of the significance of the difference, we calculate the fraction of simulations rejecting the null hypothesis (denoted by $\frej$) that the two samples are drawn from the same underlying distribution. In Table \ref{tab:test_results_pbpv}, we summarise medians and uncertainties for the resulting test statistics and $p$-values, and in Figure \ref{fig:ecdf_pbpv}, we present cumulative distributions comparing different subgroups. 

Comparing $\pkv$ distributions between subgroups suggests no difference between NS and BH binaries at a significance level of $0.01$ ($\frej=0$); we further compare BH and NS binaries for the LM and HM subgroups and still find no difference in their $\pkv$ distributions. However, we find a marked difference between the LM and HM subgroups ($\frej=100\%$), suggesting that HM binaries are from a population with lower $\pkv$ compared to the LM group. This is in line with our Maxwellian fits to the LM and HM subgroups, where the majority of the HM binaries have median $\pkv \lesssim 100\,\kms$ (Figure \ref{fig:maxwellian_distribution}), with the highest being Cen X-3 ($\pkv=108\pm 10\,\kms$), while the LM binaries have $\pkv$ up to $\sim 400~\mathrm{kms}$. This difference is mostly contributed by the NS binaries in our sample as the difference persists for NS-LM vs. NS-HM ($\frej=100\%$) but is much less significant for BH-LM vs. BH-HM ($\frej = 37\%$; Table \ref{tab:test_results_pbpv}).

\begin{table*}
    \centering
    \caption{Two-sample K-S test results comparing $\pkv$ between different classes.}
    \begin{tabular}{lcccc}
        \toprule
        Subgroups           & Sample sizes & Test statistic & \pvalue & $\frej^a$ (\%)\\
        \midrule
        NS vs. BH           & $63, 22$     & $0.20\pm 0.04$ & $0.48\pm 0.22$ & $0$ \\
        LM vs. HM           & $52, 33$     & $0.67\pm 0.02$ & $\leq 2.22\times 10^{-8}$  & $100$ \\
        NS-LM vs. BH-LM     & $37, 15$     & $0.37\pm 0.06$ & $0.10\pm 0.09$ & $4.6$ \\
        NS-HM vs. BH-HM     & $26, 7$      & $0.27\pm 0.06$ & $0.73\pm 0.20$ & $0$ \\
        NS-LM vs. NS-HM     & $37, 26$     & $0.74\pm 0.03$ & $\leq 8.78 \times 10^{-8}$  & $100$ \\
        BH-LM vs. BH-HM     & $15, 7$      & $0.66\pm 0.08$ & $\leq 0.04$ & $37.4$ \\
        \bottomrule
        \multicolumn{5}{l}{$^a$: Percentage of simulations that reject the null hypothesis at the given significance level (1\%).}
    \end{tabular}
    \label{tab:test_results_pbpv}
\end{table*}

\subsection{Mass and peculiar velocity}
\label{sec:m_vs_v}
The distinct distributions of LM vs. HM binaries suggest an anti-correlation of $\pkv$ on $\mnoncomp$, while no strong dependence of $\pkv$ on $\mcomp$, as the K-S test of NS vs. BH subgroups suggests no significant difference. With greater inertia, more massive binaries are harder to be accelerated, $\pkv$ (recoil obtained from NK) is expected to be negatively correlated with mass. Indeed, efforts have been made to find possible anti-correlation between peculiar velocity and binary component masses \citep[e.g.,][]{Fryer2001,Fryer2012,Mirabel17, Gandhi19, Atri19}; however, past studies mostly focused only BH XRBs, and the peculiar velocities are only compared with BH masses ($\mcomp$). While $\mcomp$ is close to $\mtot$ in BH-LMXBs, $\mnoncomp$ should be also be taken into account especially for binaries with massive companions (e.g., Cyg X-1). It is therefore more natural to investigate this potential correlation with $\mtot$. 

We therefore pick binaries that have either constrained values or sensible estimates on $\mcomp$ and $\mnoncomp$ for our analysis. \fr{More specifically, a binary is included if it has a constrained value on $\mcomp$ and $\mnoncomp$, or if either $\mcomp$ or $\mnoncomp$ has a lower limit while the other mass is constrained. Some binaries have no measurement of $\mcomp$ but the compact object has been confirmed to be a NS (e.g., by detection of pulsations); if such a binary has a constrained or just a lower limit on $\mnoncomp$, we also include it in the analysis and assume a canonical NS mass of $1.4\,\Msun$ for $\mcomp$ with a $20\%$ uncertainty.}  We then follow two methods to test for possible correlation: (1) fitting a log-linear model to, and (2) calculating the Pearson's ($\rhop$) and Spearman's $\rhos$ correlation coefficient for bootstrapped velocities and Monte Carlo samples of $\mtot$.

We bootstrapped 1000 velocities from the calculated $\pkv$ distribution (Sec \ref{sec:vpec}) for each of these binaries. To account for uncertainties in $\mtot$, we sample from a log-normal distribution that centres at the logarithm of the literature value with a standard deviation set to the maximum between upper and lower error; we also include binaries that only have lower limits on $\mtot$; for these lower limits, we sample from a log-uniform distribution spanning from the lower limit ($M_\mathrm{tot, lolim}$) to $2M_\mathrm{tot, lolim}$.

\begin{figure*}
    \centering
    \includegraphics[width=0.8\textwidth]{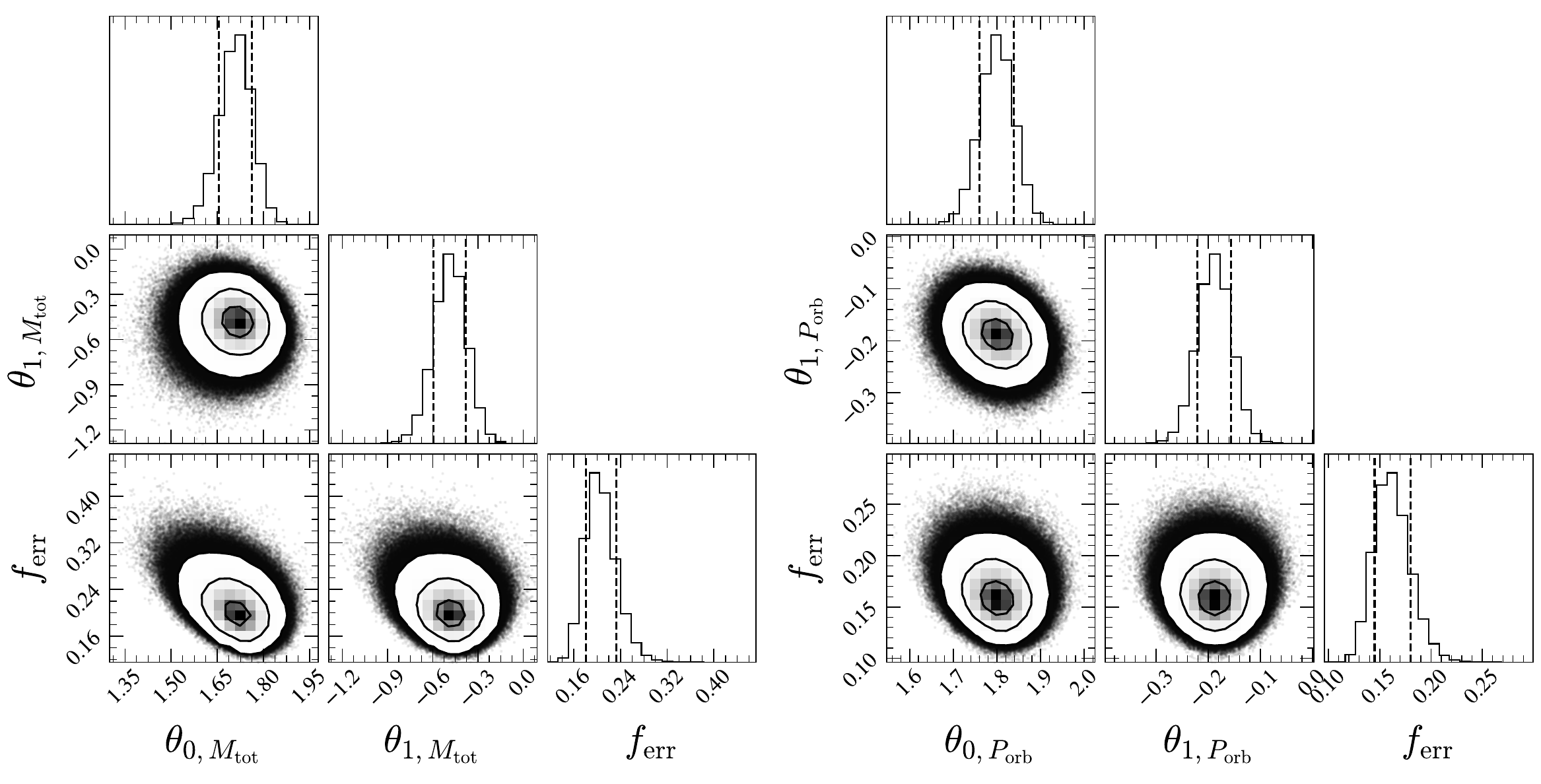}
    \caption{Corner plot made using the {\tt corner} package \citep{Foreman-Mackey16} showing posterior distributions of fit parameters of $\pkv$ vs. $\mtot$ (left) and $\pkv$ vs. $\porb$ (right). The contours in the projected distributions correspond to 1, 2, and 3\,$\sigma$ levels. The vertical dashed lines in the marginal panels indicate $68\%$ credible intervals.}
    \label{fig:linear_correlation_corner_plots}
\end{figure*}

We then follow a Bayesian approach to fit log-linear models to the 1,000 bootstrapped data set. We assume a likelihood ($\mathcal{L}$) that joins binary-specific Gaussian likelihoods $N(\mu_i, \sigma_i)$, i.e.:
\begin{equation}
    \begin{aligned}
        \mathcal{L} &= \prod_i N(\mu_i, \sigma_i) \\
        \mu_i &= \theta_{0, \mtot} + \theta_{1, \mtot} \left(\log M_{\mathrm{tot}, i} - \overline{\log\mtot}\right)\\
        \sigma_i^2 & = \sigma_{v, i}^2 + \lrb{f_\mathrm{err}\log\pkv[,\textit{i}]}^2
    \end{aligned},
    \label{eq:pkv_vs_mtot_linear_model}
\end{equation}
Here, $\theta_{0, M}$ and $\theta_{1, M}$ are the fit parameters; $\log\mtot$ in the model is offset by the mean ($\overline{\log\mtot}$) over each bootstrapped data set to remove covariance between the two coefficients; $\sigma_{v,i}$ characterises the intrinsic uncertainty in our calculated distribution, which we set to the maximum between offsets of 16th and 84th percentiles from the median (i.e., the maximum between upper and lower errors reported in Table \ref{tab:vpec_and_distances}); we further inflate the uncertainty by assuming that it is underestimated by some fractional value $f_\mathrm{err}$. We use flat priors for all hyperparameters: $\theta_{0, \mtot}\sim U(-5, 5)$, $\theta_{1, \mtot}\sim U(-5, 5)$, and $f_\mathrm{err}\sim U(0,1)$. The sampling is then performed with {\sc pymc} using the same configuration for the sampler as in Sec \ref{sec:pkv_distribution_analyses}. All chains have good convergence with $|\hat R - 1| / \hat R < 0.001$, and samples from all MCMC runs are concatenated and presented in Figure \ref{fig:linear_correlation_corner_plots}. The fit results (Table \ref{tab:linear_fit_results}) strongly suggest a negative slope ($\theta_{1, \mtot}$), with a $99.7\%$ credible interval ranging from $-0.84$ to $-0.16$; 
 the fit also suggests a moderate error fraction $f_\mathrm{err} = 0.20$, indicating some additional uncertainty to the measured velocities.

With the simulated data, we use {\sc scipy} to calculate $\rhop$ and $\rhos$ and the associated \pvalue s. As a point of comparison, we also calculate the coefficients with velocities sampled from distributions that are ``smeared" by $\ferr$ percent additional error (in Table \ref{tab:linear_fit_results}). When no additional error is included, all simulations yield a negative coefficient, hinting at a significant anti-correlation (Table \ref{tab:pearson_r_and_power_law_fit}); a more sensible measure of robustness is the \pvalue; here, we compare the \pvalue s to an a priori significance level of $1\%$ and calculate $\frej$ (fraction of simulations that reject the null hypothesis; Sec \ref{sec:KS_test}).

When no additional error is incorporated, all simulations favour an anti-correlation according to the results using $\rhop$ and $\rhos$, respectively. $\frej$ drops to a lower percentage of $85\%$ (or $84\%$ for $\rhos$) when a fraction of $\ferr$ additional errors is added to $\pkv$ which still suggests a significant negative correlation.

\begin{figure*}
    \centering
    \includegraphics[width=0.7\textwidth]{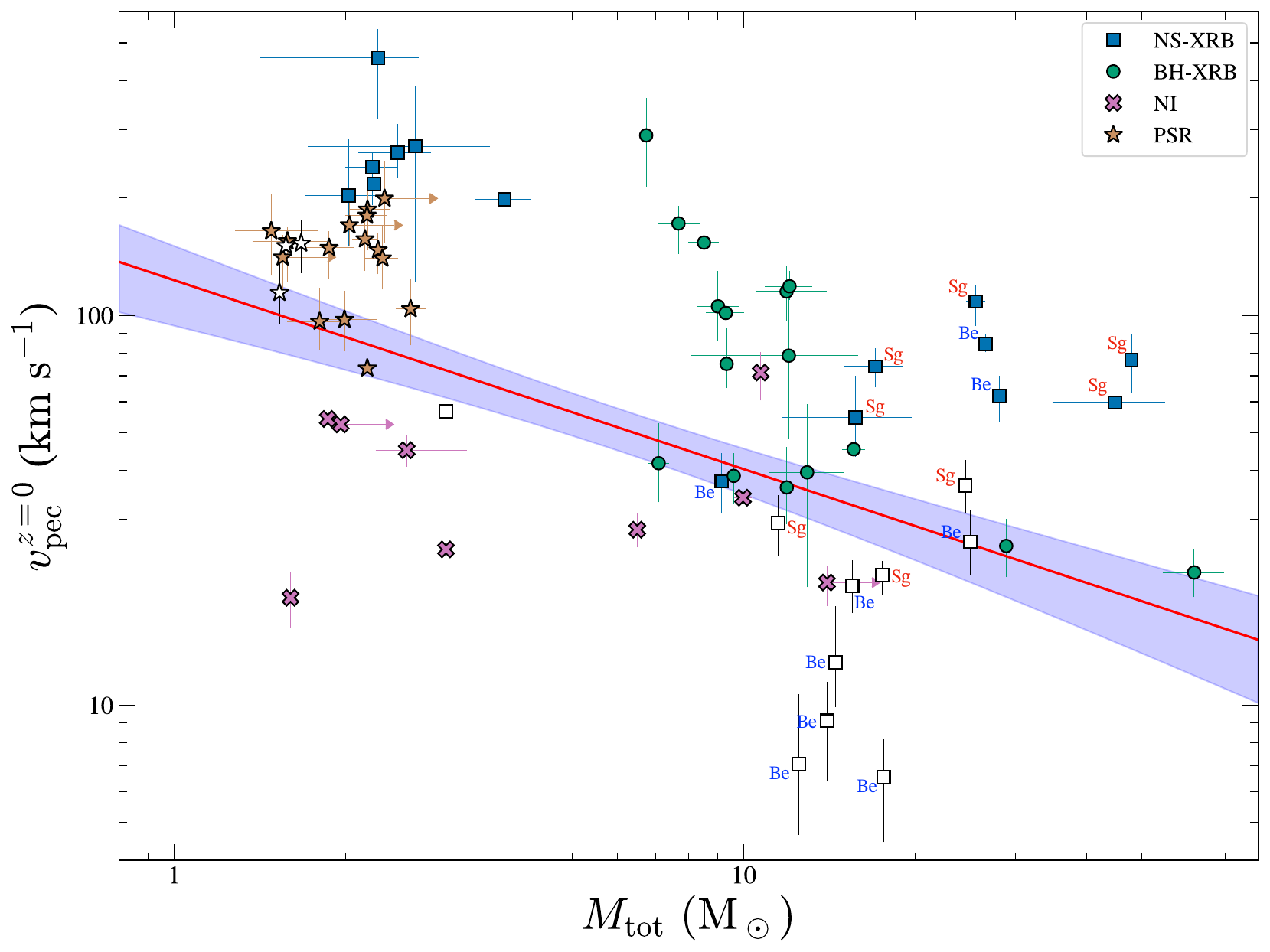}
    \caption{$\pkv$ vs. $\mtot$ for binaries with constrained or lower limits on $\mtot$ (filled markers) and those that have only approximate masses (empty markers with no error bars on $\mtot$). The error bars of $\pkv$ correspond to $16$th and $84$th percentiles. The red solid line indicates the best-fit model (Sec \ref{sec:m_vs_v}), with the blue shading indicating the $68\%$ confidence region. The NS-HMXBs are further distinguished between those that have a supergiant companion (``Sg") and a Be companion (``Be").}
    \label{fig:vpec_vs_mass}
\end{figure*}

\subsection{Orbital period and peculiar velocity}
\label{sec:porb_vs_pkv}
The binaries in our compilation all survived their NKs; their observed orbital periods ($\porb$) can give insights on how tightly bound they were and thus on distribution of NK magnitudes. We thus also include known $\porb$ in the compilation (Table \ref{tab:vpec_and_distances}). 

It is also interesting to check if $\porb$ and $\pkv$ show any evidence of dependences, so we follow a Bayesian approach similar to that in Sec \ref{sec:m_vs_v} to fit the data. The intercept and slope are denoted as $\theta_{0, \porb}$ and $\theta_{1, \porb}$.

\begin{figure*}
    \centering
    \includegraphics[width=0.7\textwidth]{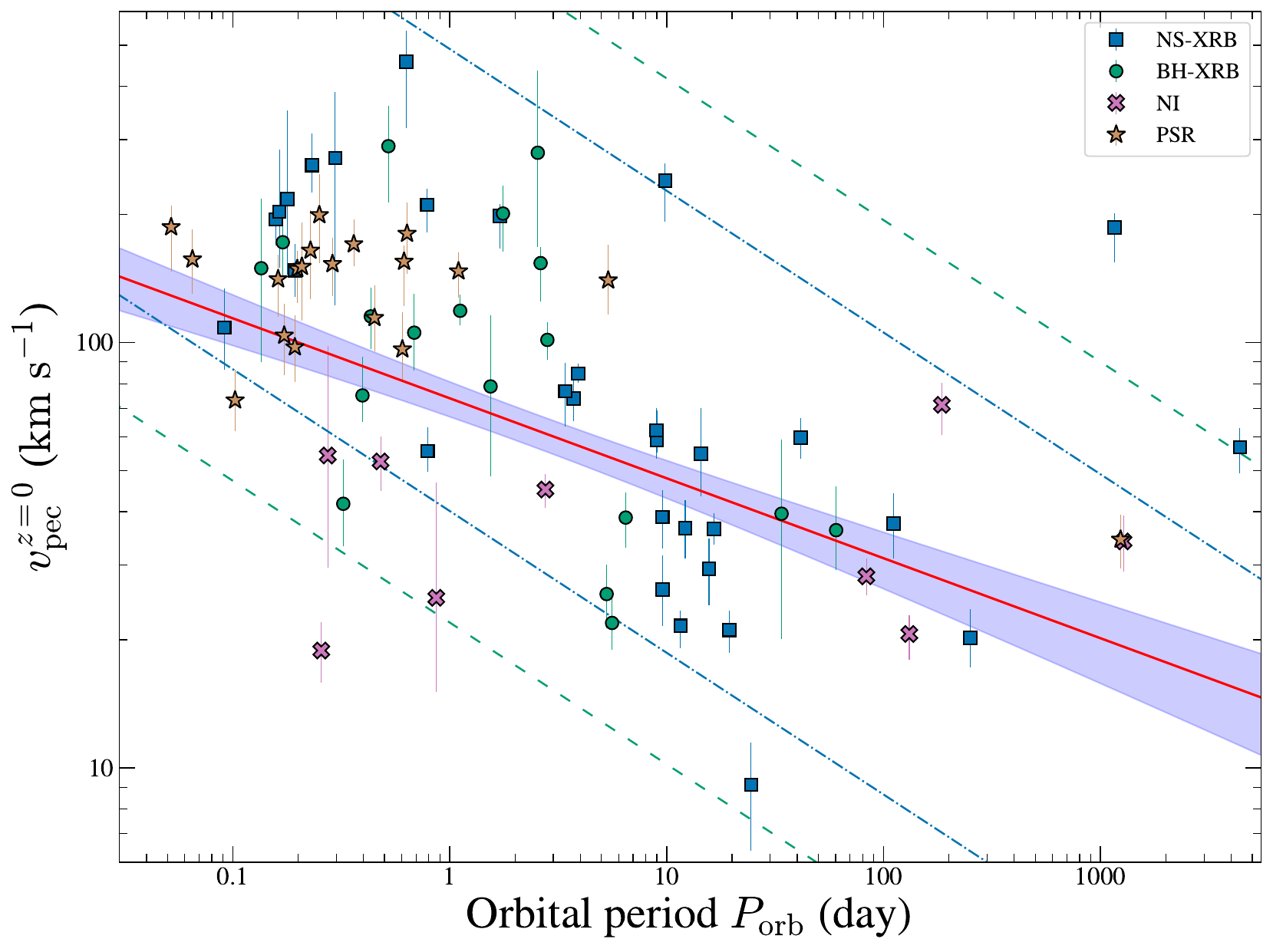}
    \caption{Scatter plot of $\pkv$ and $\porb$ for binaries with measured $\porb$. The error bars on $\pkv$ correspond to $16$th and $84$th percentiles from Table \ref{tab:vpec_and_distances}. The model with the median bootstrapped fit parameters is shown with the solid red line; the blue shading indicates the $68\%$ confidence region. The blue dashed-dotted line and green dashed line represent the \citetalias{Kalogera96} limits (as example trend lines) for a typical NS and BH-LMXB, respectively, assuming $\mcomp$ and $\Delta M$ in Table \ref{tab:k96_limits_parameters} and $\mnoncomp=0.5\,\Msun$.}
    \label{fig:pkv_vs_porb_plot}
\end{figure*}

This model is applied to simulated data generated by a similar bootstrapping process as in Sec \ref{sec:m_vs_v}, i.e., for each binary that has a well measured $\porb$, we sample 1000 velocities from the calculated distribution; the only difference here is that we do not simulate $\porb$ but use the literature $\porb$ values for all simulated velocities as the compiled $\porb$'s are all tightly-constrained (median fractional error of $5\times 10^{-6}$). The fit result suggests a negative slope of $-0.19$ with a $99.7\%$ credible interval between $-0.29$ and $-0.09$. We plot the distribution of model parameters in Figure \ref{fig:linear_correlation_corner_plots} and plot the scatters of $\pkv$ and $\porb$ in Figure \ref{fig:pkv_vs_porb_plot}.

Following the same approach, we also calculate the $\rhos$ and $\rhop$ for $\pkv$ vs. $\porb$. The correlation is significant as all simulations yield \pvalue s below $1\%$, and even with additional uncertainty in $\pkv$, the $\frej$ is still at a robust level (Table \ref{tab:pearson_r_and_power_law_fit}). We can therefore conclude that the anti-correlation between $\pkv$ and $\porb$ is statistically significant.

\begin{table}
    \caption{Log-linear fit results, with $68\%$ confidence uncertainties.}
    \centering
    \begin{tabular}{cccc}
    \toprule
        $\pkv$ vs. & $\theta_{0}$ & $\theta_{1}$ & $\ferr$ \\
    \midrule
        $\mtot$ & $1.71\pm 0.05$ & $-0.48\pm 0.11$ & $0.20\pm 0.03$ \\
        $\porb$ & $1.80 \pm 0.04$ & $-0.19 \pm 0.03$ & $0.16\pm 0.02$ \\
    \bottomrule
    \end{tabular}
    \label{tab:linear_fit_results}
\end{table}

\begin{table*}
    \caption{Pearson and Spearman's coefficients and the corresponding \pvalue s. Uncertainties correspond to the $68\%$ confidence level.}
    \centering
    \begin{tabular}{ccccccc}
    \toprule
        $\pkv$ vs. & $\rhop$ & \pvalue\ ($\rhop$) & $\frej$ (\%) & $\rhos$ & \pvalue\ ($\rhos$) & $\frej$ (\%)\\
    \midrule
    \multicolumn{7}{c}{With no additional error in $\pkv$}\\
        $\mtot$ & $-0.52\pm 0.03$ & $4.69_{-3.61}^{+15.05}\times 10^{-6}$ & $100$ & $-0.55\pm 0.03$ & $1.18_{-0.96}^{+4.91}\times 10^{-6}$ & $100$ \\
        $\porb$ & $-0.53\pm 0.02$ & $6.68^{+15.31}_{-4.69}\times 10^{-7}$ & $100$ & $-0.57\pm 0.02$ & $5.15_{-4.11}^{+17.48}\times 10^{-8}$ & $100$ \\
        \\
    \multicolumn{7}{c}{With $\ferr$ percent additional error in $\pkv$} \\
        $\mtot$ & $-0.39\pm 0.08$ & $9.96_{-9.20}^{+80.36}\times 10^{-4}$ & $85$ & $-0.39\pm 0.08$ & $9.09_{-8.49}^{+90.90}\times 10^{-4}$ & $84$ \\
        $\porb$ & $-0.46\pm 0.06$ & $2.96^{+32.04}_{-2.80}\times 10^{-4}$ & $99$ & $-0.50\pm 0.07$ & $5.81^{+108.85}_{-5.63}\times 10^{-6}$ & $99$ \\
    \bottomrule
    \end{tabular}
    \label{tab:pearson_r_and_power_law_fit}
\end{table*}

\section{Discussion}
\label{sec:discussion}
Our statistical analyses together suggest a significant anti-correlation between $\pkv$ and $\mtot$, and between $\pkv$ and $\porb$. In this section, we present some physical interpretation of these results and discuss possible observational and physical biases that could have contributed to our results.

\subsection{Peculiar velocity vs. binary total mass}
\label{sec:discussion_on_v_vs_mass}
NKs imparted on compact objects at SNe can be broadly attributed to two effects: recoil due to a sudden ejection of mass \citep[also known as the ``Blaauw kick",][]{Blaauw61} and/or any additional impulses of momentum on the compact object. The difference between the two is that the former conserves system (ejecta-binary) linear momentum, so binaries in which the companion star of the compact object is more massive would acquire a smaller $\pkv$ if the ejecta momenta are comparable among different binaries. This can, however, be complicated by the fact that the orbital speeds of the supernova progenitor can, in some cases, be faster in shorter-period binaries, which may be associated with lower mass companion stars.  An impulse of an additional momentum beyond ejecta-induced recoil, on the other hand, is equivalent to having an external force on the compact object, so this conservation is no longer met. With an additional kick, even a massive binary can end up with high $\pkv$ if the kick velocity is at a supportive magnitude and direction that keep the binary bound.

The kinematic formulation of $\pkv$ was derived by \citet{Kalogera96} (\citetalias{Kalogera96} hereafter) which incorporates the effect of both mass ejecta and additional kick velocity. According to \citetalias{Kalogera96}, a bound binary that survived the SN gets a $\pkv$ constrained between an upper and lower limit given by\footnote{Note that the equations are slightly different from the K96 version because we re-write it with the notations in this paper (Sec \ref{sec:notations}).}:

\begin{equation}
    \begin{aligned}
        \pkv[,min] / V_r &= \frac{M_1 + \Delta M}{\mtot + \Delta M} - \frac{\sqrt{2}\mcomp}{\mtot^{1/2}\lrb{\mtot + \Delta M}^{1/2}}\\
        \pkv[,max] / V_r &= \frac{M_1 + \Delta M}{\mtot + \Delta M} + \frac{\sqrt{2}\mcomp}{\mtot^{1/2}\lrb{\mtot + \Delta M}^{1/2}}
    \end{aligned},
    \label{eq:k96_limits}
\end{equation}
where $\Delta M$ is the ejecta mass; $V_r$ is the relative orbital velocity at SN. If one assumes a circular orbit, 
\begin{equation}
    V_r = 212.9 \lrb{\frac{\mtot + \Delta M}{\Msun}}^{1/3}\lrb{\frac{\porbi}{\mathrm{day}}}^{-1/3}\,\kms,
    \label{eq:V_r}
\end{equation}
where $\porbi$ is the orbital period at the instant of SN. Physically, the upper limit is reached when the NK is in the opposite direction to the orbital velocity of the BH/NS progenitor at the instant of SN while keeping the binary bound; the lower limit, on the other hand, corresponds to the minimum possible NK on the compact object to keep the binary from being disrupted by mass ejection. Eq \ref{eq:k96_limits} and \ref{eq:V_r} immediately point to an anti-correlation with $\mtot$; this simply indicates that more massive systems are harder to be accelerated by NKs.

One can also get the difference between the maximum and minimum $\pkv$:
\begin{equation}
    \begin{aligned}
        \pkv[,max] - \pkv[,min] &= \frac{2\sqrt{2}\mcomp}{\mtot^{1/2}\lrb{\mtot + \Delta M}^{1/2}}V_r \\ 
        & \propto \mcomp \mtot^{-1/2}\lrb{\mtot+\Delta M}^{-1/6}\porbi^{-1/3}
    \end{aligned}.
    \label{eq:v_max_m_v_min}
\end{equation}
This difference is related to the width of $\pkv$ distributions. 

A simple comparison can then be performed between different binary subgroups hosting NSs, given that $\mcomp$ for these binaries are very similar. From eq \ref{eq:v_max_m_v_min}, it immediately follows that NS-LMXBs exhibit a broader distribution compared to HMXBs because they have: (1) more compact orbits at SN, (2) smaller progenitor (helium core) mass at SN thus less mass was ejected (smaller $\Delta M$), and (3) apparently smaller $\mtot$ values. The progenitors of LMXBs are thought to begin with a wide orbit which then dramatically shrinks through a common envelope (CE) process \citep{Paczynski76}. The shrunken orbits greatly increase the survivability against the coming SN and lead to higher $\pkv$s. While some HMXBs could have a comparable pre-SN orbit (e.g., Cen X-3; see e.g., \citealt{vandenHeuvel72}), the $\mtot^{-1/2}\lrb{\mtot+\Delta M}^{-1/6}$ factor contributes significantly considering a factor of few or even an order of magnitude difference in $\mtot$. 

\begin{figure}
    \centering
    \includegraphics[width=\columnwidth]{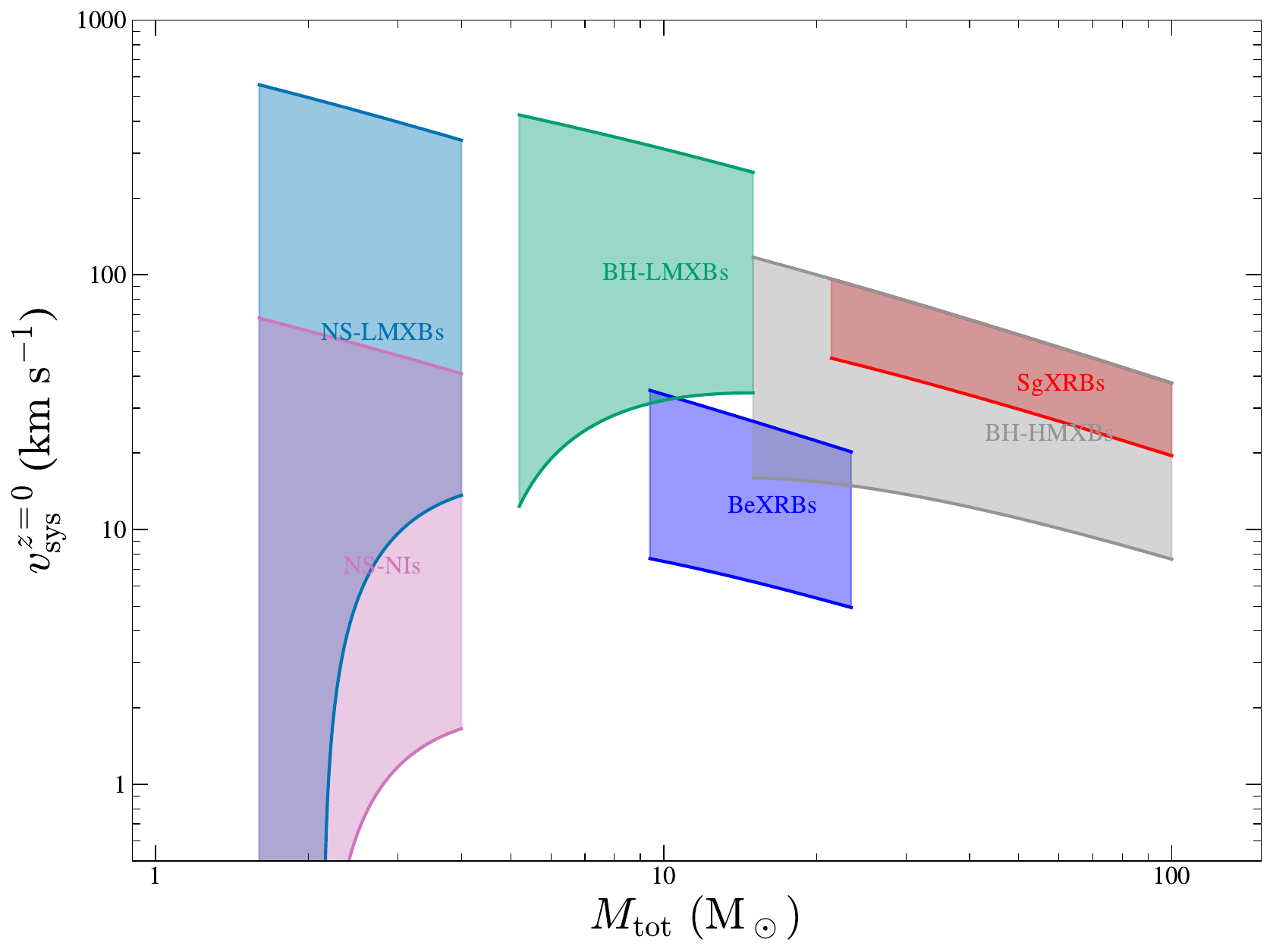}
    \caption{\citetalias{Kalogera96} limits (eq \ref{eq:k96_limits}) plotted for different binary types hosting NSs or BHs, using orbital parameters from Table \ref{tab:k96_limits_parameters}. A general decreasing trend with $\mtot$ is present despite each group having a range of possible $\pkv$ values.}
    \label{fig:pkv_vs_mtot_regions}
\end{figure}

Our compiled NS-LMXBs are within the ``allowed" range defined by the \citetalias{Kalogera96} limits (Figure \ref{fig:pkv_vs_mtot_regions}); however, we noted that all of them are above $\approx 50\,\kms$, so the lower-left of Figure \ref{fig:vpec_vs_mass} is clearly under-populated. It was pointed out by \citet{Kalogera98a} that some LMXBs could have originated from progenitors in wide orbits which then shrink if the kick velocity is at a favourable magnitude and direction; this ``direct-SN" scenario circumvents the need of a CE stage, and thus SN would occur at wide orbits, leading to low-$\pkv$ (typically between $20-50\,\kms$) binaries. Their absence could be a result of observational biases (see Sec \ref{sec:biases}). 

It is also worth noting that the NIs that host NSs have systematically lower $\pkv$ than NS-LMXBs, but the significance of this difference can only be addressed with a greater sample and more confident identifications of NS-NIs. Indeed, the nature of the compact object is yet ambiguous in some systems, as their $\mcomp$ can also overlap with the range for massive white dwarfs. Nevertheless, it is also possible that NS-NIs might have also been formed through the ``direct-SN" scenario similar to that for LMXBs, but the orbits were not perturbed sufficiently by the kick and shrunk enough for mass transfer to start. With the observed upper limit $\approx 60\,\kms$ for $\pkv$ and a $\Delta M=1\,\Msun$, the progenitor has a $\porbi \approx 560\,\mathrm{day}$ following eq \ref{eq:k96_limits}. Efficiency of such scenario is beyond the scope of this paper and can be addressed by future population synthesis studies.

One can also compare the range of $\pkv$ for subclasses of HMXBs. After the first mass-exchange stage (i.e., from the compact object progenitor to the companion), HMXBs with supergiant donors (SgXRBs) are thought to be in closer orbits (period could be as short as a few days; see e.g., \citealt{DeLoore74}), while those with Be stars (BeXRBs) could be much wider (periods of up to hundreds of days; e.g., \citealt{Habets85}). Although SgXRBs have slightly greater $\mtot$ than BeXRBs, they are not drastically different. Therefore, $\porbi$ plays a more crucial role. As a result, $\pkv$ for BeXRBs should be systematically downscaled by $V_r$ compared to SgXRBs. This distinction has been suggested by past studies \citep{vandenHeuvel00, Fortin22}, and will be tested with future statistical analyses (Nuchvanichakul et al., in prep). 

In all, the decreasing trend with increasing $\mtot$ could be roughly attributed to increasing inertia of binaries, regardless of additional kick velocity; this trend can also be guided by the \citetalias{Kalogera96} limits assuming some orbital parameters at SN for subgroups of binaries. To visualise this, in Figure \ref{fig:pkv_vs_mtot_regions}, we plot \citetalias{Kalogera96} limits for different types of binaries assuming some typical parameters at SN (Table \ref{tab:k96_limits_parameters}), where we further distinguish between SgXRBs and BeXRBs following the catalogue made by \citet{Fortin23}. Given the assumed $\porbi$ and $\Delta M$, most binaries are consistent with the \citetalias{Kalogera96} limits.

\fr{Several caveats should be considered in interpreting the results. Even though \citetalias{Kalogera96} is agnostic to detailed kick mechanism, it does rely on the assumption of a Gaussian distribution for the kick velocity components; however, it is possible that the distribution might differ for different kick mechanisms. Moreover, it should also be noted that we use the universal log-linear fit across different binary types as a complementary method to test for negative correlation, so caution should be exercised in drawing very detailed conclusions about the underlying physical processes; with discoveries of more binaries in each type, type-specific analyses (e.g., with Bayesian hierarchical modelling) would certainly be helpful to explore different kick mechanisms.}  

\begin{table}
    \centering
    \caption{Typical parameters for different binary types used to plot \citetalias{Kalogera96} limits in Figure \ref{fig:pkv_vs_mtot_regions}. }    
    \begin{tabular}{lccc}
    \toprule
      Type     & $\mcomp$ ($\Msun$) & $\Delta M$ ($\Msun$) & $\porbi$ (day)  \\
    \midrule
      NS-LMXB  & $1.4$ & $1.0$ & $1$ \\
      BH-LMXB  & $5.0$ & $6.0$ & $10$ \\
      BH-HMXB  & $5.0$ & $6.0$ & $100$ \\
      SgXRB    & $1.4$ & $5.0$ & $10$ \\
      BeXRB    & $1.4$ & $2.0$ & $300$ \\
      NS-NI    & $1.4$ & $1.0$ & $560$ \\
    \bottomrule
    \end{tabular}
    \label{tab:k96_limits_parameters}
\end{table}

\subsection{GX 1+4: a wide symbiotic X-ray binary with a high peculiar velocity}
\label{sec:syxrbs}
\fr{
GX 1+4 is a rare case of symbiotic X-ray binary where a NS is in a wide orbit ($\porb = 1160\,\mathrm{day}$) with a late-type giant \citep[SyXRBs;][]{Hinkle06}. In our compilation, it has a large $\pkv$ ($\gtrsim 100\,\kms$) so stands out in the upper right region of Figure \ref{fig:pkv_vs_porb_plot}. A high $\pkv$ could be a result of a strong NK, but this then contradicts its wide orbit that is prone to disruption by strong kicks. GX 1+4 does not have a well-measured $\gaia$\ parallax, so we adopt the distance estimate of $4.5\,\kpc$\ in \citet{Hinkle06}. Indeed, its negative parallax indicates that the true parallax is very small, so the distance should be greater than $4.5\,\kpc$\ \citep{Bailer-Jones21}, which will further increase its $\pkv$ value. As a comparison, another SyXRB, 4U 1700+24, has a wider orbit ($\porb=4391\,\mathrm{day}$; \citealt{Hinkle19}) but is less of an outlier considering that its $\pkv$ is close to the velocity dispersion of the disc population ($\approx 40\,\kms$; \citealt{Carlberg85}). 

There are multiple scenarios where wide SyXRBs can survive SNe. One possibility is that NSs in some SyXRB progenitors were formed through electron capture SNe \citep{Miyaji80, Nomoto84} or accretion-induced collapse \citep[AIC; ][]{Freire14}. Theoetrical work suggests that these events generally impart low NKs on the NSs \citep{Dessart06, Kitaura06, Gessner18}, despite the fact that population synthesis suggested that only a small fraction formed in this way \citep{Lu12}. Alternatively, the massive component (i.e., the NS progenitor) might have lost a considerable fraction of its envelope through a CE stage \citep{Taam00}, which in turn limits the amount of mass ejected in the following SNe and thus the Blauww kick. However, this is hard to reconcile with the observed high $\pkv$ values based on either a weak kick or a small mass loss. Following the formulation of \citet{Nelemans99}, after a core-collapse SN, the ejecta mass has to be less than half of the total of helium core and companion mass to not disrupt the progenitor binary; if GX 1+4 starts in a compact binary at the SN with $\porbi=10\,\mathrm{day}$, and if one assumes a helium core mass of $3\,\Msun$ and a companion mass of $1\,\Msun$ at SN, mass loss can only contribute a maximum of only $\approx 63\,\kms$ to the $\pkv$; then post-SN evolution will have to involve mechanisms that expand the orbit to match its present-day $\porb$. Fortuitous cases where the NS is accelerated by appropriately directed weak kicks might reconcile with the large $\pkv$ but should be very rare.

Alternatively, GX 1+4's proximity to the Galactic centre might suggest that it is a binary born in the bulge. In this case, it would have been dynamically perturbed by the dense stellar population there, so its $\pkv$ is not reflective of a NK. However, it is worth noting that SyXRBs with wide orbits are particularly susceptible to disruption due to dynamical encounters in dense environments \citep[see e.g.,][]{Belloni20}. There have been no confirmed symbiotic stars in comparably dense globular clusters, although there is one controversial candidate \citep{Henleywillis2018, Belloni20}.
}

\subsection{Present-day orbital period vs. peculiar velocity}
\label{sec:discussion_on_v_vs_porb}
\revised{Interpreting the anti-correlation with $\porb$ following \citetalias{Kalogera96} can be more complex. One could naively ascribe the decreasing trend in $\pkv$ to the upper \citetalias{Kalogera96} limits that scale negatively with $\porbi$ (see Figure \ref{fig:pkv_vs_porb_plot}); however, transposing from $\porbi$ to the present-day $\porb$ requires knowledge of (1) the kick magnitude and direction on the compact object, and (2) post-SN orbital evolution. \citet[][B95, hereafter]{Brandt95} have conducted a systematic investigation of pre- and post-SN orbital parameters and found that most binaries have a post-SN orbital period ($\porbp$) close to $\porbi$ within a factor of a few (fig 4a in \citetalias{Brandt95}). Additionally, \citetalias{Brandt95} points out that among binaries that survived SNe, the ones with high $\vpec$ values are preferentially those perturbed by NKs in approximately the {\em opposite} direction to the instantaneous orbital velocity of the compact objects; this can essentially decelerate the compact object and shrink the orbits, resulting in a decreasing trend with $\porbp$ (e.g., fig 5b in \citetalias{Brandt95}).

After SNe, there are various mechanisms that can further change the orbital period. While $\porbp$ should be reasonably close to their present-day values for the HM binaries considering their young ages, the orbit of LM binaries (especially LMXBs) is dependent on the synergy of mass transfer and/or processes that dissipate angular momentum. The former sets in when the mass donor fills its Roche lobe due to nuclear evolution (sub-giant donors, e.g., Cyg X-2) or orbit shrinkage induced by angular momentum loss (main-sequence donors); the orbit tends to expand on thermal timescale under stable and conservative mass transfer. However, to maintain the donor's contact with its Roche lobe, mechanisms are needed to extract angular momentum from the system to counteract orbit expansion. The current consensus is that angular momentum can be dissipated via gravitational wave radiation \citep{Faulkner71, Verbunt93} and/or magnetic braking \citep{Rappaport83}; the former only dominates in ultra-compact X-ray binaries, while magnetic braking is considered effective in the majority of LMXBs. Over these processes, orbits can expand or shrink at the bifurcation period \citep[$\approx 1\,\mathrm{day}$;][]{Tutukov85, Pylyser88}. Depending on the initial (at the onset of mass transfer) state of the donor and magnetic braking prescription, the overall mass transfer phase can modify the orbital period by up to $\approx$ 2 orders of magnitude \citep[see e.g.,][]{Ergma96, Podsiadlowski02, Van19}. 

Despite this multitude of uncertainty, Figure\,\ref{fig:pkv_vs_porb_plot} still shows a significant anti-correlation. So while different evolutionary trajectories probably account for much of the scatter visible in the figure, these do not erase underlying properties linking binaries across the dynamic range of our sample. Figure\,\ref{fig:pkv_vs_porb_plot} thus provides a useful constraint for population synthesis models to satisfy. The strongest governing link is likely the system binding energy, with wider binaries being loosely bound and thus being more likely to be disrupted when $\pkv$ is higher. Testing this hypothesis, however, will require proper characterisation of XRB selection, combined with \gaia's selection function \citep[cf. ][]{Cantat-Gaudin2023} which is beyond the scope of the present work. 
}

\subsection{Possible biases}
\label{sec:biases}

The $\pkv$-$\mtot$ and $\pkv$-$\porb$ correlations could have been contributed by selection biases on our sample. It should be firstly noted that our compilation contains only NSs and BHs in \emph{present-day} bound systems, which is not necessarily a representation of the binary demography at the instant of SNe. In part, either mass ejection or additional kicks would typically bias against wide binaries, which were more prone to being disrupted; NSs in very close pre-SN orbits, on the other hand, could be re-directed to collide with the companion and form Thorne-\.Zytkow objects \citep{Thorne75, Thorne77}.

Wide binaries are also prone to observational biases considering the difficulty of measuring long $\porb$. There is an obvious paucity of binaries above $\porb\approx 100\,\mathrm{day}$ in Figure \ref{fig:pkv_vs_porb_plot}: six binaries are present including two wide SyXRBs (Sec \ref{sec:syxrbs}), the non-interacting binary Gaia BH2 \citep{El-Badry23b}, the massive binary pulsar PSR B1259$-$63, and two HMXBs (1A 0535+262 and X Persi). Compact objects in NIs can only be confirmed by robust constraints on mass function, which in turn asks for multi-epoch observations spanning at least one full orbital period; for very long $\porb$'s, this could be time-expensive. Dedicated radial velocity campaign can therefore only be performed for a handful of well-justified cases, despite the fact that there might be a uncharted population of wide NIs \citep[e.g.,][]{El-Badry23a, El-Badry23b}. Wide binaries can also be sought by searches for astrometric wobble with, e.g. \gaia\ \citep{Gould, Barstow}.

There might also be more wide BeXRBs to be discovered \citep{Vinciguerra2020}. BeXRBs are mostly transient systems that exhibit periodic X-ray outbursts (Type I outbursts) that can be orders of magnitude brighter than their quiescence. Because these outbursts happen at roughly $\porb$, they can be used to measure their $\porb$. However, this relies on the consecutiveness of outburst detection, which is less feasible if $\porb$ is very long. Moreover, long $\porb$ might also restrict X-ray detectability, typically for wind-fed HMXBs. Identification of a NS in HMXBs can be assisted by detections of (accretion-induced) X-ray pulsations. However, at a given wind mass loss rate of the mass donor, X-ray luminosities of such binaries scale inversely with orbital separation \citep{Bondi44}, so wide systems would be more difficult to detect. Contrary to wind-fed HMXBs, thinner stellar wind in wide binaries could in favour discovery of more OB-pulsar binaries like PSR B1259$-$63. The NS nature in such binaries can be confirmed by detection of radio pulsations \citep{Johnston92, Lyne15}, which can otherwise be greatly smeared out by stronger stellar wind in more compact systems. 

\revised{We have excluded binaries in globular clusters from our sample; however, it is possible that some of the LMXBs in our sample were formed in globular clusters but were later dynamically ejected. This can account for typically (ultra-)compact X-ray binaries that are more likely to survive dynamical interactions in dense environment. Combining an escape velocity of $50\,\kms$ (velocity relative to the cluster) and cluster velocity dispersion of 100--120\,$\kms$ \citep{Vasiliev19} results in $\vpecp$ over $100\,\kms$; therefore, the high-velocity part of our sample might have been contaminated, although simulations only suggest a small number of such binaries \citep{Kremer18}. Future metalicity measurements will provide information to further identify/discover these ejected systems.}

\begin{figure}
    \centering
    \includegraphics[width=\columnwidth]{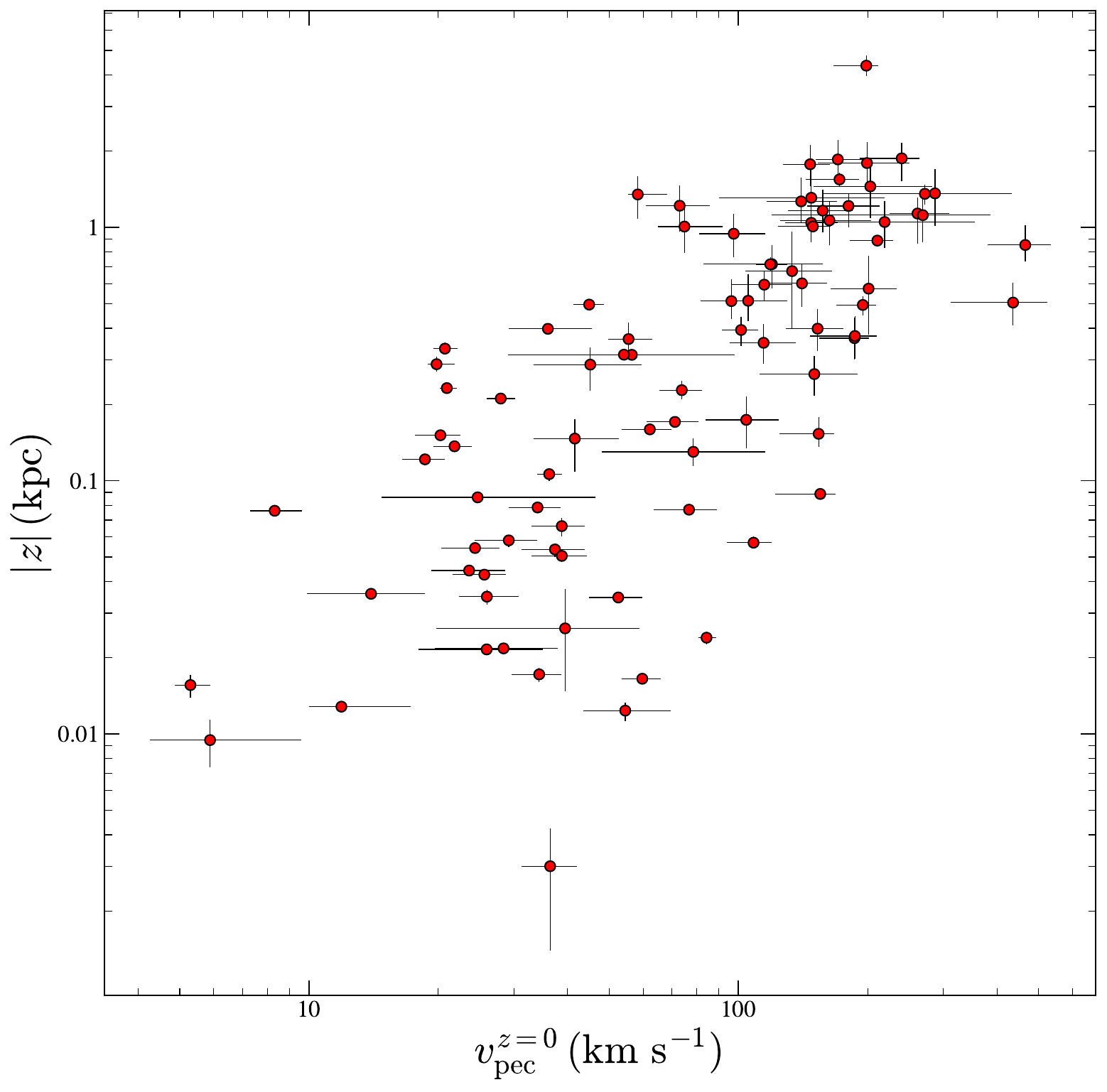}
    \caption{Galactic height $\galacticheight$ vs. $\pkv$ showing a clear positive correlation for binaries in Table \ref{tab:vpec_and_distances}. Error bars correspond to $16$th and $84$th percentiles. }
    \label{fig:z_vs_pkv}
\end{figure}

Another observational bias involves the Galactic height ($\galacticheight$) of binaries off the plane. We select the binaries based on robust identification of compact objects. This would require optical/near-IR visibility of the luminous component and would therefore bias against potential low-$\pkv$ systems as they are less likely to reach large $\galacticheight$ as the high-$\pkv$ ones (Figure \ref{fig:z_vs_pkv}) and therefore are more obscured by Galactic extinction \citep{Gandhi20, Jonker21}. This bias could be typically strong against LM binaries that have faint components, e.g., LM-NIs, and is even worse for the low-$\pkv$ LMXBs that formed through direct-SN (\citealt{Kalogera98a}; Sec \ref{sec:discussion_on_v_vs_mass}), which could have populated the lower-left region of the $\pkv$-$\porb$ plot. It will be interesting to constrain the extent to which these low-velocity binaries are affected by observational biases while exploring possible physical mechanisms that contribute to the lack of such binaries.

\section{Conclusion}
\label{sec:conclusion}
We compile a sample of \finalsamplesizewapprox\ binaries that host NSs and BHs and have measured astrometry and systemic radial velocities. We calculate their present-day peculiar velocities ($\vpecp$) and their extrapolated peculiar velocity at the Galactic disc ($\pkv$) which are treated as the potential velocity at birth incorporating uncertainties in astrometric parameters and Galactic constants. These calculated velocity samples can be compared to simulated distributions in future population synthesis studies.

We performed some statistical modelling the overall distribution and comparison between distributions of different subgroups. Our findings suggest no statistically significant difference between BH and NS binaries; however, we find that HM binaries are restricted to $\pkv$ below $\approx 100\,\kms$ while LM binaries exhibit a much wider distribution that goes up to $\approx 400\,\kms$. The difference is mostly contributed by NS binaries, while comparing BH-LM and HM binaries only suggest a moderate significance. 

Fitting a log-linear model of $\pkv$ vs. $\mtot$ results in a negative slopes at over $99.7\%$ confidence, while a similar log-linear fit to $\pkv$ vs. $\porb$ also gives a significant negative slope (at 99.7\% confidence). Both fits suggest an non-negligible fraction of additional uncertainty, with which both anti-correlations are slightly weakened but are still at robust levels. The anti-correlation between $\pkv$ with $\mtot$ follows the kinematic formulation derived by \citet{Kalogera96}. Interpretation of the correlation with $\porb$ is complicated by uncertainty in underlying kick velocity distribution and binary evolution models, which would require more detailed population synthesis studies.

\section*{Acknowledgements}

\revised{We thank the anonymous reviewer for their helpful feedback}. This work has made use of data from the European Space Agency (ESA) mission {\it Gaia} (\url{https://www.cosmos.esa.int/gaia}), processed by the {\it Gaia}
Data Processing and Analysis Consortium (DPAC,
\url{https://www.cosmos.esa.int/web/gaia/dpac/consortium}). Funding for the DPAC
has been provided by national institutions, in particular the institutions
participating in the {\it Gaia} Multilateral Agreement. This research has also made use of the SIMBAD database, 
operated at CDS, Strasbourg, France. PG, YZ, and CDB acknowledge UKRI and STFC for support.

\section*{Data Availability}
The \gaia\ astrometry data are available by querying the \gaia\ archive. The calculated $\vpecp$ and $\pkv$ samples are available upon request.



\bibliographystyle{mnras}
\bibliography{ref_new} 



\appendix
\section{Galactic positions and proper motions of the binaries}
\begin{figure*}
    \centering
    \includegraphics[width=0.7\textwidth]{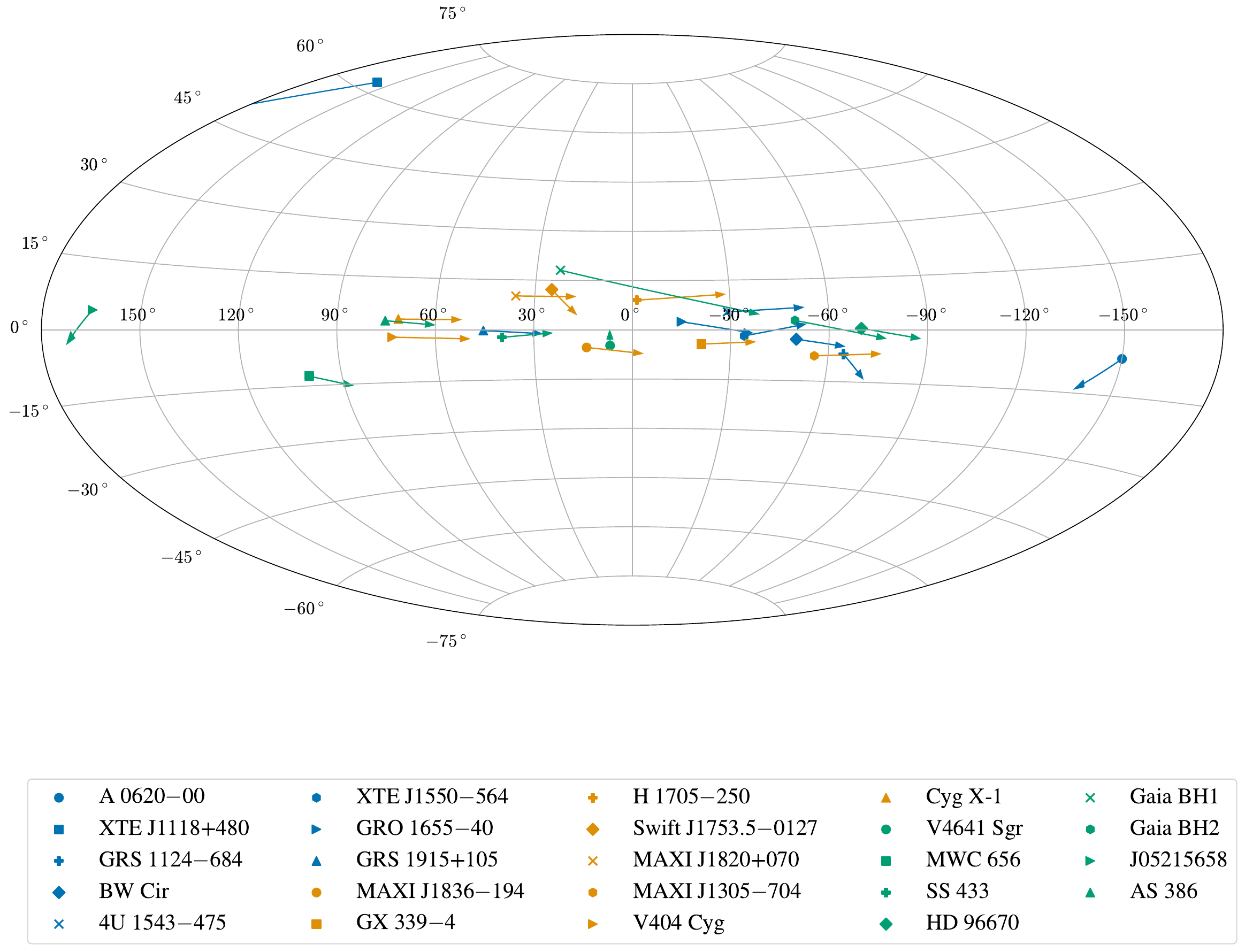}
    \caption{Galactic map of binaries that host a BH. Displacement due to proper motion is indicated by the length and direction of the arrows.}
    \label{fig:galactic_map_BH}
\end{figure*}

\begin{figure*}
    \centering
    \includegraphics[width=0.7\textwidth]{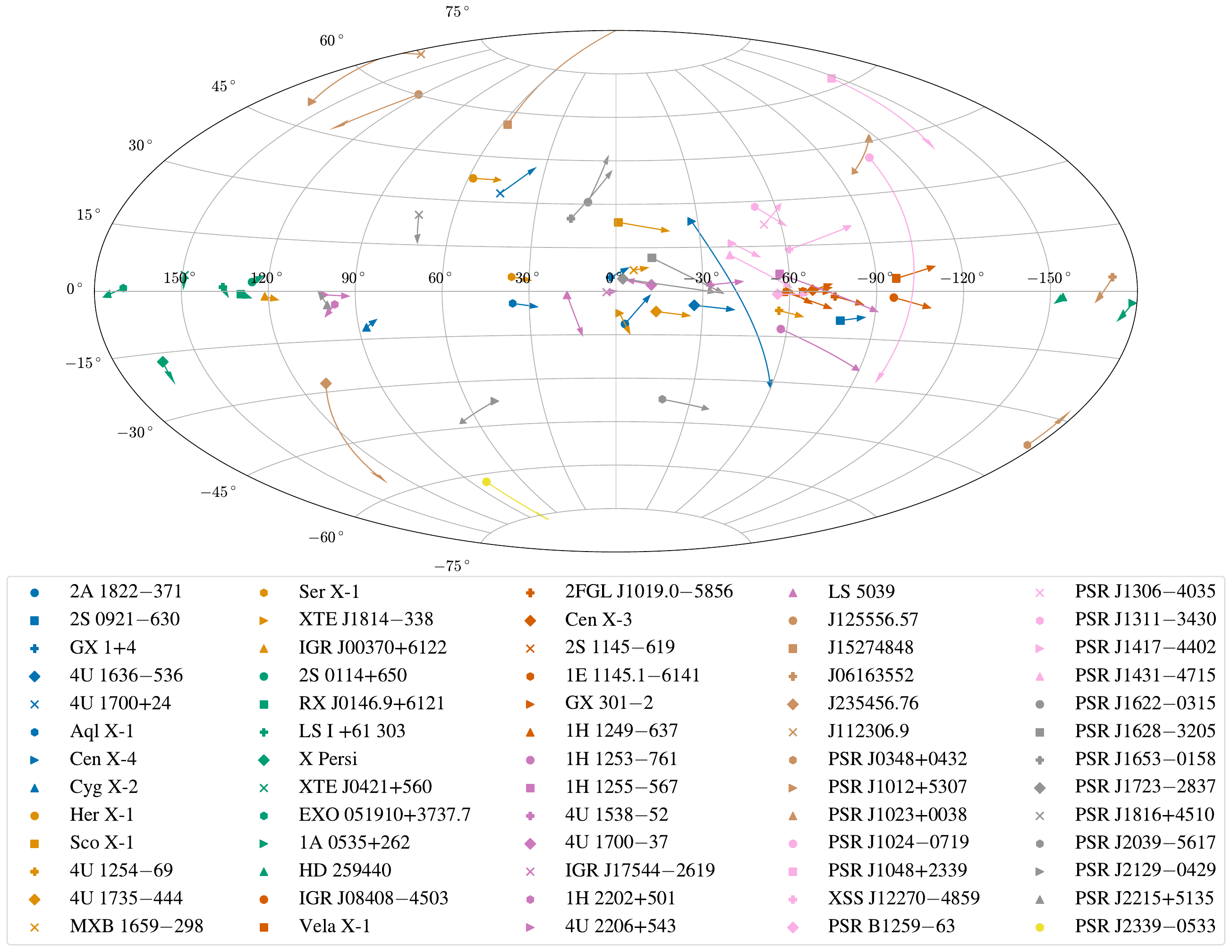}
    \caption{The same as Figure \ref{fig:galactic_map_BH} but plotted for binaries hosting a NS.}
    \label{fig:galactic_map_NS}
\end{figure*}

Galactic positions and proper motion vectors of the binaries in our compilation are shown in Figure \ref{fig:galactic_map_BH} and \ref{fig:galactic_map_NS}.

\section{Peculiar velocity distributions}
\begin{figure}
    \centering
    \includegraphics[width=\columnwidth]{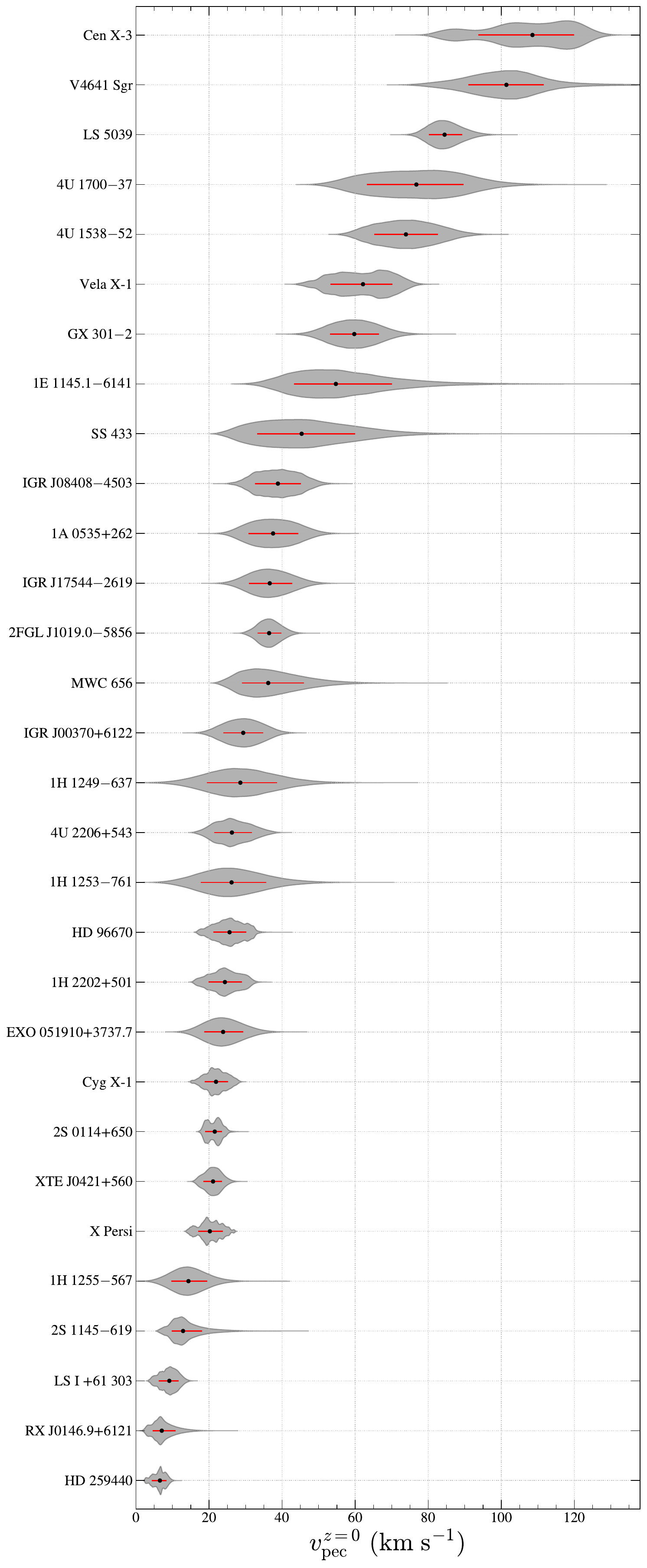}
    \caption{Probability density distribution of $\pkv$ of HMXBs, sorted by their medians. The median is marked by a filled black circle, and the range corresponding to the 16th and 84th percentiles are indicated with a red horizontal line in each distribution.}
    \label{fig:hmxb_vpec_violin_plot}
\end{figure}

\begin{figure}
    \centering
    \includegraphics[width=\columnwidth]{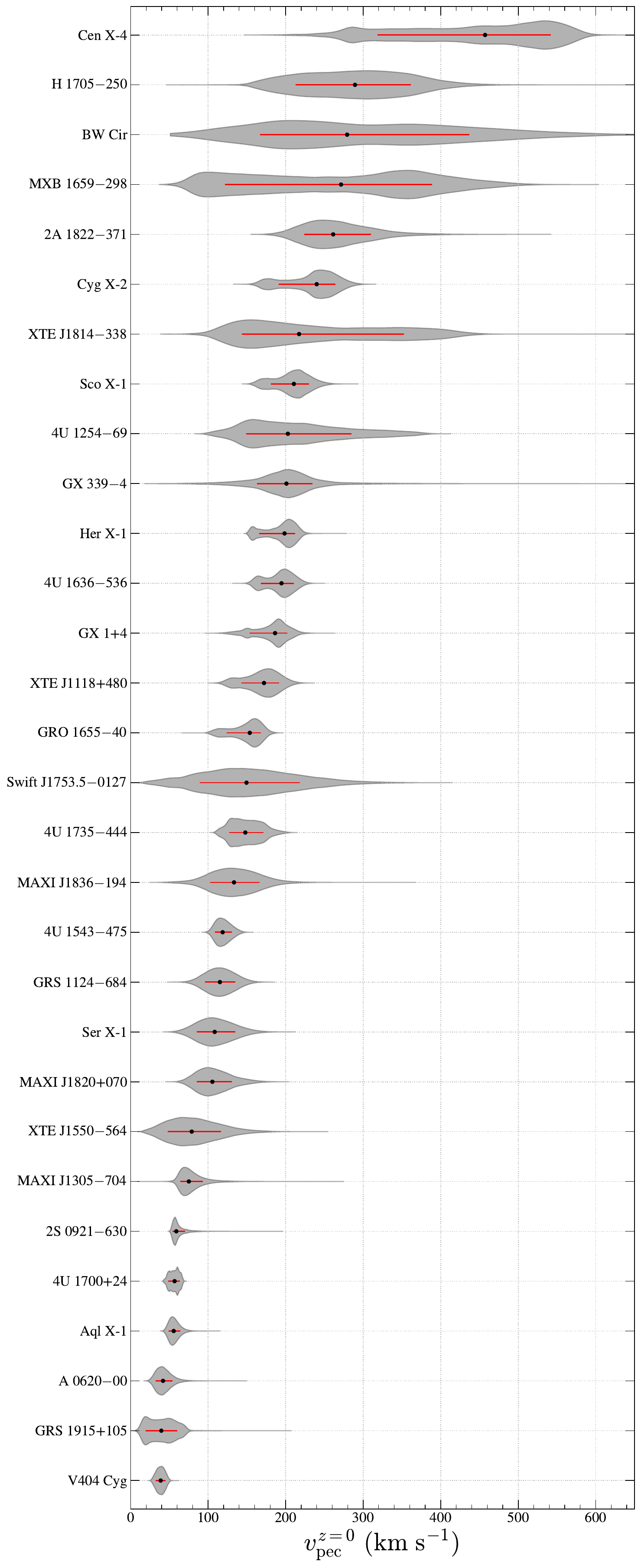}
    \caption{The same as Figure \ref{fig:hmxb_vpec_violin_plot}, but plotted for $\pkv$ of LMXBs.}
    \label{fig:lmxb_vpec_violin_plot}
\end{figure}

\begin{figure}
    \centering
    \includegraphics[width=\columnwidth]{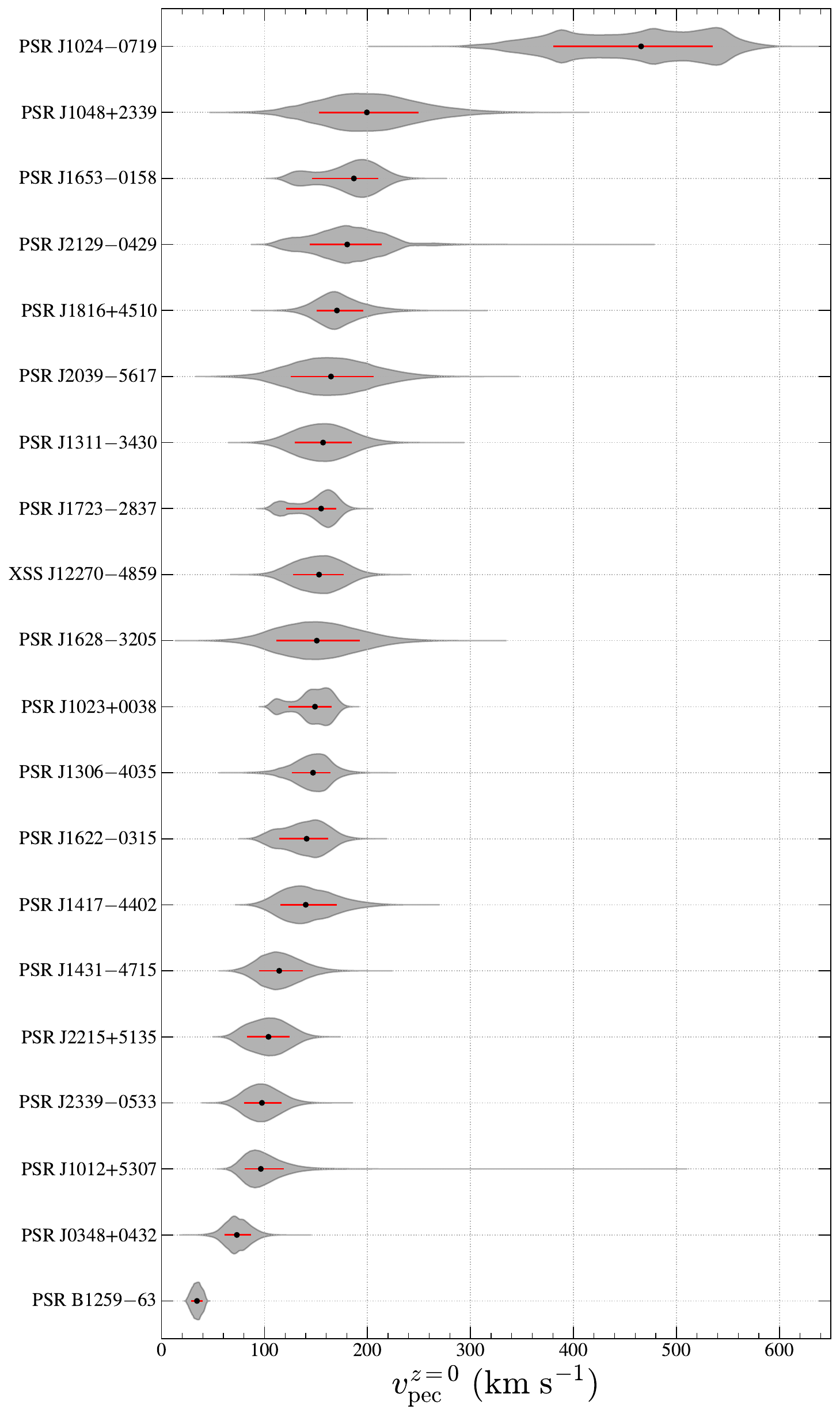}
    \caption{The same as Figure \ref{fig:hmxb_vpec_violin_plot}, but plotted for $\pkv$ of PSRs.}
    \label{fig:psr_vpec_violin_plot}
\end{figure}

\begin{figure}
    \centering
    \includegraphics[width=\columnwidth]{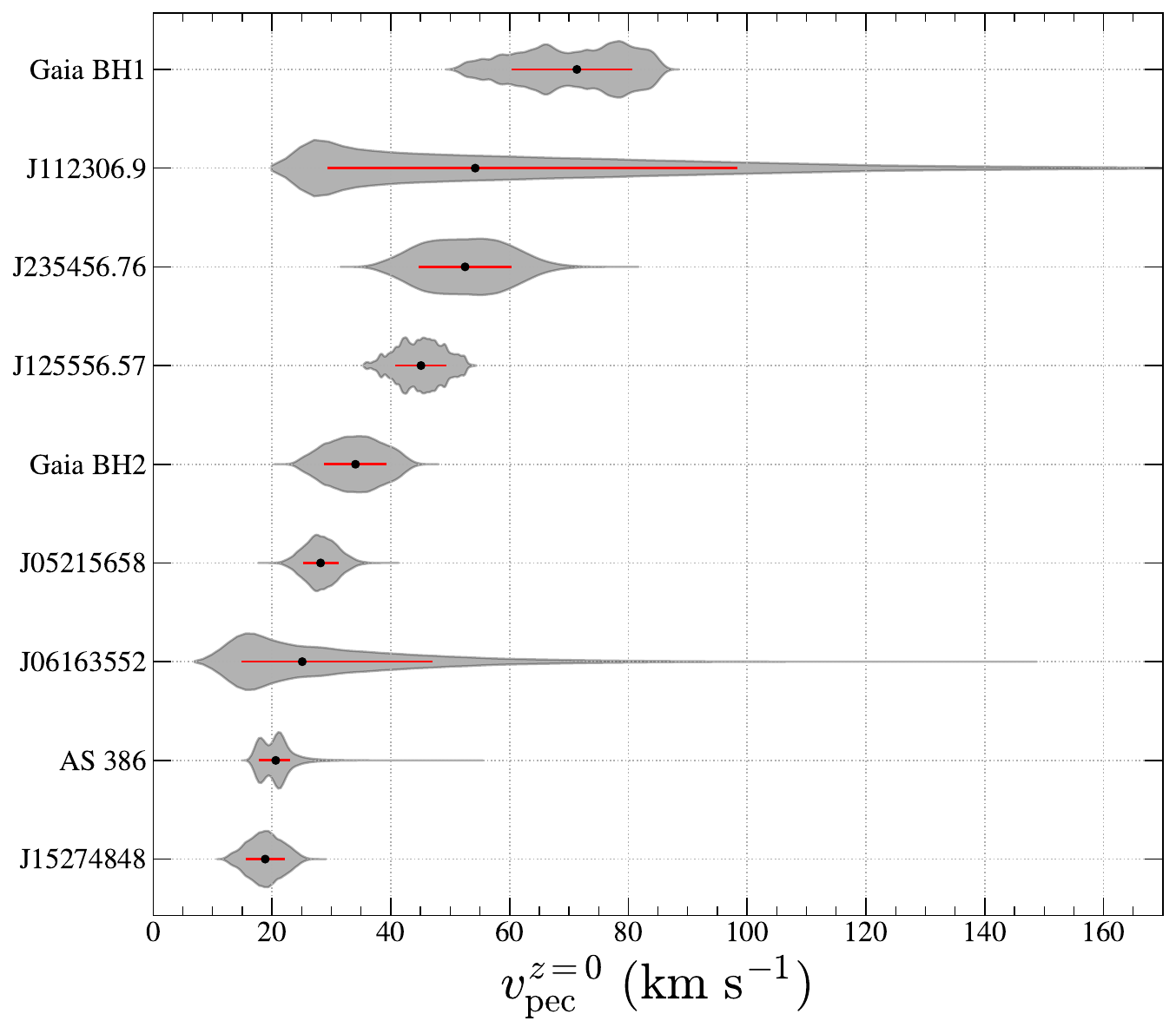}
    \caption{The same as Figure \ref{fig:hmxb_vpec_violin_plot}, but plotted for $\pkv$ of non-interacting (NI) binaries.}
    \label{fig:ni_vpec_violin_plot}
\end{figure}

In this section, we present distributions of potential peculiar velocity at birth ($\pkv$) for binaries in our compilation in Figure \ref{fig:lmxb_vpec_violin_plot}, \ref{fig:hmxb_vpec_violin_plot}, \ref{fig:psr_vpec_violin_plot}, and \ref{fig:ni_vpec_violin_plot}.

\section{Literature distances used for modeling the prior}
Here we summarise a list of literature distances ($d_\mathrm{lit}$) that are measured separately from \gaia\ in Table \ref{tab:d_lit_table}. These distances are used for fitting the scaling parameter ($L$) of the exponential prior (eq 4) and for calculating $\vpecp$ when \gaia\ parallax is poorly constrained.

\begin{table*}
    \centering
    \caption{Literature distances ($d_\mathrm{lit}$)}
    \begin{adjustbox}{max width=0.9\textwidth}
    \begin{threeparttable}
        \begin{tabular}{lllllllll}
            \toprule
            Name & $d_\mathrm{lit}$ & Ref & Name & $d_\mathrm{lit}$ & Ref & Name & $d_\mathrm{lit}$ & Ref \\
                 & (kpc) & & & (kpc) & & & (kpc) & \\
            \midrule
				IGR J00291+5934         &  $4.2\pm 0.5$        & [1]   & 				Aql X-1                 &  $5.2\pm 0.8$        & [17]  & 				4U 0115+634             &  $7.0\pm 0.3$        & [73]   \\
				4U 0919$-$54            &  $4.2\pm 1.3$        & [2]   & 				4U 1916$-$05            &  $6.8\pm 1.0$        & [9]   & 				V 0332+53               &  $7.5\pm 1.5$        & [74]   \\
				2S 0921$-$630           &  $8.5^\ast$          & [3]   & 				XTE J2123$-$058         &  $8.5\pm 2.5$        & [36]  & 				4U 0352+309             &  $0.7\pm 0.3$        & [75]   \\
				XTE J0929$-$314         &  $\geq 7.4$          & [4]   & 				Cyg X-2                 &  $11.0\pm 2.0$       & [9]   & 				RX J0440.9+4431         &  $3.3\pm 0.5$        & [76]   \\
				4U 1254$-$69            &  $13.0\pm 3.0$       & [5]   & 				GRO J0422+32            &  $2.5\pm 0.3$        & [37]  & 				A 0535+262              &  $2.0\pm 0.7$        & [77]   \\
				Cen X-4                 &  $1.4\pm 0.3$        & [6]   & 				A 0620$-$00             &  $1.1\pm 0.1$        & [38]  & 				XTE J0658$-$073         &  $3.9\pm 0.1$        & [78]   \\
				Cir X-1                 &  $9.3\pm 0.9$        & [7]   & 				MWC 656                 &  $2.6\pm 1.0$        & [39]  & 				3A 0726$-$260           &  $4.6\pm 1.3$        & [79]   \\
				UW CrB                  &  $\geq 5.0$          & [8]   & 				GRS 1009$-$45           &  $3.8\pm 0.3$        & [40]  & 				RX 0812.4$-$3114        &  $8.6\pm 1.8$        & [80]   \\
				4U 1608$-$52            &  $3.2\pm 0.3$        & [9]   & 				HD 96670                &  $2.8\pm 0.8$        & [41]  & 				GS 0834$-$430           &  $4.0\pm 1.0$        & [81]   \\
				Sco X-1                 &  $2.8\pm 0.3$        & [10]  & 				XTE J1118+480           &  $1.7\pm 0.1$        & [42]  & 				Vela X-1                &  $1.9\pm 0.2$        & [82]   \\
				4U 1626$-$67            &  $9.0\pm 4.0$        & [11]  & 				GRS 1124$-$684          &  $5.0\pm 0.7$        & [43]  & 				GRO J1008$-$57          &  $5.8\pm 0.5$        & [73]   \\
				4U 1636$-$536           &  $6.0\pm 0.5$        & [12]  & 				MAXI J1305$-$704        &  $7.7\pm 1.6$        & [44]  & 				1A 1118$-$615           &  $5.0\pm 2.0$        & [83]   \\
				Her X-1                 &  $6.6\pm 0.4$        & [13]  & 				MAXI J1348$-$630        &  $2.1\pm 0.6$        & [45]  & 				1E 1145.1$-$6141        &  $8.5\pm 1.5$        & [84]   \\
				MXB 1659$-$298          &  $9.0\pm 2.0$        & [9]   & 				BW Cir                  &  $\geq 25.0$         & [46]  & 				1H 1249$-$637           &  $0.4\pm 0.0$        & [85]   \\
				4U 1700+24              &  $0.4\pm 0.0$        & [14]  & 				Swift J1357.2$-$0933    &  $\geq 6.0$          & [47]  & 				1H 1253$-$761           &  $0.2\pm 0.0$        & [86]   \\
				4U 1702$-$429           &  $4.2\pm 0.1$        & [9]   & 				4U 1543$-$475           &  $7.5\pm 0.5$        & [17]  & 				1H 1255$-$567           &  $0.1\pm 0.0$        & [86]   \\
				4U 1705$-$32            &  $13.0\pm 2.0$       & [15]  & 				MAXI J1543$-$564        &  $\geq 8.5$          & [48]  & 				GX 304$-$1              &  $2.4\pm 0.5$        & [87]   \\
				4U 1705$-$44            &  $5.8\pm 0.2$        & [9]   & 				XTE J1550$-$564         &  $4.5\pm 0.5$        & [49]  & 				2S 1417$-$62            &  $6.2\pm 4.8$        & [88]   \\
				SAX J1712.6$-$3739      &  $5.1\pm 0.5$        & [16]  & 				4U 1630$-$472           &  $10.0^\ast$         & [50]  & 				SAX J1452.8$-$5949      &  $9.0\pm 3.0$        & [89]   \\
				1H 1715$-$321           &  $6.0\pm 0.9$        & [17]  & 				XTE J1650$-$500         &  $2.6\pm 0.7$        & [51]  & 				4U 1538$-$52            &  $6.4\pm 1.0$        & [90]   \\
				1RXS J171824.2$-$402934 &  $6.5\pm 0.5$        & [15]  & 				GRO 1655$-$40           &  $3.2\pm 0.2$        & [52]  & 				IGR J16318$-$4848       &  $3.5\pm 2.6$        & [91]   \\
				GRO J1719$-$24          &  $2.4\pm 0.4$        & [18]  & 				MAXI J1659$-$152        &  $8.6\pm 3.7$        & [53]  & 				IGR J16465$-$4507       &  $13.7\pm 9.9$       & [92]   \\
				GX 1+4                  &  $4.3^\ast$          & [19]  & 				GX 339$-$4              &  $\geq 5.0$          & [54]  & 				OAO 1657$-$415          &  $6.4\pm 1.5$        & [93]   \\
				4U 1728$-$34            &  $4.8\pm 0.3$        & [20]  & 				H1705$-$250             &  $8.6\pm 2.1$        & [17]  & 				XTE J1739$-$302         &  $2.4\pm 0.6$        & [94]   \\
				SLX 1735$-$269          &  $6.0\pm 1.2$        & [21]  & 				GRS 1716$-$249          &  $2.4\pm 0.4$        & [55]  & 				GRO J1750$-$27          &  $18.0\pm 4.0$       & [95]   \\
				4U 1735$-$444           &  $8.5\pm 1.3$        & [22]  & 				XTE J1720$-$318         &  $6.5\pm 3.5$        & [56]  & 				IGR J17544$-$2619       &  $3.0\pm 1.0$        & [96]   \\
				GRS 1741.9$-$2853       &  $9.0\pm 0.5$        & [23]  & 				GRS 1739$-$278          &  $7.2\pm 1.2$        & [57]  & 				SAX J1802.7$-$2017      &  $12.4\pm 0.1$       & [97]   \\
				GX 3+1                  &  $6.1\pm 0.1$        & [24]  & 				XTE J1752$-$223         &  $6.0\pm 2.0$        & [58]  & 				SAX J1818.6$-$1703      &  $2.7\pm 0.3$        & [80]   \\
				SAX J1747.0$-$2853      &  $7.5\pm 1.3$        & [25]  & 				Swift J1753.5$-$0127    &  $6.0\pm 2.0$        & [59]  & 				AX J1820.5$-$1434       &  $4.8\pm 1.2$        & [98]   \\
				IGR J17473$-$2721       &  $6.4\pm 0.9$        & [26]  & 				V4134 Sgr               &  $6.5\pm 2.5$        & [60]  & 				LS 5039                 &  $2.5\pm 0.1$        & [99]   \\
				SAX J1750.8$-$2900      &  $5.2\pm 0.1$        & [9]   & 				XTE J1818$-$245         &  $3.5\pm 0.8$        & [61]  & 				AX J1841.0$-$0536       &  $3.2\pm 1.5$        & [100]  \\
				SAX J1808.4$-$3658      &  $3.5\pm 0.1$        & [27]  & 				V4641 Sgr               &  $6.2\pm 0.7$        & [62]  & 				AX J1845.0$-$0433       &  $6.4\pm 0.8$        & [80]   \\
				SAX J1810.8$-$2609      &  $4.9\pm 0.3$        & [28]  & 				MAXI J1820+070          &  $3.0\pm 0.3$        & [63]  & 				IGR 18483$-$0311        &  $2.8\pm 0.1$        & [97]   \\
				GX 13+1                 &  $7.0\pm 1.0$        & [29]  & 				XTE J1859+226           &  $16.5\pm 2.5$       & [64]  & 				XTE J1858+034           &  $10.5\pm 3.5$       & [101]  \\
				XMMU J181227.8$-$181234 &  $14.0\pm 2.0$       & [30]  & 				XTE J1908+094           &  $6.5\pm 3.5$        & [65]  & 				IGR J19140+0951         &  $3.6\pm 0.0$        & [97]   \\
				GX 17+2                 &  $9.8\pm 0.4$        & [9]   & 				Cyg X-1                 &  $2.2\pm 0.2$        & [66]  & 				XTE 1946+274            &  $9.5\pm 2.9$        & [102]  \\
				XTE J1814$-$338         &  $8.0\pm 1.6$        & [31]  & 				GRS 1915+105            &  $8.8\pm 1.8$        & [67]  & 				GRO J2058+42            &  $9.0\pm 1.3$        & [103]  \\
				2A 1822$-$371           &  $2.5\pm 0.5$        & [32]  & 				4U 1957+11              &  $9.0\pm 6.0$        & [68]  & 				Cep X-4                 &  $3.8\pm 0.6$        & [104]  \\
				GS 1826$-$238           &  $5.7\pm 0.2$        & [33]  & 				GS 2000+251             &  $2.7\pm 0.7$        & [17]  & 				4U 2206+543             &  $3.1\pm 0.1$        & [105]  \\
				Ser X-1                 &  $7.7\pm 0.9$        & [9]   & 				Cyg X-3                 &  $7.4\pm 1.1$        & [69]  & 				SAX J2239.3+6116        &  $4.9\pm 0.8$        & [106]  \\
				XTE J1856+053           &  $\geq 1.0$          & [34]  & 				V404 Cyg                &  $2.4\pm 0.1$        & [70]  & 				AS 386                  &  $2.4\pm 0.3$        & [107]  \\
				HETE J1900.1$-$2455     &  $3.6\pm 0.5$        & [9]   & 				2S 0053+604             &  $0.1\pm 0.0$        & [71]  & 				J05215658               &  $3.2\pm 0.8$        & [108]  \\
				4U 1905+000             &  $8.0\pm 1.0$        & [35]  & 				2S 0114+650             &  $4.5\pm 1.5$        & [72]  & 				             &      &  \\
            \bottomrule
        \end{tabular}
        \begin{tablenotes}
           \item References: [1] \citet{DeFalco17}, [2] \citet{Cornelisse02}, [3] \citet{Cowley82}, [4] \citet{Marino17}, [5] \citet{intZand03}, [6] \citet{GonzalezHernandez05}, [7] \citet{Heinz15}, [8] \citet{Hakala05}, [9] \citet{Galloway08}, [10] \citet{Bradshaw99}, [11] \citet{Chakrabarty98}, [12] \citet{Galloway06b}, [13] \citet{Reynolds97}, [14] \citet{Masetti02}, [15] \citet{intZand05}, [16] \citet{Lin20}, [17] \citet{Jonker04}, [18] \citet{Ling05}, [19] \citet{Hinkle06}, [20] \citet{Galloway03}, [21] \citet{Molkov05}, [22] \citet{Watts08}, [23] \citet{Pike21}, [24] \citet{Kuulkers00}, [25] \citet{Werner04}, [26] \citet{Chen10}, [27] \citet{Galloway06a}, [28] \citet{Natalucci00}, [29] \citet{Bandyopadhyay99}, [30] \citet{Goodwin19}, [31] \citet{Strohmayer03}, [32] \citet{Mason82}, [33] \citet{Chenevez16}, [34] \citet{Sala08}, [35] \citet{Chevalier90}, [36] \citet{Tomsick01}, [37] \citet{Gelino03}, [38] \citet{Cantrell10}, [39] \citet{Casares12}, [40] \citet{Gelino02}, [41] \citet{Gomez21}, [42] \citet{Gelino06}, [43] \citet{Wu16}, [44] \citet{MataSanchez21}, [45] \citet{Chauhan21}, [46] \citet{Casares09}, [47] \citet{Charles19}, [48] \citet{Stiele12}, [49] \citet{Orosz11}, [50] \citet{Seifina14}, [51] \citet{Homan06}, [52] \citet{Hjellming95}, [53] \citet{Kuulkers13}, [54] \citet{Heida17}, [55] \citet{dellaValle94}, [56] \citet{Chaty06b}, [57] \citet{Greiner96}, [58] \citet{Ratti12}, [59] \citet{CadolleBel07}, [60] \citet{Angelini03}, [61] \citet{CadolleBel09}, [62] \citet{MacDonald14}, [63] \citet{Atri20}, [64] \citet{Corral-Santana11}, [65] \citet{Chaty06a}, [66] \citet{Miller-Jones21}, [67] \citet{Reid14b}, [68] \citet{Maccarone20}, [69] \citet{McCollough16}, [70] \citet{Miller-Jones09}, [71] \citet{Megier09}, [72] \citet{Farrell08}, [73] \citet{Riquelme12}, [74] \citet{Negueruela99}, [75] \citet{Lyubimkov97}, [76] \citet{Reig05}, [77] \citet{Steele98}, [78] \citet{McBride06}, [79] \citet{Corbet84}, [80] \citet{Coleiro13}, [81] \citet{Israel00}, [82] \citet{Sadakane85}, [83] \citet{Janot-Pacheco81}, [84] \citet{Densham82}, [85] \citet{VanLeeuwen07}, [86] \citet{Chevalier98}, [87] \citet{Parkes80}, [88] \citet{Grindlay84}, [89] \citet{Oosterbroek99}, [90] \citet{Reynolds92}, [91] \citet{Filliatre04}, [92] \citet{Clark10}, [93] \citet{Chakrabarty02}, [94] \citet{Negueruela06}, [95] \citet{Lutovinov19}, [96] \citet{Martinez-Nunez17}, [97] \citet{Torrejon10}, [98] \citet{Segreto13}, [99] \citet{Casares05}, [100] \citet{Nespoli08}, [101] \citet{Tsygankov21}, [102] \citet{Wilson03}, [103] \citet{Wilson05}, [104] \citet{Bonnet-Bidaud98}, [105] \citet{Hambaryan22}, [106] \citet{Reig17}, [107] \citet{Khokhlov18}, [108] \citet{Thompson19} 
            \item $^\ast$: Distances that have no literature constraints.
        \end{tablenotes}
    \end{threeparttable}
    \end{adjustbox}
    \label{tab:d_lit_table}
\end{table*}

\onecolumn
\section{Notations and terminology}
\label{sec:notations}
This section lists salient notations and terminology used in this work.
\begin{itemize}
    \item [] Natal kick (kick): A general term used to refer to the impulsive acceleration imparted on a BH or NS.
    \item [] Blaauw kick: A kick due to mass ejection.
    \item [] Additional kick: Kicks that are caused by mechanisms other than mass ejection. 
    \item [] $\dlit$: Literature distance.
    \item [] $\dpost$: Parallax-inferred distance.
    \item [] $\gamma$: Systemic radial velocity. 
    \item [] $\mcomp$: Mass of the compact object.
    \item [] $\mnoncomp$: Mass of the non-degenerate companion. In XRBs, this refers to the donor mass.
    \item [] $\mtot$: Binary total mass. $\mtot = \mcomp + \mnoncomp$.
    \item [] $\Delta M$: Ejecta mass. This is the amount of mass that the helium core ejected at SN.
    \item [] $\sigma_v$: Scale parameter that characterises the Maxwellian distribution. For two-component Maxwellian model, $\sigma_{v,1}$ and $\sigma_{v,2}$ are used.
    \item [] $\frej$: Fraction of simulations that reject the null hypothesis of a test.
    \item [] $\ferr$: Fraction of additional error (see eq 8)
    \item [] $\fdiv$: Fraction of chains that diverge ($|\hat R - 1| / R \geq 0.01$)
    \item [] $\loo$: The weight value associated with the leave-one-out cross-validation (LOO).
    \item [] $\porb$: Present-day orbital period.
    \item [] $\porbi$: Orbital period at the instant of a SN.
    \item [] $\porbp$: Orbital period right after a SN.
    \item [] $\rhop$: The Pearson correlation coefficient.
    \item [] $\rhos$: The Spearman's rank correlation coefficient.
    \item [] $\hat R$: The Gelman-Rubin diagnostic used to test convergence of MCMC chains.
    \item [] $\vpec$: Peculiar velocity in general. This is the velocity of an object relative to Galactic rotation.
    \item [] $\vpecp$: Present-day peculiar velocity. The $\vpec$ calculated using the current astrometric parameters of the object.
    \item [] $\pkv$: Potential birth peculiar velocity. This is the $\vpec$ at Galactic disc ($z=0$).
    \item [] $V_r$: Relative orbital velocity at the instant of SN (see eq 10).
\end{itemize}
\bsp	
\label{lastpage}
\end{document}